\documentclass[fleqn,usenatbib]{mnras}

\usepackage{newtxtext,newtxmath}
\usepackage{deluxetable}

\usepackage[T1]{fontenc}

\DeclareRobustCommand{\VAN}[3]{#2}
\let\VANthebibliography\thebibliography
\def\thebibliography{\DeclareRobustCommand{\VAN}[3]{##3}\VANthebibliography}

\usepackage{graphicx}	
\usepackage{amsmath}	
\usepackage{orcidlink}  
\usepackage{threeparttable} 
\usepackage{lineno}     
\usepackage{soul}       
\usepackage{soul}       

\title[The `10 keV feature']{An investigation of the `10 keV feature' in the spectra of Accretion Powered X-ray Pulsars with \textit{NuSTAR}}

\author[Hemanth M et al.]{
Hemanth Manikantan,$^{1}$\thanks{E-mail: hemanthm@rri.res.in (RRI)}${\orcidlink{0000-0001-9404-1601}}$
Biswajit Paul,$^{1}$
Vikram Rana$^{1}$ ${\orcidlink{0000-0003-1703-8796}}$
\\
$^{1}$Astronomy and Astrophysics Dept., Raman Research Institute, CV Raman Avenue, Sadashivanagar, Bangalore 560080, India\\
}

\date{Accepted XXX. Received YYY; in original form ZZZ}

\pubyear{2023}

\begin{document}
\label{firstpage}
\pagerange{\pageref{firstpage}--\pageref{lastpage}}
\maketitle

\begin{abstract}
Some of the accreting X-ray pulsars are reported to exhibit a peculiar spectral feature at $\sim$10 keV, known as the ``\textit{10 keV feature}". The feature has been {characterized} as either an emission line or an absorption line, and its origin is unknown. It has been {found} in multiple observations of the same source by different observatories, but not all the observations of any particular source consistently showed the presence of it. In this work, we have carried out a systematic investigation for the presence of the ``\textit{10 keV feature}" using data from \textit{NuSTAR}, a low background spectroscopic observatory having uninterrupted wide band coverage on either side of 10 keV. We performed a systematic spectral analysis on 58 archival \textit{NuSTAR} observations of 30 bright X-ray pulsars. {The 3$-$79 keV spectral continua of these selected sources were fitted with a model chosen on the basis of its fitting quality in 3$-$15 keV and model simplicity, and then inspected for the presence of the ``\textit{10 keV feature}". Our analysis indicates the presence of such a feature in 16 out of 58 the \textit{NuSTAR} observations of 11 different sources and is fitted with a gaussian absorption model centred around 10 keV. Our analysis also suggests that such a feature could be wrongly detected if flare data is not analyzed separately from persistent emission.}
\end{abstract}



\section{Introduction} \label{sec:intro}

The spectrum of Accreting X-ray Pulsars (XRPs) generally have a smooth broad-band continuum with a high energy cut-off along with some additional features like narrow K-shell emission lines of iron, thermal soft excess, cyclotron absorption line, etc. (See \citealt{intro_white_1983} and \citealt{nagase_1989PASJ...41....1N}). The X-ray spectral continuum carries information related to the photon generation mechanism and physical conditions at the X-ray emission site. Generally, we parameterize the X-ray spectral continuum using a set of simplified empirical models. The spectral continuum of XRPs is generally described by a power-law having high energy cutoff \citep{nagase_1989PASJ...41....1N}. The lack of `perfect' theoretical models for representing the observed XRP spectra is owed to the complexity involved with the consistent calculation of the coupled radiation transfer and magneto-hydrodynamics equations of the accretion flow (the accretion flow dynamics and radiation transfer are inter-dependent) at the location of origin of emergent spectra (\citealt{intro_isenberg_1998}, \citealt{intro_orlandini_2006AdSpR..38.2742O}). In addition to the power-law continuum, the physical origins of most of the observed XRP spectral features are well established, viz., the narrow fluorescence emission lines especially from iron, Cyclotron Resonance Scattering Feature (CRSF) due to electron-cyclotron resonance in the super-strong magnetic field near the surface of neutron star (B\textsubscript{NS} $\sim10^{12}$ Gauss) (\citealt{intro_harding_2003pasb.conf..127H}, \citealt{intro_staubert_2019A&A...622A..61S}), photoelectric absorption of the soft X-rays by the interstellar medium and(or) circum-stellar/circum-binary matter, absorption edges, soft excess of thermal origin due to hard X-rays reprocessed by the inner accretion disk (\citealt{Paul_2002_softexcess}, \citealt{Hickox_2004}), etc. 

High signal-to-noise ratio (SNR) data and the broad-band spectral coverage delivered by present-day X-ray space observatories facilitate detailed examination of characteristic features in XRP spectra. In this work, we explore the mysterious feature found in XRP spectra, introduced as a distinct spectral feature by \cite{into_coburn_2002} called as ``\textit{The feature at 10 keV}" or the ``\textit{10 keV feature}" (hereafter \textit{TKF}). \textit{TKF} appears as a bump or depression or wiggle around 10 keV in the residuals to the best fit spectral model and is generally modelled as a gaussian emission or gaussian absorption or a combination of these two models. This spectral feature has been observed in multiple XRP sources across different observatories viz., \textit{Ginga} (\citealt{intro_mihara_2002}), \textit{RXTE} \citep{into_coburn_2002}, \textit{BeppoSAX} \citep{xtej1946_doroshenko_2017_refId0}, \textit{Suzaku} \citep{intro_elizabeth_2010} and \textit{NuSTAR} \citep{intro_furst_2014AAS}. The appearance of \textit{TKF} in spectra from multiple observatories rules out the possibility of its instrumental origin. The most suspected reason for the origin of \textit{TKF} is the common usage of over-simplified and imperfect phenomenological spectral continuum models to describe the high statistics X-ray spectral data from the sensitive modern observatories (See \citealt{into_coburn_2002}, \citealt{herx1_vasco_2013_refId0}, and \citealt{intro_doroshenko_2017ASS}). Reports of detection of \textit{TKF} are sporadic, which impacts the selection of sources in an organized manner for its study. For instance, \textit{TKF} may be reported in the spectrum of a particular source with data from a specific instrument (observatory) but may not be reported in the spectrum from another instrument. No particular pattern has been observed in the occurrence of \textit{TKF} until now.

The broad-band (3$-$79 keV) focussing X-ray telescope \textit{Nuclear Spectroscopic Telescope Array} (\textit{NuSTAR}) is an instrument well suited to study the \textit{TKF}, as it covers the 10 keV spectral region by a factor of a few on either side with good spectral resolution and low background levels. Additionally, there is the availability of a large archival \textit{NuSTAR} spectral data of multiple XRPs. We selected some of the bright XRPs for which \textit{NuSTAR} archival observations were available and then surveyed the available literature for spectral analysis studies of these chosen sources with different instruments. All the previous report(s) of detection of \textit{TKF} in these sources were noted down. We then tried to model the available \textit{NuSTAR} spectra of these XRPs with a suitable continuum model and check for the presence of \textit{TKF} in them.

Since \cite{into_coburn_2002} presented \textit{TKF} as a distinct spectral feature, it has been extensively reported in spectral analysis of XRPs. Generally, an extra model component with a gaussian absorption profile or a gaussian emission profile or a combination of the two is used to account for the residuals remaining at $\sim10$ keV following the spectral fits. \cite{into_coburn_2002} also pointed out the residuals resembling \textit{TKF} that were visible in spectral fits to \textit{Ginga} observations of XRPs \citep{intro_mihara_2002}. We surveyed the literature containing spectral analysis of XRPs in the following manner. First, the X-ray observatories covering the spectral band 5-20 keV and having an adequate spectral resolution to resolve \textit{TKF} (which usually turns up as a broad feature) are screened. This confines the survey to the Large Area Counter instrument on \textit{Ginga} (\textit{Ginga}-LAC) spanning 1.5--37 keV with $\sim18\%$ spectral resolution \citep{Ginga_1989PASJ...41..345T}, the Proportional Counter Array (\textit{RXTE}-PCA) covering 2--60 keV with $\sim18\%$ spectral resolution and High Energy X-ray Timing Experiment (\textit{RXTE}-HEXTE) covering 15--250 keV with $\sim15\%$ spectral resolution onboard \textit{Rossi X-Ray Timing Explorer} \citep{RXTE_996SPIE.2808...59J}, the High Pressure Gas Scintillation Proportional Counter (HPGSPC) \citep{bepposax_hpgspc_refId0} and Phoswich Detection System (PDS) \citep{bepposax-pds} instruments on-board \textit{Beppo-SAX} \citep{bepposax_1997A&AS..122..299B} covering 4--120 keV, the Joint European X-Ray Monitor (JEM-X) instrument on-board \textit{INTEGRAL} covering 4--35 keV with $\sim10\%$ spectral resolution \citep{jemx_2003A&A...411L.231L}, \textit{Suzaku}-XIS (X-ray Imaging Spectrometer) with 0.2--12 keV coverage and  $\sim2\%$ spectral resolution, and \textit{Suzaku}-HXD (Hard X-ray Detector) with 10--70 keV coverage and spectral resolution of 3.0 keV (\citealt{Suzaku_XIS_2007PASJ...59S..23K} and \citealt{Suzaku_HXD2007PASJ...59S..35T} respectively), \textit{NuSTAR} \citep{itro_nustar_2013ApJ} and AstroSAT-\textit{LAXPC} (Large Area X-ray Proportional Counters) covering 3--80 keV with $\sim10\%$ spectral resolution \citep{astrosat_laxpc_2017ApJS..231...10A}. For the selected bright XRPs, we searched for the reported spectral analysis results using data from the aforementioned X-ray observatories. 

\begin{figure}
    \centering
    \includegraphics[width=\linewidth]{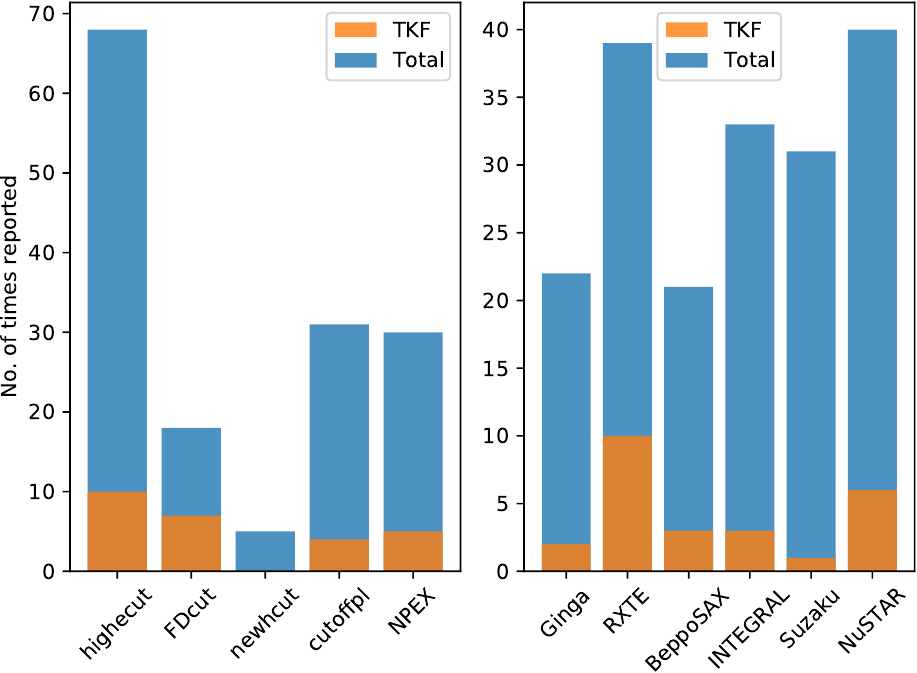}
    \caption{The bar chart shows the specifics of the performed survey (Table~\ref{tab:survey}). The distribution of reports of \textit{TKF} for different continuum models, and from different observatories\protect\footnotemark are shown.}
    \label{fig:survey}
\end{figure}

A summary of the published results in the context of the \textit{TKF} is given in Table~\ref{tab:survey}. Out of the 30 selected sources given in the table, \textit{TKF} has been reported at least once in 9 sources viz., Vela X--1, Her X--1, XTE J1946+274, 4U 1907+09, 4U 1538-52, Cep X--4, SMC X--1, GX 304--1 and EXO 2030+375. A graphical depiction of the survey is shown in Fig.~\ref{fig:survey}. It is most commonly observed with \textit{RXTE}-PCA observatory and for power-law continuum having a high energy cutoff \texttt{highecut}\footnote{The model is explained in Section. \ref{sec:tas}} (See Fig.~\ref{fig:survey}). However, analysis of the same data by different groups also sometimes has given contrasting results regarding \textit{TKF} (See for example \citealt{cepx4_vybornov2017luminosity}, \citealt{cepx4_furst_2015ApJ...806L..24F} and \citealt{cepx4_bhargava_2019MNRAS}).

\begin{table*}
    \centering
    \caption{For each source spectrum from the particular observatory, the spectral model used to fit the continuum in the literature is given. If a \textit{TKF} was used for the spectral fit, then the cell is given in bold fonts. Whether the \textit{TKF} is modelled like a dip (gaussian absorption) or a hump (gaussian emission) is shown with letters A (absorption) and E (emission), respectively. If the authors have not reported the presence of \textit{TKF} for any observation, but residuals resembling \textit{TKF} are visible in their best fit model, it is indicated with the letter R (residuals).}
    \vspace{-6pt}
    \tiny
    \begin{tabular}{lcccccccc}
    \hline
    \hline
    { Source} &{ Ginga} &{ RXTE} &{ BeppoSAX} 
    &{INTEGRAL} &{Suzaku} &{NuSTAR} &{Astrosat} &{NuSTAR} ({\textit{This work}})\\
    \hline
    \bf Her X--1 &PL$^{4}$ &HEC$^{14}$ &BPLHEC$^{13}$ &HEC$^{12}$ &NPEX$^{15}$ &{ \bf HEC$^{11}$ (A)} &HEC$^{110}$ &{NPEX, HEC}\\
            &NPEX$^{71}$ &{ \bf HEC$^{16}$ (E)} &\nodata &\nodata &\nodata &\nodata &\nodata &{\bf FDC (A)}\\
            &\nodata & HEC$^{26}$ (R) &\nodata &\nodata &\nodata &\nodata &\nodata &\nodata\\
    \bf Vela X--1 &NPEX$^4$ &{ \bf NPEX$^5$ (A)} &{ \bf NPEX$^1$ (E)} &CPL$^6$ &compTT$^9$ &{ \bf FDC$^8$ (A)} &\nodata &{ \bf compTT (A) }\\
                    &PL$^{3}$  &HEC$^{26}$    &NPEX$^{2}$   &CPL$^{7}$    &NPEX$^{10}$   &\nodata    &\nodata &{CPL, compTT, HEC}\\
                    &\nodata &FDC$^{52}$    &compTT$^{51}$ &\nodata    &\nodata    &\nodata   &\nodata &\nodata\\
    \bf XTE J1946+274 &\nodata &{ \bf FDC$^{105}$ (A)} &{ \bf FDC$^{112}$ (E)} &{ \bf FDC$^{105}$ (A)} &HEC$^{84}$ &{ \bf HEC,FDC,NPEX,compTT$^{34}$ (A)} &\nodata &{ \bf HEC (A)}\\
                &\nodata &NPEX$^{111}$ &\nodata    &\nodata    &\nodata   &\nodata &\nodata &\nodata\\
                &\nodata &HEC$^{26}$ &\nodata    &\nodata    &\nodata   &\nodata &\nodata &\nodata\\
    KS 1947+300 &\nodata &compTT$^{69}$ &compTT$^{68}$ &HEC$^{58}$ &CPL$^{113}$ &CPL$^{67}$ &\nodata &{ HEC, NPEX}\\
    \bf 4U 1907+09 &{ \bf NPEX$^{4}$ (A)} &{ \bf HEC$^{26}$ (A)} &HEC$^{29}$ &CPL$^{30}$ &{ \bf FDC,HEC,NPEX$^{27}$ (E)} &{ \bf HEC, CPL$^{157}$ (A)} &HEC$^{81}$ &{ \bf compTT (A)}\\
                &\nodata &\nodata &\nodata &\nodata &NPEX$^{84}$ &\nodata &\nodata &\nodata\\
    \bf 4U 1538-52& NPEX$^4$ (R) &HEC$^{46}$ &HEC$^{48}$ &{ \bf CPL$^{30}$ (E)} &\nodata &HEC$^{19}$ &FDC$^{17}$ &{ NPEX}\\
                &NPEX$^{71}$ &{ \bf CPL$^{18}$ (A)} &\nodata &\nodata &\nodata &\nodata &\nodata &\nodata\\
                &HEC$^{47}$ &{ \bf HEC$^{26}$ (A)} &\nodata &\nodata &\nodata &\nodata &\nodata &\nodata\\
    \bf Cep X--4 &NPEX$^{71}$ &{ \bf FDC$^{97}$ (E)} &\nodata &\nodata &NPEX$^{95}$ &FDC$^{93}$ &\nodata &{ compTT, NPEX}\\
            &HEC$^{94}$ &\nodata &\nodata &\nodata &\nodata &FDC$^{96}$ &\nodata &\nodata\\
            &HEC$^{114}$ &\nodata &\nodata &\nodata &\nodata &{ \bf FDC$^{98}$ (E)} &\nodata &\nodata\\
    4U 1626-67 &\nodata &HEC$^{26}$ &HEC$^{51}$ &\nodata &NPEX$^{66}$ &HEC$^{83}$ &\nodata &{ FDC}\\
                &\nodata &\nodata &HEC$^{64}$ &\nodata &HEC$^{82}$ &\nodata &\nodata &\nodata\\
                &\nodata &\nodata &CPL$^{65}$ &\nodata &NPEX$^{80}$ &\nodata &\nodata &\nodata\\
    SMC X--2 &\nodata &\nodata &\nodata &\nodata &\nodata &NPEX$^{36}$ &\nodata &{ compTT, NPEX}\\
    IGR J17544--2619 &\nodata &\nodata &\nodata &\nodata &\nodata &nthcomp$^{99}$ &\nodata &{ NPEX}\\
    IGR J16393--4643 &\nodata &\nodata &\nodata &\nodata &HEC$^{104}$ &CPL$^{103}$ &\nodata &{ \bf NPEX (A)}\\
    2S 1553--542 &\nodata &BPL$^{115}$ &\nodata &\nodata &\nodata &CPL$^{116}$ &\nodata &{ FDC}\\
    RX J0520.5--6932 &\nodata &\nodata &\nodata &\nodata &\nodata &FDC$^{28}$ &\nodata &{ \bf CPL,NPEX (A)}\\
    Cen X--3 &HEC$^{33}$ &HEC$^{26}$ (R) &compTT$^{51}$ &PCH$^{54}$ &FDC$^{84}$ &NHC$^{50}$ &\nodata &{ \bf HEC (A)}\\
            &\nodata &\nodata &PCH$^{32}$ (R) &PCH$^{55}$ &NHC$^{50}$ &\nodata &\nodata &\nodata\\
            &\nodata &\nodata &HEC$^{63}$ &\nodata &\nodata &\nodata &\nodata &\nodata\\
    GX 301--2 &NPEX$^{4}$ &HEC$^{25}$ &NHC$^{23}$ &FDC$^{22}$ &FDC$^{21}$ &NPEX$^{20}$ &\nodata &{ NPEX}\\
            &\nodata &HEC$^{26}$ &\nodata &FDC$^{24}$ &\nodata &\nodata &\nodata &\nodata\\
            &\nodata &\nodata &\nodata &HEC$^{54}$ &\nodata &\nodata &\nodata &\nodata\\
    XTE J1829--098 &\nodata &\nodata &\nodata &\nodata &\nodata &CPL$^{43}$ &\nodata &{ NHC}\\
    V 0332+53 &HEC$^{39}$ &CPL$^{38}$ &\nodata &compTT$^{37}$ &\nodata &compTT$^{56}$ &\nodata &{ HEC, NHC, NPEX, compTT}\\
            &NPEX$^{4}$ &CPL$^{41}$ &\nodata &CPL$^{40}$ &\nodata &compTT$^{57}$ &\nodata &\nodata\\
            &NPEX$^{71}$ &\nodata &\nodata &HEC$^{42}$ &\nodata &\nodata &\nodata &\nodata\\
            &\nodata &\nodata &\nodata &HEC$^{54}$ &\nodata &\nodata &\nodata &\nodata\\
    XTE J1858+034 &\nodata &HEC$^{31}$ &\nodata &HEC$^{32}$ &\nodata &compTT$^{33}$ &\nodata &{ NPEX}\\
    4U 1700-67 &HEC$^{115}$ &2PL$^{117}$ &HEC$^{141}$ &\nodata &NPEX$^{116}$ &NHC$^{118}$ &\nodata &{ \bf NPEX (A)}\\
        &\nodata &\nodata &\nodata &\nodata &NHC$^{116}$ &\nodata &\nodata &\nodata\\
    LMC X--4 &HEC$^{120}$ &HEC$^{123}$ &HEC$^{122}$ &HEC$^{125}$ &HEC$^{127}$ &compTT$^{128}$ &\nodata &{ \bf NPEX (A)}\\
    &NPEX$^{4}$ &\nodata &PL$^{124}$ &HEC$^{126}$ &\nodata &\nodata &\nodata &{NPEX, FDC, compTT}\\
    &HEC$^{121}$ &\nodata &\nodata &\nodata &\nodata &\nodata &\nodata &\nodata\\
    IGR J17329-2731 &\nodata &\nodata &\nodata &PL$^{119}$ &\nodata &HEC$^{119}$ &\nodata &{NPEX}\\
    \bf SMC X--1 &HEC$^{133}$ &HEC$^{129}$ &HEC$^{132}$ &\nodata &CPL$^{134}$ &{ \bf FDC$^{130}$ (E)} &\nodata &{ HEC, FDC}\\
    &\nodata &\nodata &CompTT$^{132}$ &\nodata &\nodata &NPEX$^{131}$ &\nodata &\nodata\\
    GRO J1008--57 &\nodata &CPL$^{138}$ &\nodata &CPL$^{136}$ &NPEX$^{135}$ &NPEX$^{137}$ &\nodata &{ NPEX}\\
    &\nodata &\nodata &\nodata &CPL$^{138}$ &CPL$^{138}$ &2compTT$^{140}$ &\nodata &\nodata\\
    \bf GX 304--1 &\nodata &HEC$^{146}$ &\nodata &HEC$^{145}$ &FDC$^{141}$ &2compTT$^{144}$ &\nodata &{ PL}\\
    &\nodata &FDC,NPEX$^{141}$ &\nodata &CPL$^{147}$ &NPEX$^{141}$ &\nodata &\nodata &\nodata\\
    &\nodata &{ \bf HEC,CPL$^{148}$ (E)} &\nodata &\nodata &NPEX$^{142}$ &\nodata &\nodata &\nodata\\
    &\nodata &\nodata &\nodata &\nodata &HEC$^{143}$ &\nodata &\nodata &\nodata\\
    1A 0535+26 &\nodata &CPL$^{85}$ &\nodata &CPL$^{85}$ &CPL$^{85}$ &CPL$^{86}$ &\nodata &{ FDC, PL, 2compTT}\\
    &\nodata &HEC$^{87}$ &\nodata &HEC$^{87}$ &NPEX$^{84}$ &2compTT$^{89}$ &\nodata &\nodata\\
    &\nodata &PL$^{88}$ &\nodata &HEC$^{91}$ &\nodata &CPL$^{90}$ &\nodata &\nodata\\
    GRO J2058+42 &\nodata &\nodata &\nodata &\nodata &\nodata &FDC$^{153}$ &FDC$^{154}$ &{ \bf compTT (A)}\\
            &\nodata &\nodata &\nodata &\nodata &\nodata &compTT$^{155}$ &\nodata &{NHC}\\
    1E 1145.1--6141 &\nodata &HEC$^{74}$&\nodata &CPL$^{72}$ &\nodata &CPL$^{73}$ &\nodata &{ CPL}\\
    OAO 1657--415 &HEC$^{70}$ &\nodata &CPL$^{45}$ &HEC$^{62}$ &HEC$^{49}$ &CPL$^{59}$ &PL$^{61}$ &{ CPL}\\
            &\nodata &\nodata &\nodata &\nodata &HEC$^{53}$ &CPL$^{60}$ &\nodata &\nodata\\
    \bf EXO 2030+375 &\nodata &{\textbf{HEC}$^{76}$ \bf(A)} &\nodata &{\textbf{HEC}$^{77}$ \bf(A,E)} &HEC$^{78}$ &{\textbf{CPL}$^{75}$ \bf(A)} &HEC$^{79}$ &{ NPEX}\\
            &\nodata &HEC$^{92}$ &\nodata &compTT$^{106}$ &HEC$^{101}$ &HEC$^{79}$ &\nodata &{ \bf CPL (A), NPEX (A)}\\
            &\nodata &HEC$^{156}$ &\nodata &HEC$^{107}$ &\nodata &HEC$^{100}$ &\nodata &\nodata\\
    IGR J19294+1816 &\nodata &PL$^{151}$ &\nodata &PL$^{44}$ &\nodata &compTT$^{109}$ &HEC$^{108}$ &{ NPEX, FDC}\\
    &\nodata &PL$^{152}$ &\nodata &\nodata &\nodata &\nodata &\nodata &\nodata\\
    \hline
    \hline
    \end{tabular}
    \label{tab:survey}
    \begin{tablenotes}
        \tiny
        \item \texttt{NOTE--} PL: power-law, 2PL: double power-law, HEC: PL with \texttt{highecut}, FDC: PL with Fermi-Dirac cutoff, NHC: \texttt{newhcut}, CPL: \texttt{cutoffpl}, NPEX: Negative Positive Exponential, BPL: Broken power-law, BPLHEC: Broken power-law with \texttt{highecut}, compTT: comptonization model, 2compTT: double compTT   \hspace{20pt}\texttt{NOTE--} 4U 1700--37 is not an established XRP, but it is included because of the reported CRSF. 
        \item \texttt{References--} $^{1}$\cite{velax1_barbera_refId0}  $^{2}$\cite{velax1_Orlandini:1997hj}
      $^{3}$\cite{velax1_choi_1996ApJ...471..447C}  $^{4}$\cite{intro_mihara_2002}  $^{5}$\cite{velax1_kreykenbohm_2002_refId0}  $^{6}$\cite{velax1_wang_2014MNRAS.440.1114W}  $^{7}$\cite{velax1_schanne_2007ESASP.622..479S}  $^{8}$\cite{velax1_Furst1_2013}  $^{9}$\cite{velax1_maitra_2013ApJ...763...79M}  $^{10}$\cite{velax1_odaka_2013ApJ...767...70O}  $^{11}$\cite{herx1_furst1_2013ApJ}  $^{12}$\cite{herx1_klochkov_2007ESASP.622..461K}  $^{13}$\cite{herx1_dalfiume_1998_Fiume:1997ag}  $^{14}$\cite{herx1_gruber_2001ApJ...562..499G}  $^{15}$\cite{herx1_enoto_2008PASJ...60S..57E}  $^{16}$\cite{herx1_vasco_2013_refId0}  $^{17}$\cite{4u1538-52_varun_2019MNRAS.484L...1V}  $^{18}$\cite{4u1538-52_roca_2009_refId0}  $^{19}$\cite{4u1538-52_hemphill_2019ApJ...873...62H}  $^{20}$\cite{301-2_furst_2018A_refId0}  $^{21}$\cite{301-2_Suchy_2012}  $^{22}$\cite{3012_doroshenko_2010_refId0} $^{23}$\cite{3012_barbera_2005_refId0}  $^{24}$\cite{301-2_kreykenbohm_2004ESASP.552..333K}  $^{25}$\cite{3012_umukherjee_2004_refId0}  $^{26}$\cite{into_coburn_2002}  $^{27}$\cite{4u1907+09_rivers_2010ApJ...709..179R}
      $^{28}$\cite{0520_tendulkar_2014ApJ...795..154T}
      $^{29}$\cite{4u1907+09_cusumano_2000AdSpR..25..409C}  $^{30}$\cite{4u1907+09_hemphil_2013ApJ...777...61H}  $^{31}$\cite{uddipan_1858034_2006JApA...27...25M}  $^{32}$\cite{doroshenko_1858034_2008ARep...52..138D}  $^{33}$\cite{cenx3_nagase_1992ApJ...396..147N} $^{34}$\cite{devaraj_xtej1946} $^{36}$\cite{SMCX-2_jaiswal_2016MNRAS.461L..97J} $^{37}$\cite{v0332_zachary_10.1093/mnras/stx384}   $^{38}$\cite{v0332_lutovinov_2015MNRAS.448.2175L}   $^{39}$\cite{v0332_makishima_1990ApJ...365L..59M}  $^{40}$\cite{v0332_kreyken_2006ESASP.604..275K}    $^{41}$\cite{v0332_tsygankov_10.1111/j.1365-2966.2006.10610.x}  $^{42}$\cite{v0332_ferrigno_2016} $^{43}$\cite{shtykovsky_2019MNRAS.482L..14S} $^{44}$\cite{igrj19294_roy_jayashree} $^{45}$\cite{oao_1657_bepposax} $^{46}$\cite{4u1538-52_mukherjee_2007astro.ph..2142M} $^{47}$\cite{4u1538-52_clark_1990ApJ...353..274C}  $^{48}$\cite{4u1538-52_robba_2000AIPC..510..213R} $^{49}$\cite{oao1657_suzaku_gaurava}   $^{50}$\cite{cenx3_gunjan_2021MNRAS.500.3454T}  $^{51}$\cite{intro_doroshenko_2017ASS} $^{52}$\cite{velax1_kreykenbohm1998} $^{53}$\cite{oao1657_suzaku_pragati}  $^{54}$\cite{filippova_2005AstL...31..729F} $^{55}$\cite{burderi_2000ApJ...530..429B}  $^{56}$\cite{v0332_vybornov_2018}    $^{57}$\cite{v0332_doroshenko_2017MNRAS.466.2143D}   $^{58}$\cite{ks1947+300_tsygankov_2005AstL...31...88T} $^{59}$\cite{oao1657_nustar_enzo} $^{60}$\cite{oao1657_nustar_prince} $^{61}$\cite{oao1657_astrosat_gaurava} $^{62}$\cite{oao1657_integral} $^{63}$\cite{barbera_2004ESASP.552..337L} $^{64}$\cite{1626_orlandini_1998ApJ...500L.163O} $^{65}$\cite{1626_owens1997complex} $^{66}$\cite{1626_iwakiri2012possible}  $^{67}$\cite{ks1947+300_furst_2014ApJ...784L..40F}  $^{68}$\cite{ks1947+300_naik_2006ApJ...647.1293N}  $^{69}$\cite{ks1947+300_galloway_2004ApJ...613.1164G} $^{70}$\cite{oao1657_ginga} $^{71}$\cite{herx1_makishima_1999ApJ...525..978M} $^{72}$\cite{1e1145_integral_2008AA...479..533F} $^{73}$\cite{1e1145_nustar_bcp} $^{74}$\cite{1e1145_rxte_Ray_2002} $^{75}$\cite{exo2030_bcp} $^{76}$\cite{exo2030_RXTE_Wilson_2008} $^{77}$\cite{exo2030_integral} $^{78}$\cite{exo2030_suzaku_Naik_2015} $^{79}$\cite{exo2030_astrosat_nustar_gjaiswal} $^{80}$\cite{4u1626-67_Iwakiri_2019}  $^{81}$\cite{4u1907+09_varun_2019ApJ...880...61V} $^{82}$\cite{1626_camero20124u} $^{83}$\cite{1626_d'ai2017broad} $^{84}$\cite{Maitra_2013-xtej1946}
      $^{85}$\cite{1a0535_caballero2013double}
      $^{86}$\cite{1a0535_ballhausen_2017}
      $^{87}$\cite{caballero20070535+}
      $^{88}$\cite{1a0535_Rothschild_2013}
      $^{89}$\cite{1a0535_tsygankov2019MNRAS.487L..30T}
      $^{90}$\cite{1a0535_Mandal_2022}
      $^{91}$\cite{1a0535_Sartore_2015} $^{92}$\cite{exo2030_rxte_prahlad}
      $^{93}$\cite{cepx4_furst_2015ApJ...806L..24F} $^{94}$\cite{cepx4_koyama_1991ApJ...366L..19K} $^{95}$\cite{cepx4_jaisawal2015detection} $^{96}$\cite{cepx4_bhargava_2019MNRAS} $^{97}$\cite{cepx4_mcbride2007cyclotron} $^{98}$\cite{cepx4_vybornov2017luminosity} $^{99}$\cite{17544_bhalerao_2015MNRAS.447.2274B} $^{100}$\cite{exo2030_nustar_furst} $^{101}$\cite{exo2030_suzaku_Naik_2013}  $^{103}$\cite{16393_bodaghee_2016ApJ...823..146B}  $^{104}$\cite{16393_Islam_2015MNRAS.446.4148I}  $^{105}$\cite{xtej1946p274_muller_2021_refId0} $^{106}$\cite{exo2030_integral_camero} $^{107}$\cite{exo2030_integral_nunez} $^{108}$\cite{igrj19294_astrosat_raman} $^{109}$\cite{igtj19294_nustar_tsygankov}  $^{110}$\cite{herx1_bala_2020MNRAS.497.1029B}  $^{111}$\cite{xtej1946+274_heindl_2001ApJ...563L..35H}  $^{112}$\cite{xtej1946_doroshenko_2017_refId0}  $^{113}$\cite{ks1947_ballhausen_2016_refId0}  $^{114}$\cite{cepx4_mihara_1991ApJ...379L..61M}  $^{115}$\cite{1553_pahari_2012MNRAS.423.3352P}  $^{116}$\cite{1553_tsygankov_2016MNRAS.457..258T} $^{116}$\cite{4u1700_116_2015MNRAS.448..620J} $^{117}$\cite{4u1700_117_2016ApJ...821...23S} $^{118}$\cite{4U1700_118_2020MNRAS.493.3045B} $^{119}$\cite{igrj17329_119_bozzo2018igr} $^{120}$\cite{lmcx4_120_levine1991lmc} $^{121}$\cite{lmcx4_121_woo1996orbital} $^{122}$\cite{lmcx4_122_la20010} $^{123}$\cite{lmcx4_123_naik2003spectral} $^{124}$\cite{lmcx4_124_naik2004timing} $^{125}$\cite{lmcx4_125_lutovinov2004variability} $^{126}$\cite{lmcx4_126_tsygankov2005long} $^{127}$\cite{lmcx4_127_hung2010suzaku} $^{128}$\cite{lmcx4_128_shtykovsky2017nustar} $^{129}$\cite{smcx1_129_inam2010analysis} $^{130}$\cite{smcx1_130_pike2019observing} $^{131}$\cite{smcx1_131_brumback2020modeling} $^{132}$\cite{smcx1_132_naik2004bepposax} $^{133}$\cite{smcx1_133_woo1995wind} $^{134}$\cite{smcx1_134_pradhan2020superorbital} $^{135}$\cite{groj1008_135_yamamoto2014firm} $^{136}$\cite{groj1008_136_wang2014temporal} $^{137}$\cite{groj1008_137_bellm2014confirmation} $^{138}$\cite{groj1008_138_kuhnel2013gro} $^{139}$\cite{groj1008_139_naik2011suzaku} $^{140}$\cite{groj1008_140_lutovinov2021srg} $^{141}$\cite{gx304_141_yamamoto2011discovery} $^{142}$\cite{gx304_142_jaisawal2016suzaku} $^{143}$\cite{gx304_143_pradhan2021comprehensive} $^{144}$\cite{gx304_144_tsygankov2019dramatic} $^{145}$\cite{gx304_145_klochkov2012outburst} $^{146}$\cite{gx304_146_devasia2011timing} $^{147}$\cite{gx304_147_malacaria2015luminosity} $^{148}$\cite{gx304_148_rothschild2017discovery} $^{149}$\cite{4u1700_149_haberl1992ginga} $^{150}$\cite{4u1700_150_reynolds1999bepposax} $^{151}$\cite{igrj19294_rodriguez} $^{152}$\cite{igrj19294_roy_jayashree} $^{153}$\cite{groj2058_nustar_kabiraj2020broad} $^{154}$\cite{groj2058_astrosat_kallol} $^{155}$\cite{groj2058_nustar_molkov} $^{156}$\cite{exo2030_rxte_reig}
      $^{157}$\cite{tobrej2023high_1907nustar}
      
    \end{tablenotes}
\end{table*}

\footnotetext{Remark: Reports with residuals resembling \textit{TKF} visible in the best fit model were counted for the bar chart on the right-hand side of the figure.}

Many of the results mentioned before include combined spectral analysis using data from more than one instrument (except \textit{Ginga}-LAC, \textit{NuSTAR}, and \textit{AstroSat}-LAXPC), often with a gap around 10-15 keV which is above the energy range of the imaging instruments and below the energy range of the scintillator detectors. We have analyzed archival data of { 58} \textit{NuSTAR} observations of { 30} bright XRPs to probe the \textit{TKF}. The \textit{NuSTAR} observations cover either side of the 10 keV with a single instrument, thus reducing systematic effects of cross-normalization of different instruments.

We have {organized} this paper {in} the following way: In Section~\ref{instr}, a brief description of \textit{NuSTAR} instrument, the Data reduction steps followed, and a {log} of the \textit{NuSTAR} observations that we have used for spectral analysis in this work are given. Section~\ref{sec:tas} explains the details of the spectral analysis performed. The results of the analysis and its discussion are given in Section~\ref{disc}.

\section{Instrument, Observations and data-reduction} \label{instr}

\textit{NuSTAR} is an X-ray imaging spectroscopic observatory that covers the energy band by a factor of a few on either side of 10 keV with the same telescope and detector, making it an ideal instrument to probe the \textit{TKF}. \textit{NuSTAR} is the first hard X-ray focusing telescope and it provides spectral coverage in the 3$-$79 keV bandpass. It uses a pair of co-aligned Wolter-I {telescopes} to focus the source X-rays separately onto a pair of CdZnTe pixel detector arrays placed at 10 m focal length, referred to as FPMA (Focal Plane Module A) and FPMB. Each FPM detector array has four $2\times2\times0.2$ cm detectors. The spectral resolution below 50 keV is 400 eV FWHM, which is excellent for studying the relatively broad \textit{TKF} feature (For detailed information on the \textit{NuSTAR} observatory, refer \citealt{itro_nustar_2013ApJ}). In addition, the uninterrupted broadband coverage from the instrument aids in modelling the spectral continuum in the best possible manner.

We followed the standard data reduction steps from the \textit{NuSTAR} Data Analysis Software Guide\footnote{\url{https://heasarc.gsfc.nasa.gov/docs/nustar/analysis/nustar_swguide.pdf}} to extract the calibrated and screened source and background products. The task {\small nupipeline v0.4.6} of {\small HEASoft v6.25}\footnote{\url{https://heasarc.gsfc.nasa.gov/docs/software/heasoft/}} was first run on the raw data to obtain the filtered and calibrated data. Calibration database {\small CALDB v1.0.2}\footnote{\url{https://heasarc.gsfc.nasa.gov/docs/heasarc/caldb/nustar/}} was used for this purpose. Using {\small DS9}\footnote{\url{https://sites.google.com/cfa.harvard.edu/saoimageds9}}, circular source and background region files are generated. The task \texttt{nuproducts} was then run to extract the source and background energy spectra as well as the Redistribution Matrix File (RMF) and Auxiliary Response File (ARF). The extracted source spectrum was checked against the background spectrum, and the background-dominated energy range was excluded from the spectral analysis. A set of two spectra from two detector modules FPMA and FPMB were obtained for each observation. The spectra were then re-binned with the tool {\small ftgrouppha}\footnote{\url{https://heasarc.gsfc.nasa.gov/lheasoft/help/ftgrouppha.html}} using the optimal binning scheme by \cite{2016kaastra}\footnote{We verified that the best fit spectral parameters have no significant variation when using this binning scheme, compared to grouping with a minimum of 25 counts per bin.}. Both the FPMA and FPMB spectra are fitted together in {\small XSPEC v12.10.1} \citep{xspec_Arnaud_1996ASPC..101...17A}, with an instrument relative normalization constant parameter left free to vary and all other model parameters tied together across both the spectra. 

A catalogue of the sources and corresponding \textit{NuSTAR} observation IDs used in this work is given in Table~\ref{tab:obs_catalogue}.

\begin{table*}
    \caption{\textit{NuSTAR} Observations catalogue with the details of the source and observation IDs used in this work. The observation IDs having flare in the light curve have the letter `\textit{F}' super-scripted.}
    \footnotesize
    \centering
    \begin{tabular}{clc|llc|c}
        \hline
        \hline
         Obs. Sn. &Source &\textit{NuSTAR} Obs. ID &start (DATE-OBS) &stop (DATE-END) &exposure &FPMA count rate \\
          & & & & &(ks) &(cts s$^{-1}$)\\
         \hline
         1 &Her X--1 &30002006002 &2012-09-19 08:26:07 &2012-09-20 01:06:07 &27.7 &34\\
         2 &	    &30002006005 &2012-09-22 04:26:07 &2012-09-22 18:36:07 &21.9 &90\\
         3 &        &10202002002 &2016-08-20 04:31:08 &2016-08-21 06:21:08 &36.6 &80\\
         4 &Vela X--1&10002007001 &2012-07-09 13:16:07 &2012-07-09 23:06:07 &10.8 &135\\
         5 &        &30002007002 &2013-04-22 12:16:07 &2013-04-22 20:11:07 &7.14 &67\\
         6 &        &30002007003 &2013-04-22 20:11:07 &2013-04-23 19:06:07 &24.5 &43\\
         7 &	    &90402339002$^{F}$ &2019-01-10 04:11:09 &2019-01-11 02:26:09 &36.0 &41\\
         8 &XTE J1946+274 &90401328002 &2018-06-24 11:26:09 &2018-06-25 15:36:09 &47.2 &36\\
         9 &KS 1947+300 &80002015002 &2013-10-21 18:36:07 &2013-10-22 05:46:07 &18.36 &72\\
         10 &           &80002015004 &2013-11-22 21:26:07 &2013-11-23 09:31:07 &18.59 &82\\
         11 &           &80002015006 &2013-12-09 15:51:07 &2013-12-10 07:41:07 &27.29 &65\\
         12 &4U 1907+09 &30401018002 &2018-08-01 12:41:09 &2018-08-03 07:41:09 &78.86 &7\\
         13 &4U 1538-52 &30201028002 &2016-08-11 19:11:08 &2016-08-12 19:16:08 &11.1 &7\\
         14 &Cep X--4 &80002016002 &2014-06-18 22:01:07 &2014-06-19 20:11:07 &40.4 &44\\
         15 & &80002016004 &2014-07-01 10:16:07 &2014-07-02 07:26:07 &41.16 &12\\
         16 &4U 1626-67 &30101029002 &2015-05-04 12:26:07 &2015-05-05 20:41:07 &65.0 &15\\
         17 &SMC X--2  &90102014002 &2015-09-25 21:51:08 &2015-09-26 11:21:08 &24.51 &19\\
         18 &        &90102014004 &2015-10-12 21:41:08 &2015-10-13 12:06:08 &23.06 &9\\
         19 &        &90101017002 &2015-10-21 21:31:08 &2015-10-22 11:16:08 &26.72 &6\\
         20 &IGR J17544--2619 &30002003003 &2013-06-19 09:31:07 &2013-06-19 23:41:07 &26.3 &26\\
         21 &IGR J16393--4643 &30001008002 &2014-06-26 02:21:07 &2014-06-27 05:31:07 &50.57 &0.54\\
         22 &2S 1553--542 &90101002002 &2015-04-03 11:36:07 &2015-04-04 01:46:07 &27.42 &19\\
         23 &RX J0520.5--6932 &80001002002 &2014-01-22 20:16:07 &2014-01-23 11:36:07 &27.75 &14\\
         24 &                &80001002004 &2014-01-24 23:56:07 &2014-01-25 18:31:07 &33.2 &16\\
         25 &Cen X--3 &30101055002 &2015-11-30 18:11:08 &2015-12-01 05:01:08 &21.4 &65\\
         26 &GX 301--2 &30001041002 &2014-10-29 07:41:07 &2014-10-29 22:01:07 &38.2 &12\\
         27 &         &30101042002$^{F}$ &2015-10-04 08:01:08 &2015-10-04 22:41:08 &35.7 &19\\\
         28 &XTE J1829--098 &90401332002 &2018-08-16 02:11:09 &2018-08-16 18:01:09 &27.8 &6\\
         29 &V0332+53 &80102002008 &2015-09-30 22:51:08 &2015-10-01 09:46:08 &18.1 &37 \\
         30 &       &80102002010 &2015-10-04 23:36:08 &2015-10-05 11:51:08 &20.1 &21 \\
         31 &       &90202031002 &2016-07-30 18:01:08 &2016-07-31 06:31:08 &25.2 &12 \\
         32 &       &90202031004 &2016-07-31 18:06:08 &2016-08-01 06:46:08 &25.0 &10 \\
         33 &XTE J1858+034 &90501348002  &2019-11-03 06:56:09 &2019-11-04 08:31:09 &43.7 &17 \\
         34&4U 1700--37 &30101027002$^{F}$ &2016-03-01 10:21:08 &2016-03-02 07:31:08 &38.0 &23\\
         35&LMC X--4 &30102041002$^F$ &2015-10-30 01:01:08 &2015-10-30 12:46:08 &24.5 &13\\
         36&  &30102041004 &2015-11-04 19:46:08 &2015-11-05 05:56:08 &21.9 &10\\
         37&  &30102041006 &2015-11-11 11:16:08 &2015-11-11 22:36:08 &22.9 &3\\
         38&  &30102041008$^F$ &2015-11-27 09:16:08 &2015-11-27 20:31:08 &20.3 &10\\
         39&IGR J17329-2731 &90301012002 &2017-08-29 15:36:09 &2017-08-30 02:36:09 &20.8 &3\\
         40&SMC X--1 &30202004002 &2016-09-08 21:26:08 &2016-09-09 09:06:08 &22.5 &21\\
         41&  &30202004004 &2016-09-19 07:11:08 &2016-09-19 19:06:08  &21.2 &15\\
         42&  &30202004006 &2016-10-01 00:56:08 &2016-10-01 12:01:08 &20.4 &1.52\\
         43&  &30202004008 &2016-10-24 19:31:08 &2016-10-25 07:36:08 &20.8 &24\\
         44&GRO J1008--57 &80001001002 &2012-11-30 08:41:07 &2012-11-30 17:31:07 &12.4 &227\\
         45&GX 304--1 &90401326002 &2018-06-03 05:56:09 &2018-06-04 12:06:09 &58.1 &0.21\\
         46 &1A 0535+26 &80001016002 &2015-02-11 04:11:07&2015-02-11 16:21:07 &21.4 &36\\
         47 & &80001016004 &2015-02-14 01:36:07 &2015-02-14 17:31:07 &29.7 &14\\
         48 & &90401370001 &2018-12-26 02:41:09 &2018-12-27 12:11:09 &54.9 &68\\
         49 &GRO J2058+42 &90501313002 &2019-03$-$25 07:06:09 &2019-03$-$25 19:26:09 &20.4 &38\\
         50 & &90501313004 &2019-04-11 00:11:09 &2019-04-11 22:11:09 &38.6 &42\\
         51 & &90501336002 &2019-08-28 12:36:09 &2019-08-29 09:26:09 &38.5 &4\\
         52 &1E 1145.1--6141 &30501002002 &2019-07-23 10:56:09 &2019-07-24 07:26:09 &44.2 &7.3\\
         53 &OAO 1657--415 &30401019002 &2019-03$-$25 07:06:09 &2019-06-13 08:01:09 &74.7 &7.6\\
         54 &EXO 2030+375 &90201029002 &2016-07-25 08:36:08 &2016-07-26 17:11:08 &56.7 &2.6\\
         55 & &90701336002 &2021-11-08 03:31:09 &2021-11-08 17:56:09 &23.5 &71\\
         56 & &80701320002 &2021-08-30 00:56:09 &2021-08-31 03:56:09 &32.4 &194\\
         57 &IGR J19294+1816 &90401306002 &2018-03-02 21:56:09 &2018-03-03 20:06:09 &40.6 &0.06\\
         58 & &90401306004 &2018-03-16 03:11:09 &2018-03-17 01:31:09 &40.4 &2.1\\
         \hline
         \hline
    \end{tabular}
    \label{tab:obs_catalogue}
\end{table*}

\section{Time-averaged spectroscopy}\label{sec:tas}
We have performed the time-averaged spectral analysis on the \textit{NuSTAR} pointed observations of sources listed in Table~\ref{tab:obs_catalogue} in the 3$-$79 keV spectral band unless otherwise specified. The light curves of some of the observations showed the presence of flares (See Table~\ref{tab:obs_catalogue}). In those cases, the flare and out-of-flare spectra were separately generated and then analysed\footnote{We consider any spike in the light curve as a flare if the ratio of the peak count rate of the spike to the average count rate outside the spike is greater than 3. }. The spectra {were} fitted with a composite model of the following form
 \break
\begin{eqnarray}\label{eq:mod-validation}
\frac{dN}{dE} = e^{-\sigma_{\rm{abs}(E)}} {\times \left(fe^{-\sigma_{\rm{abs},\rm{local}}(E)}\ + (1-f)\right) \times}\ [\ \textrm{continuum}(E)\ &\nonumber\\ \times\ {\textrm{gabs}_{\textrm{TKF}}}(E)\ +\ \textrm{gaussian}_{\textrm{Fe}}(E)\ +\ {\textrm{bbody(E)}\ ] \times \ \textrm{gabs}_{\textrm{CRSF}}(E)}
\end{eqnarray}

$e^{-\sigma_{abs}(E)}$ denotes absorption by the interstellar medium (ISM) and {was} modelled with the Tuebingen-Boulder ISM absorption model (\texttt{tbabs}) using ISM abundances from \citealt{abund_wilm_2000ApJ} and photoelectric absorption cross-sections from \citealt{xsect_vern_1996ApJ}. For the observations in which the galactic absorption column density towards the source could not get constrained by the fit, the absorption column density {was} frozen to the galactic line of sight value obtained from the tool \textit{w3nh}\footnote{\url{https://heasarc.gsfc.nasa.gov/cgi-bin/Tools/w3nh/w3nh.pl}}. { Local absorption was modelled by multiplying the spectrum with a partial covering absorption model of the form $fe^{-\sigma_{\rm{abs},\rm{local}}(E)}\ + (1-f)$. The implication of this modified Tuebingen-Boulder ISM absorption model (\texttt{tbpcf}) is that only a fraction $f$ of the total X-ray flux is absorbed, while the remaining fraction $(1-f)$ of the flux travels unabsorbed \citep[See][]{Maitra_2013-xtej1946}. The redshift parameter of \texttt{tbpcf} model was fixed to 0. The soft excess was modelled with a blackbody spectrum model \textit{bbody}. \textit{TKF} was modelled with a gaussian absorption model \texttt{gabs}.}

The spectral continuum is modeled with a physical comptonization model like \texttt{compTT} \citep{intro_compTT_1994ApJ...434..570T} or  one of the following phenomenological power-law {based} continuum models:\break

\begin{equation}
    {\texttt{powerlaw} =  A\times E^{-\Gamma}} \nonumber\\
\end{equation}

\begin{equation}
    \texttt{cutoffpl} = A\times E^{-\Gamma} \mathrm{exp}\left(-\frac{E}{E_\mathrm{cut}}\right)\nonumber\\
\end{equation}

\begin{equation}
    \texttt{powerlaw} \times {\texttt{mplcut} =}
    \begin{cases}
    A\times E^{-\Gamma}{\mathrm{exp}\left(-\frac{(E-E_\mathrm{cut})^2}{(0.1E_\mathrm{cut})^2}\right)} &E<E_\mathrm{cut}\\
    A\times E^{-\Gamma}\mathrm{exp}\left(-\frac{E-E_\mathrm{cut}}{E_\mathrm{fold}}\right){\mathrm{exp}\left(-\frac{(E-E_\mathrm{cut})^2}{(0.1E_\mathrm{cut})^2}\right)} &E\geq E_\mathrm{cut}\nonumber\\
    \end{cases}
\end{equation}

\begin{equation}
    \texttt{powerlaw} \times \texttt{FDcut} =
    \begin{cases}
    A\times E^{-\Gamma} &E<E_\mathrm{cut}\\
    A\times E^{-\Gamma}\times1/\left[1+\mathrm{exp}\left(-\frac{E-E_\mathrm{cut}}{E_\mathrm{fold}}\right)\right] &E\geq E_\mathrm{cut}\nonumber\\
\end{cases}  
\end{equation}

\begin{equation}
    \texttt{powerlaw} \times \texttt{newhcut} =
    \begin{cases}
    A\times E^{-\Gamma} &E\leq E_\mathrm{cut}-\Delta E\\
    c_0 + c_1E +c_2E^2 +c_3E^3 &\scriptstyle  E_\mathrm{cut}-\Delta E < E < E_\mathrm{cut}+\Delta E\\
    A\times E^{-\Gamma}\mathrm{exp}\left(-\frac{E-E_\mathrm{cut}}{E_\mathrm{fold}}\right) &E\geq E_\mathrm{cut}+\Delta E\nonumber\\
    \end{cases}
\end{equation}

\begin{equation}
   \texttt{NPEX} = (A_1\times E^{2} + A_2\times E^{-\Gamma})\mathrm{exp}\left(-\frac{E}{E_\mathrm{cut}}\right)\nonumber\\
\end{equation}

\texttt{highecut} \citep{intro_white_1983} represents a power-law continuum modified by an exponential roll-off above energy E\textsubscript{cut}. \texttt{mplcut} \citep{into_coburn_2002} is a slightly modified version of \texttt{highecut} wherein the abrupt break in continuum at E\textsubscript{cut} is smoothed out with a gaussian absorption model (\texttt{gabs} in {\small XSPEC}). Similarly, \texttt{FDcut} \citep{intro_tanaka_1986LNP...255..198T} has a Fermi-Dirac distribution-like function and \texttt{newhcut} \citep{intro_buderi_2000ApJ...530..429B} has a third-order polynomial function to smoothen the kink at E\textsubscript{cut}\footnote{We have consistently used $\Delta E$=5 keV for the \texttt{newhcut} model throughout the analysis.}. \texttt{NPEX} \citep{intro_mihara_2002} is a combination of two cutoff power-law functions, with the photon-index of one of the power-law components fixed to 2. \texttt{NPEX} mimics saturated inverse Compton scattering. CRSF is modelled with a gaussian absorption profile (\texttt{gabs}) and iron fluorescence line with simple gaussian (\texttt{gaussian}). \texttt{newhcut} and \texttt{FDcut} are local models installed into {\small XSPEC}. All other spectral models used in this work are available in {\small XSPEC}\footnote{https://heasarc.gsfc.nasa.gov/xanadu/xspec/manual/Models.html}.

{\subsection{Spectral analysis and Model selection Methodology}\label{ref:subsec:analysis-method}}

{Our aim is to search for the best fitting composite spectral model for each \textit{NuSTAR} observation (Table~\ref{tab:obs_catalogue}) and check if it contains the \textit{TKF} model component. We employed a two-step spectral fitting process invoking the Akiake Information Criterion \citep{akaike1998information} for finding the best fitting model. The AIC score for a model fit is given by

\begin{equation*}
    {\rm{AIC}} = 2n - 2\mathrm{ln}(L_{\rm{max}}) = 2n + \chi^2_{\rm{min}}
\end{equation*}

Here, $L_{\rm{max}}$ is the maximum value of likelihood and $-2\mathrm{ln}(L_{\rm{max}}) = \chi^2_{\rm{min}}$ is the equivalent minimum value of the fit statistic, and $n$ is the number of model parameters. As evident from the equation, AIC score will be low for a model that fits the data well, but it increases with the complexity of the model (based on the number of model parameters). We compared the AIC scores of all the composite models derived from equation~\ref{eq:mod-validation} after they were fitted to the spectrum of each observation, and chose the best fitting model based on the lowest AIC score.}

{Since the primary interest of this work is to identify the presence/absence of a spectral feature near 10 keV, we employed a two-step spectral fitting procedure so that importance is given to the 10 keV spectral band in the selection of the best fitting composite model.

\subsubsection{Training and Validation}

We performed an initial manual fit on the broadband spectrum of each observation (Table~\ref{tab:obs_catalogue}) to identify the presence of iron fluorescence line(s) and CRSF(s) in the spectrum. The spectrum from each observation in 3$-$15 keV (the \textit{TKF} band) and 3$-$79 keV (broadband) was treated as two separate data sets, and we named them the `Training' data set and the `Validation' data set, respectively. 

The 3$-$15 keV training data set spectra were fitted with every possible composite model combination derived from
\break
\begin{eqnarray}\label{mod-train}
\frac{dN}{dE} = e^{-\sigma_{abs}(E)} \times \left(fe^{-\sigma_{\rm{abs},\rm{local}}(E)}\ + (1-f)\right) \times &\nonumber\\ \left[ \textrm{continuum}(E)\times \rm{gabs}_{TKF} +\ \textrm{gaussian}_{\textrm{Fe}}(E)\ +\ \textrm{bbody}(E) \right]
\end{eqnarray}

The combinations were identified, ensuring that the continuum, \texttt{tbabs} and \texttt{gaussian} (if iron emission line is present) were included in all of them. Since the cutoff energy of most pulsars is beyond 15 keV, continuum models without a high energy cutoff (\texttt{powerlaw}, \texttt{cutoffpl}, \texttt{NPEX} and \texttt{compTT}) were only used for training. A total of 32 composite model candidates (without high energy cutoff continuum models) were tested on each training set.

The performance of each candidate model on the training set was evaluated based on AIC. Only those models were considered in which all the spectral parameters were within the globally accepted ranges. The accepted ranges of various spectral model parameters were chosen as follows: galactic absorption column density as per the source, partial covering absorption column density less than $10^{25}$ atoms cm$^{-2}$, power-law index between 0 and 3, high energy cutoff between 0 and 40 keV, T$_0$, kT and $\tau$ of \texttt{compTT} between 0.01-5 keV, 2-100 keV and 0.01-100, respectively, the line centre and line width of iron fluorescence emission between 6.2-6.8 keV and 0-1 keV, respectively, temperature kT of \texttt{bbody} between 0.01 and 3 keV, and the centre and width of the \texttt{gabs} used to model the \textit{TKF} between 9-12 keV and 0-3 keV, respectively. 

The best fitting training model so identified (having the lowest AIC), along with other models within 5\% AIC score (or sometimes 10\% AIC in case the 5\% does not contain multiple candidates) were selected for fitting the 3$-$79 keV broadband validation data set. The validation was performed after including the extra \texttt{gabs} or \texttt{cyclabs} model component(s) to account for CRSF(s). If any of the training-screened composite models contain a simple \texttt{powerlaw} continuum, in addition to it, the same composite model after modifying with different kinds of high energy cutoff (\texttt{mplcut, newhcut, newhcut}) was also tested on the validation data set. The best-fitting composite model on the validation data set was selected based on the lowest AIC score and best fitting spectral parameters within the accepted ranges. 

However, it was observed in some spectral data sets that none of the best-fitting training models would fit the validation data set well. In those cases, the training was skipped and single-step validation was performed using all the possible 56 composite model combinations derived from equation~\ref{mod-train} (including the high energy cutoff continuum models) and the best-fitting model was selected following the same model screening procedure.

We report the presence of \textit{TKF} only when all the best fitting composite model(s) on the validation data set within the top 5\% AIC contain \textit{TKF} model component.}

The rest of this section briefs the details of each source and explains the spectral analysis performed on their \textit{NuSTAR} observations. In each observation, the background-dominated spectral range is excluded from the fitting process. {The figure showing the best fitting model (the unfolded spectrum and model in E$\frac{dN}{dE}$ as solid curves, the contribution of additive model components as dotted curves) and residuals to the best fit model for one observation containing \textit{TKF} and one observation not containing \textit{TKF}, based on availability, of each source are also shown. In case \textit{TKF} is required in the fit, a third panel showing the significance of \textit{TKF} is also included. The data points from the FPMA instrument are shown in black and FPMB is shown in red.} The fit statistic is $\chi^2$ distributed in all the observations except Obs. Sn.45. The tables containing best fit spectral parameter values for each observation are given {across four tables in} Appendix. \ref{Tables}.

\subsection{Her X--1}\label{subsec:herx1}
Hercules X--1 is a moderately bright, persistent, eclipsing intermediate mass X-ray binary (IMXB) pulsar with 1.24 s spin period \citep{herx1_tanabaum_1972ApJ...174L.143T} located 6.6 kpc away, hosting a NS and a 2.3 M$_\odot$ optical companion (Hz Her) in a low eccentricity orbit. Nearly edge-on view of the binary manifests as periodic $\sim1.7$ d intensity variation in the X-ray light curve in the form of eclipses. In addition, it also exhibits a 35 d intensity modulation due to absorption by a precessing warped accretion disk \citep{herx1_giacconi_1973ApJ...184..227G}. The first ever report of CRSF in an XRP spectrum was with the balloon observations of Her X--1 \citep{herx1_trumper_1978ApJ...219L.105T}. The broadband spectrum of Her X--1  is usually modelled with an absorbed power-law continuum with iron fluorescence lines, modified by CRSF at $\sim$ 38 keV \citep[See][]{herx1_xiao_2019_201929}.

\begin{figure}
    \centering
    \includegraphics[width=\linewidth]{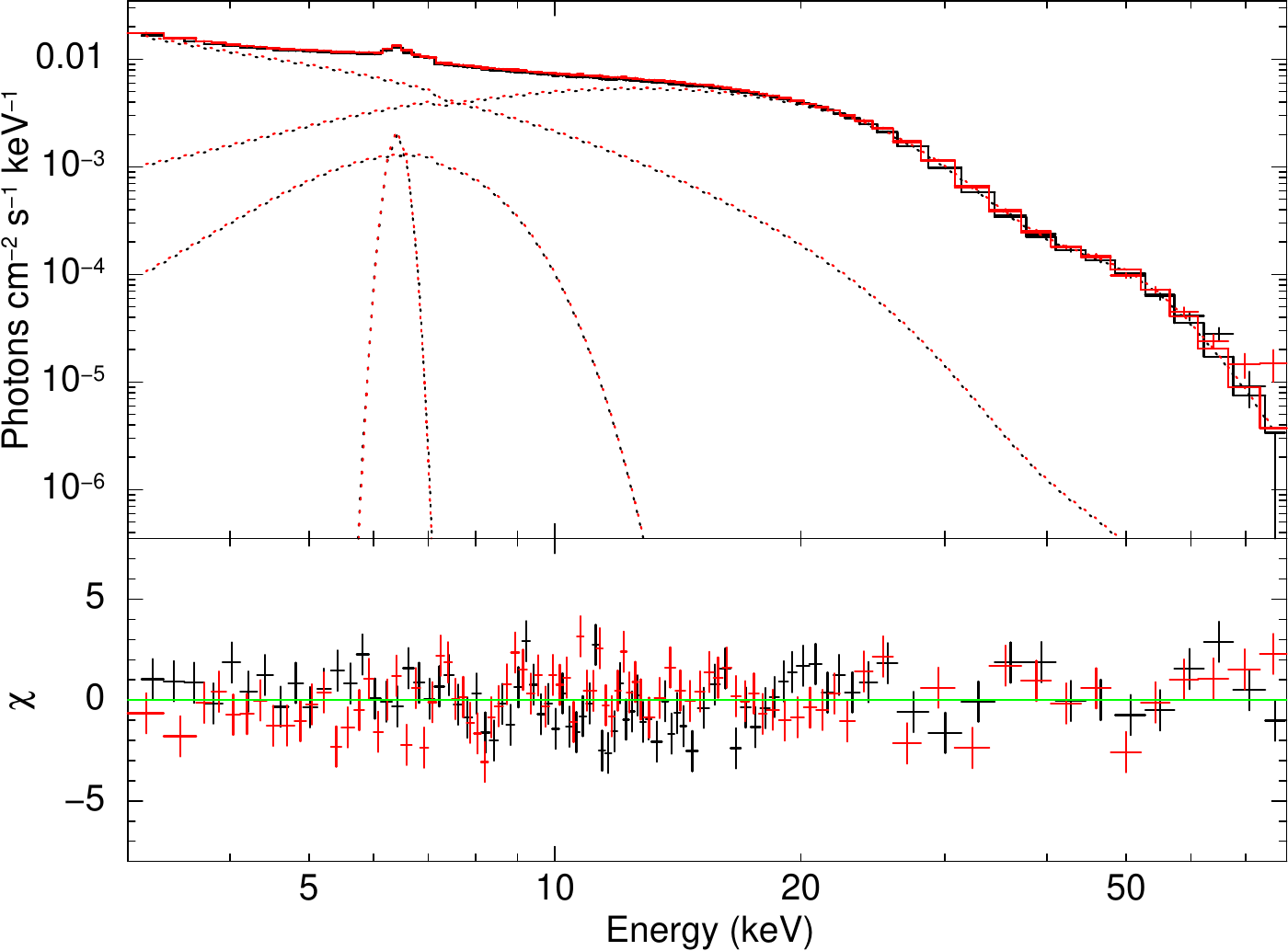}
    \includegraphics[width=\linewidth]{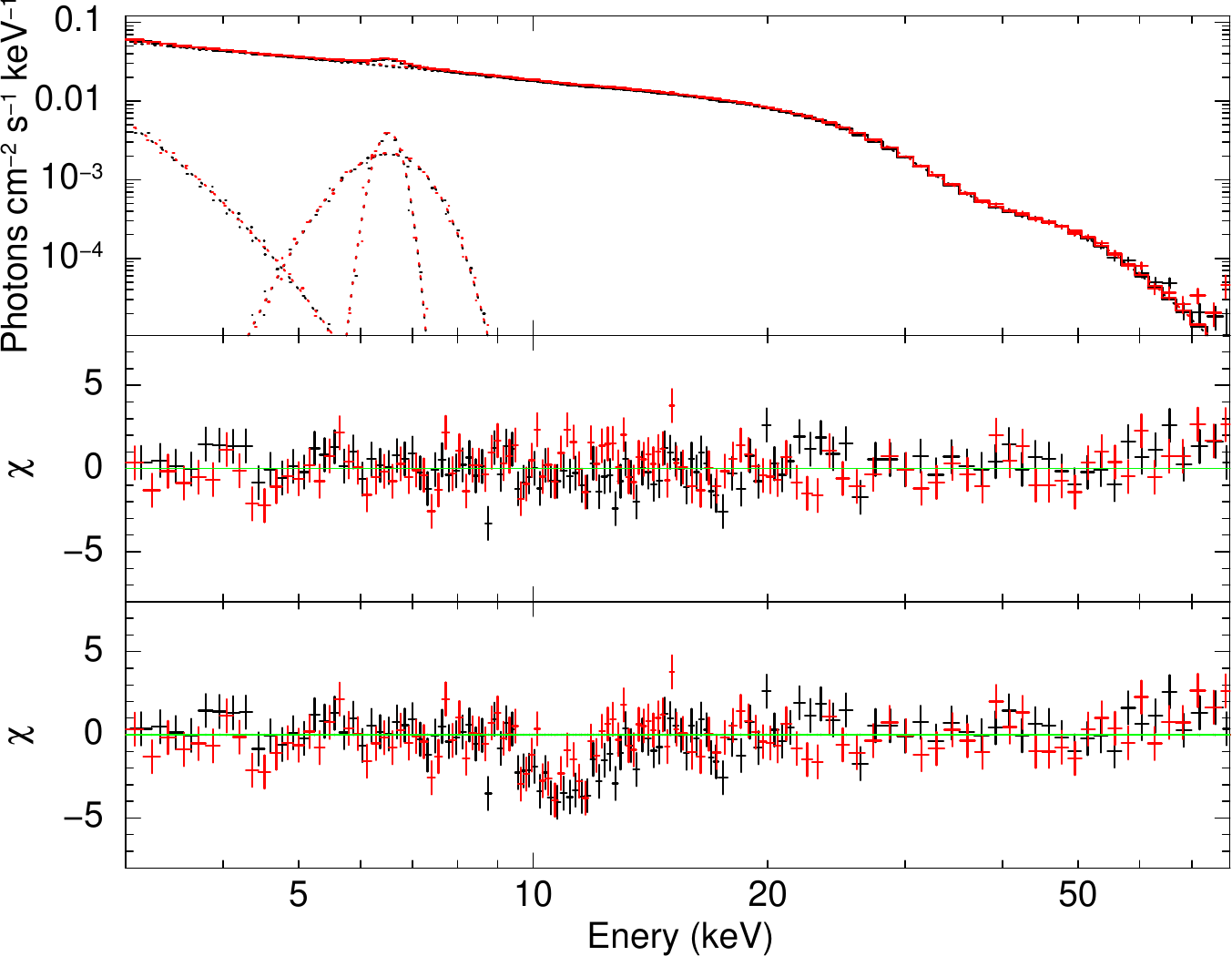}
    \caption{Top: Spectral fit on Her X--1 Obs. Sn.1 with \texttt{NPEX} continuum model, having no \textit{TKF}. The top panel shows the best-fit model and the bottom panel shows the residuals to the best-fit model. Bottom: Spectral fit on Her X--1 Obs. Sn.2. with \texttt{FDcut} continuum model, having \textit{TKF}. The top panel shows the best-fit model, the middle panel shows the residuals to the best-fit model, and the bottom panel shows the residuals when the strength of the \texttt{gabs} component modelling \textit{TKF} is set to 0.}
    \label{fig:herx1}
\end{figure}

{ We analysed the spectra from the observations Obs. Sn.1, 2 and 3 (Table~\ref{tab:obs_catalogue}). Obs. Sn.1 spans the eclipse and out of eclipse orbital phases of the binary. Therefore, the time average spectrum and out-of-the-eclipse phase spectrum of Obs. Sn.1 were analyzed. Galactic absorption towards the source could not be constrained by the fit, therefore we fixed it to the galactic value of $1.5\times10^{20}$ atoms cm\textsuperscript{-2}. The iron emission line region is complex, and we modelled it with the combination of a narrow and a broad \texttt{gaussian} (adopted from \citealt{herx1_furst1_2013ApJ}). CRSF at $\sim38$ keV was fitted with a \texttt{gabs} model.

None of the best fitting training models could fit the validation data sets of Obs. Sn.2 and 3. Therefore, single-step fitting was performed on Obs. Sn.2 and 3 skipping training. The best-fitting models on the time-averaged and out-of-eclipse spectrum of Obs. Sn.1 does not contain \textit{TKF}. The best fitting models on Obs. Sn.2 and 3 contain \textit{TKF}.

The spectral parameters of the best fitting models for all three observations of Her X--1 are given in Tables ~\ref{tab:cyc1} and \ref{tab:tkf1}. The spectral fit for the time-averaged spectrum of Obs. Sn. 1 (having no \textit{TKF}) and Obs. Sn. 2 (having \textit{TKF}) are shown in Fig.~\ref{fig:herx1}.}

\subsection{Vela X--1}\label{sec:vx1}
Vela X--1 (4U 0900-40) is a bright and persistent eclipsing high mass X-ray binary (HMXB) {pulsar} hosting a NS and a B-type Super-giant stellar companion, located 1.9 kpc away. The XRP has a spin period of $\sim$283 s, and the binary has an orbital period of $\sim$8.9 d. Even though the typical X-ray luminosity of the pulsar is about $4\times10^{36}$ erg s\textsuperscript{-1}, the relatively compact orbital separation keeps the pulsar embedded in the clumpy stellar wind of the companion, and this shows up as strong variability in XRP luminosity. CRSF has been detected in the source spectrum at two energies, with the fundamental line at $\sim25$ keV and its harmonic at $\sim55$ keV (See \citealt{velax1_kreykenbohm_2002_refId0} and references therein, \citealt{velax1_maitra_2013ApJ...763...79M}, and  \citealt{velax1_Furst1_2013}).

\begin{figure} 
    \centering
    \includegraphics[width=\linewidth]{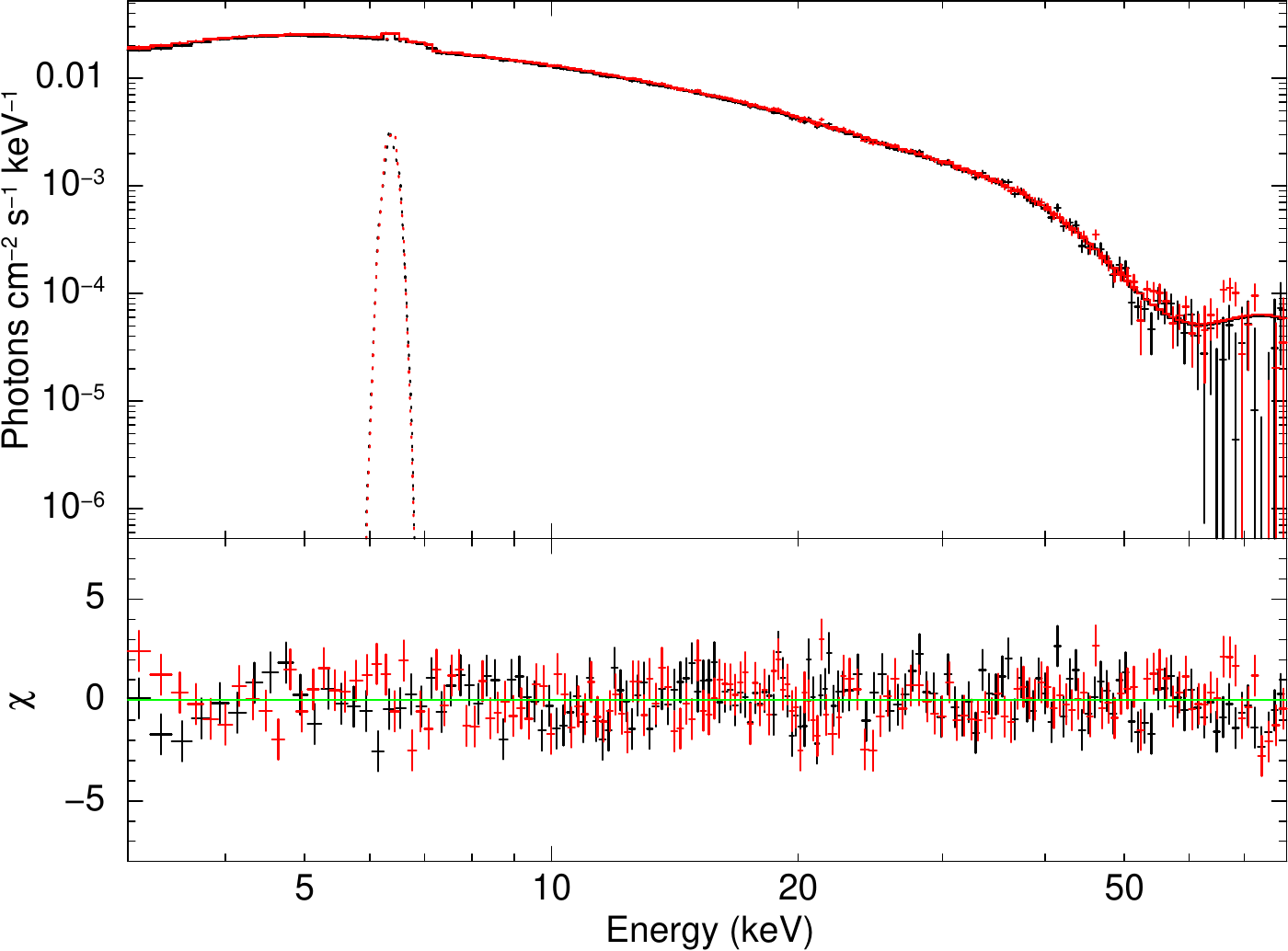}
    \includegraphics[width=\linewidth]{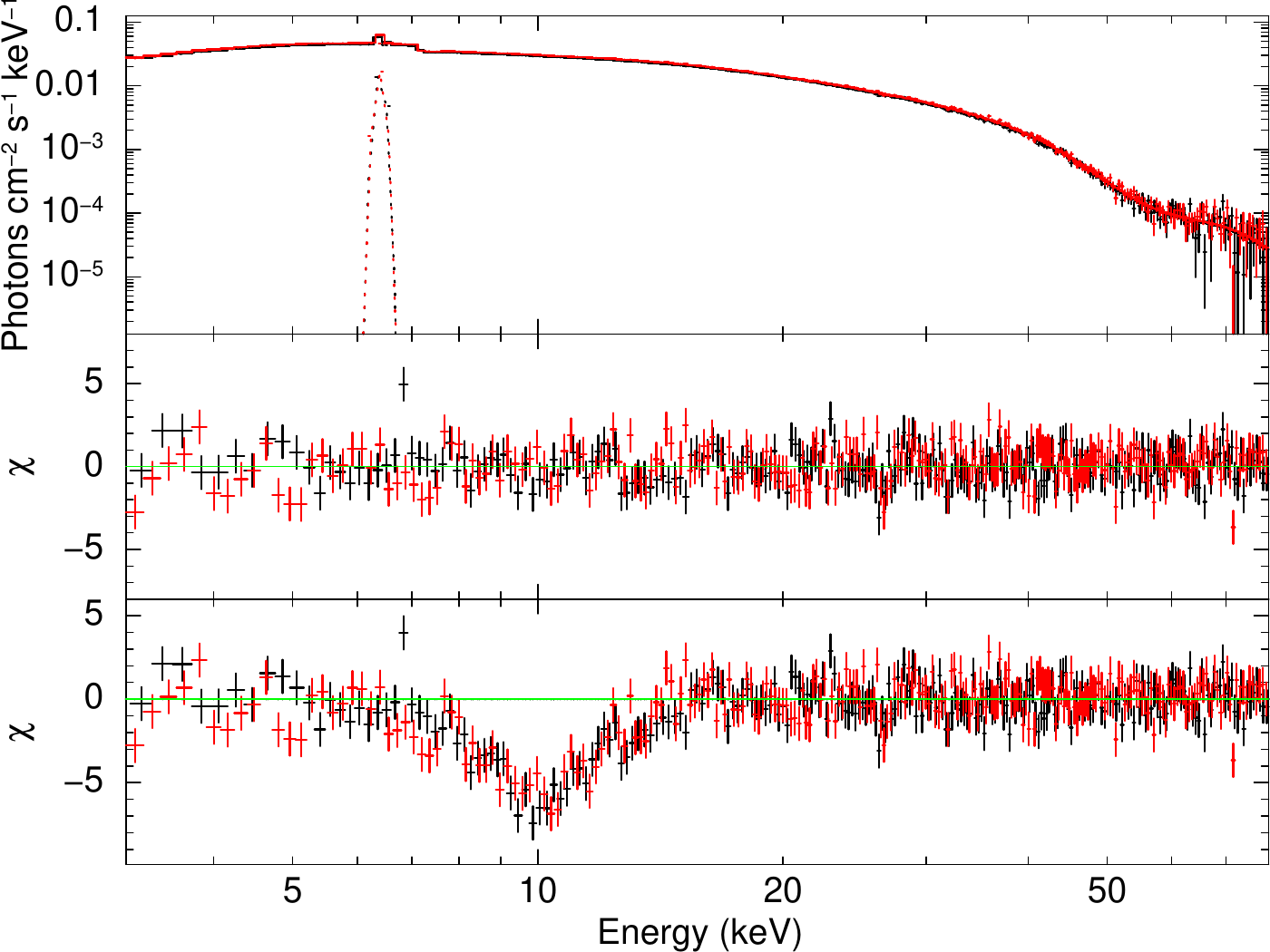}
    \caption{Top: Spectral fit on Vela X--1 Obs. Sn.5 with \texttt{cutoffpl} continuum model, having no \textit{TKF}. The top panel shows the best-fit model and the bottom panel shows the residuals to the best-fit model. Bottom: Spectral fit on Vela X--1 Obs. Sn.4. with \texttt{compTT} continuum model, having \textit{TKF}. The top panel shows the best-fit model, the middle panel shows the residuals to the best-fit model, and the bottom panel shows the residuals when the strength of the \texttt{gabs} component modelling \textit{TKF} is set to 0.}
    \label{fig:velax1}
\end{figure}

{We analyzed the spectra from four observations Obs. Sn.4, 5, 6, and 7 (Table~\ref{tab:obs_catalogue}). Because of the presence of flare in Obs. Sn.7, we analysed the flaring state spectra of Obs. Sn.7. We could not get satisfactory fits for the time-averaged and out-of-flare-state spectra of Obs. Sn.7 due to the presence of a narrow absorptive feature around 7 keV. Galactic absorption towards the source could not be constrained by the fit, therefore we fixed it to the galactic value of 3.7$\times10^{21}$ atoms cm$^{-2}$. Iron line emission was fitted with a \texttt{gaussian} model. The fundamental CRSF at $\sim$25 keV and its harmonic at $\sim$50 keV were fitted with two \texttt{gabs} model components. The width of CRSF was not constrained by the fit in Obs. Sn.4, it was thus fixed to the average width obtained from other observations. The width of CRSF was also not constrained by the fit in the best fitting model containing \textit{TKF} in Obs. Sn.6, it was fixed to the width from the best fitting model on Obs. Sn.6 not containing \textit{TKF}.

The best fitting models on Obs. Sn.5 and 7 do not contain \textit{TKF}, while the best fitting model on Obs. Sn.4 contains \textit{TKF}. Even though the best fitting model Obs. Sn.6 contains \textit{TKF}, there also exists a similar well-fitting model in which \textit{TKF} is absent.

The spectral parameters of the best fitting models for all four observations of Vela X--1 are given in Tables \ref{tab:cyc1} and \ref{tab:tkf1}. The spectral fit for the time-averaged spectrum of Obs. Sn. 4 (having TKF) and Obs. Sn.2 (having no TKF) are shown in Fig.~\ref{fig:velax1}.}

\subsection{XTE J1946+274}\label{subsec:XTEJ1946p274}
XTE J1946+274 is a transient HMXB pulsar at a distance of $9.5\pm2.9$ kpc, in which the binary hosts the NS and a Be-type stellar companion. The pulsar has a spin period of about 15.8 s, and the binary has an orbital period of about 169 d \citep[See][and references therein]{xtej1946_doroshenko_2017_refId0}. Interaction between the circumstellar equatorial disk of the Be companion and the pulsar at the periastron in a moderately eccentric orbit results in violent X-ray outbursts which appear as variability in the long term X-ray light curve \citep{xtej1946274_Wilson_2003}. CRSF has been reported in the source spectrum at $\sim38$ keV (\citealt{xtej1946+274_heindl_2001ApJ...563L..35H}, \citealt{Maitra_2013-xtej1946}, \citealt{xtej1946_doroshenko_2017_refId0}, \citealt{devaraj_xtej1946}) and also at $\sim25$ keV \citep{xtej1946p274_muller_2021_refId0}.

\begin{figure} 
    \centering
    \includegraphics[width=\linewidth]{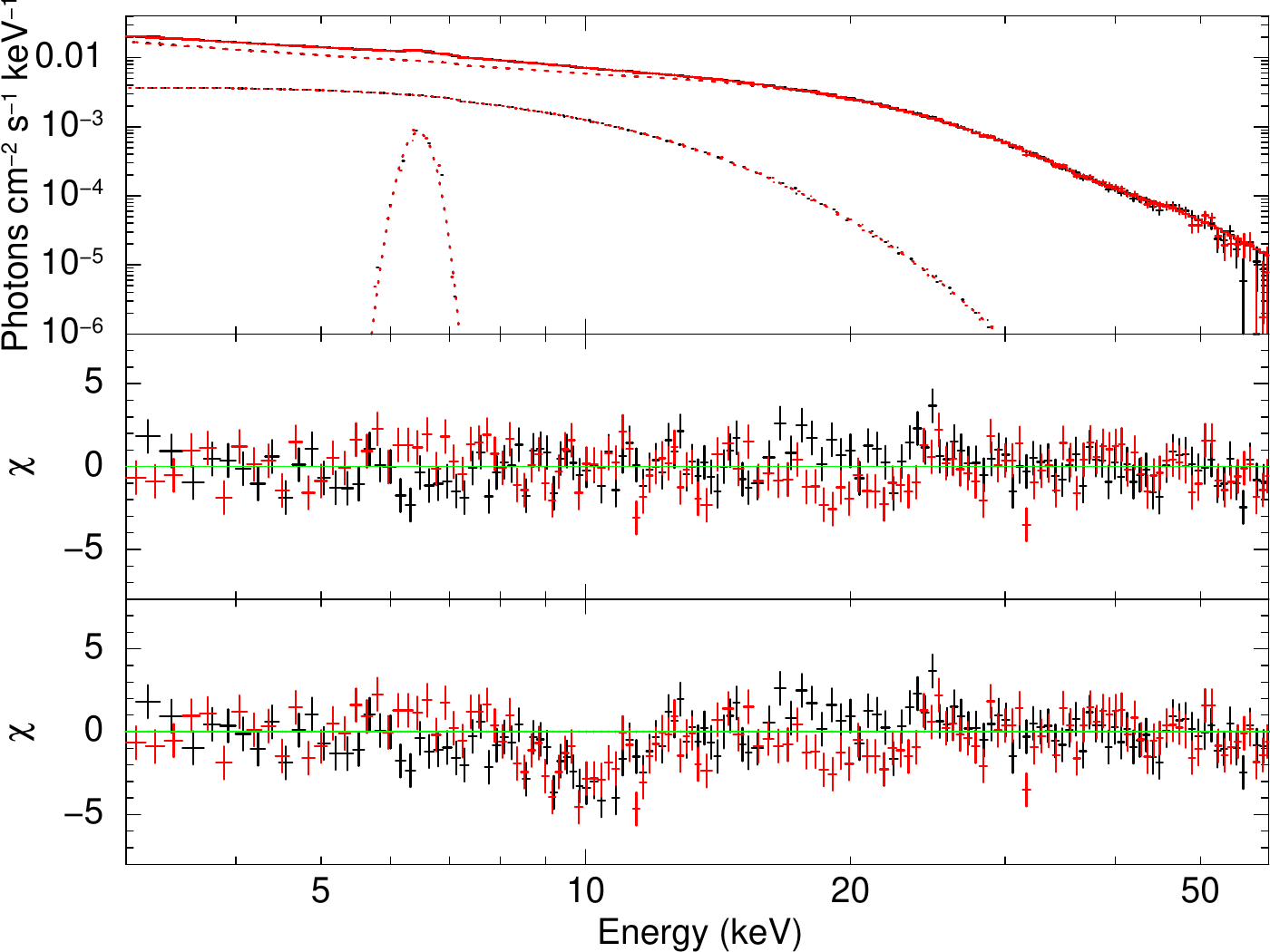}
    \caption{Spectral fit on XTE J1946+274 Obs. Sn.8 using \texttt{mplcut} model, having \textit{TKF}. The top panel shows the best-fit model, the middle panel shows the residuals to the best-fit model, and the bottom panel shows the residuals when the strength of the \texttt{gabs} component modelling \textit{TKF} is set to 0.}
    \label{fig:xtej1946p274}
\end{figure}

{ We analyzed the 3$-$60 keV spectrum from Obs. Sn.8 (Table~\ref{tab:obs_catalogue}). Iron fluorescence line was present, which was fitted with a \texttt{gaussian}. CRSF centred around 40 keV was fitted with \texttt{gabs}. 

The best fitting model on Obs. Sn.8 contains \textit{TKF}.

The spectral parameters of the best fitting models for Obs. Sn.8 of XTE J1946+274 are given in Tables \ref{tab:cyc1} and \ref{tab:tkf1}. The spectral fit for the time-averaged spectrum of Obs. Sn.8 (having TKF) is shown in Fig.~\ref{fig:xtej1946p274}}

\subsection{KS 1947+300}\label{subsec:ks1947p300}
KS 1947+300 is a transient XRP in an HMXB hosting the NS and Be-companion star in an almost circular orbit. The binary orbital period is about 41.5 d, and the pulsar has a spin period of 18.8 s \citep[][and references therein]{ks1947+300_furst_2014ApJ...784L..40F}. CRSF was first reported in the \textit{NuSTAR} spectrum of the source by \cite{ks1947+300_furst_2014ApJ...784L..40F}, but this discovery is debated \citep{ks1947+300_doroshenko_2020MNRAS.493.3442D}.

\begin{figure} 
    \centering
    \includegraphics[width=\linewidth]{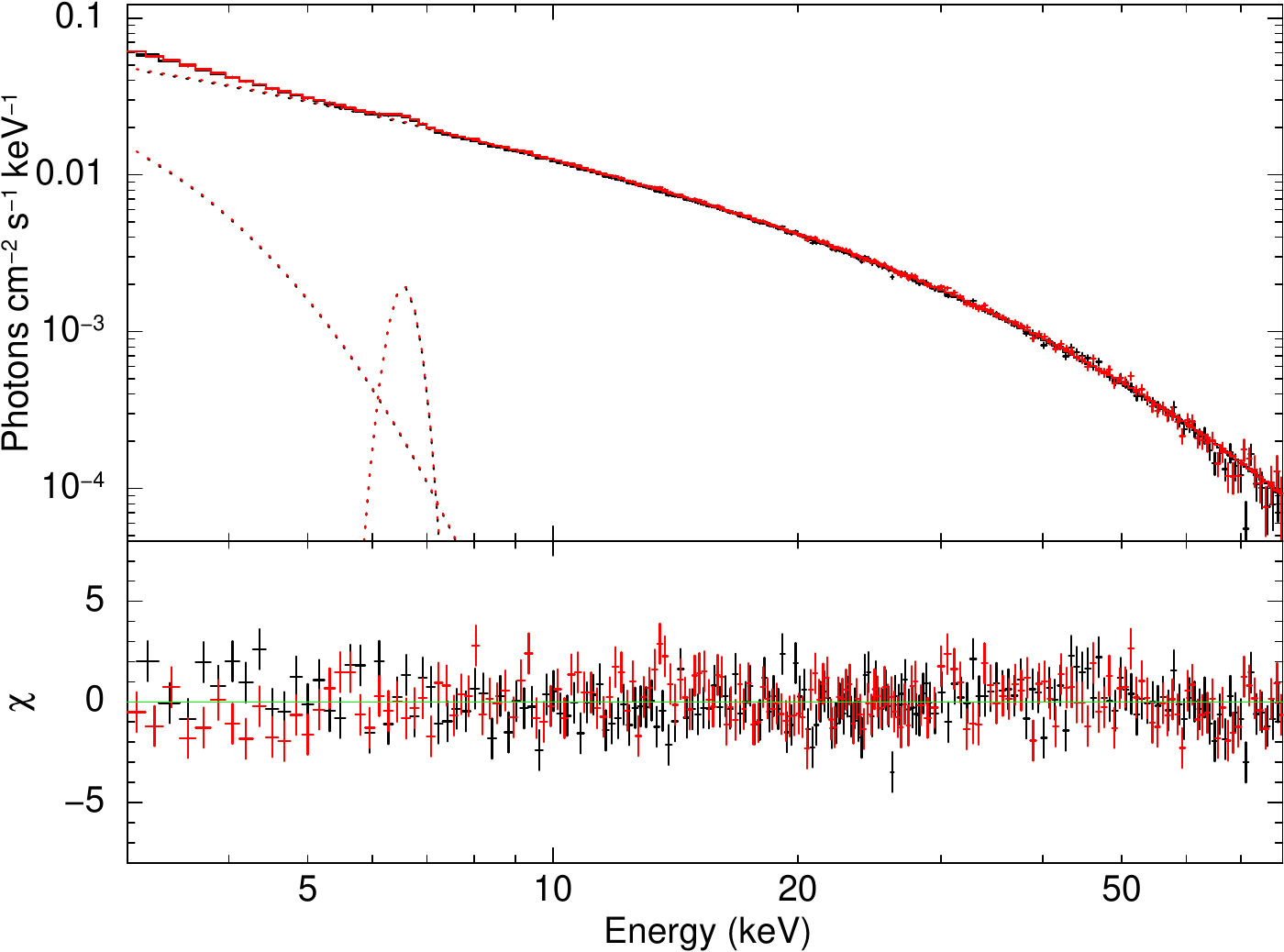}
    \caption{Spectral fit on KS 1947+300 Obs. Sn.9 using \texttt{mplcut} model, having no \textit{TKF}/CRSF. The top panel shows the best-fit model and the bottom panel shows the residuals to the best-fit model.}
    \label{fig:1947p300}
\end{figure}

{ 
\cite{ks1947+300_furst_2014ApJ...784L..40F} reported that the absorption feature detected in Obs. Sn.10 (for \texttt{highecut} model) could be CRSF. However, in a revisit of the same data and earlier \textit{BeppoSAX} observations, \cite{ks1947+300_doroshenko_2020MNRAS.493.3442D} reported that a two-component \texttt{compTT} continuum model does not show such a prominent absorption feature and hence questioning the claim of CRSF.

 We analyzed the spectra from three observations Obs Sn.9, 10 and 11 (Table~\ref{tab:obs_catalogue}), the same ones reported in \cite{ks1947+300_furst_2014ApJ...784L..40F} and \cite{ks1947+300_doroshenko_2020MNRAS.493.3442D}. The three observations covered the increase, peak and decline phase of the 2013 outburst of the source. Galactic absorption column density could be constrained by the fit. Iron fluorescence line was fitted with a \texttt{gaussian}. 

None of the best fitting training models could fit the validation data sets of Obs. Sn.9 and 10. Therefore, single-step fitting was performed on Obs. Sn.9 and 10 skipping training. The best models for validation were based on the top 10\% AIC on the training data set of Obs. Sn.11 because only one model was in the top 5\% AIC. The best fitting models on Obs. Sn.9 and 10 does not contain \textit{TKF}. The best-fitting model from training-validation on Obs. Sn.11 also does not contain \textit{TKF}. We did not include the claimed \textit{CRSF} feature around 10 keV during the fitting process and the best fitting models on all the three observations does not contain an absorptive feature around 10 keV. Thus, our analysis indicates towards the result of \cite{ks1947+300_doroshenko_2020MNRAS.493.3442D} that the CRSF feature is not present.

The spectral parameters of the best fitting models for all three observations of KS 1947+300 are given in Tables \ref{tab:cyc1} and \ref{tab:tkf1}. The spectral fit for the time-averaged spectrum of Obs. Sn. 9 (having no \textit{TKF}/CRSF) is shown in Fig.~\ref{fig:1947p300}.}

\subsection{4U 1907+09}\label{subsec:1907}
4U 1907+09 (X 1908+075) is a persistent HMXB pulsar with the NS accreting matter from the wind of an O-type Super-giant (\citealt{1907_cox_2005A&A...436..661C}, \citealt{1907_marshall_10.1093/mnras/193.1.7P}). The HMXB has a relatively short orbital period of about $8$ d, and the pulsar has a spin period of about 437.5 s \citep{1907_makishima_1984PASJ...36..679M}. The fundamental CRSF at 18 keV and its harmonic at 36 keV have been reported in the source \citep{4u1907+09_cusumano_2000AdSpR..25..409C}.

\begin{figure}
    \centering
    \includegraphics[width=\linewidth]{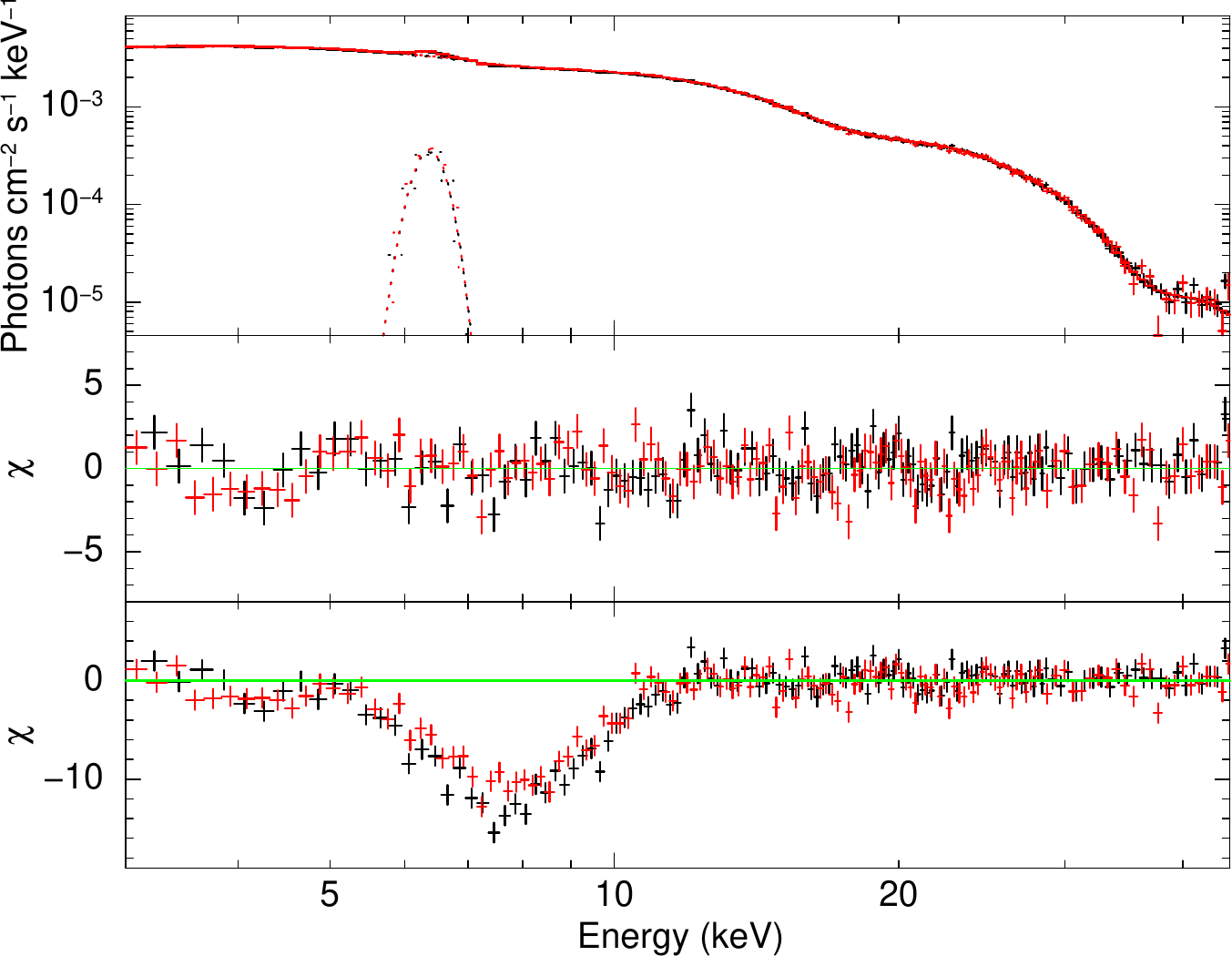}
    \caption{Spectral fit on 4U 1907+09 Obs. Sn.12 with \texttt{CompTT} continuum model, having \textit{TKF}. The top panel shows the best-fit model, the middle panel shows the residuals to the best-fit model, and the bottom panel shows the residuals when the strength of the \texttt{gabs} component modelling \textit{TKF} is set to 0.}
    \label{fig:1907}
\end{figure}

{We analyzed the 3$-$45 keV spectrum from one \textit{NuSTAR} observation Obs. Sn.12 (Table~\ref{tab:obs_catalogue}). Iron fluorescence line was present, which was fitted with a \texttt{gaussian}. We used two \texttt{gabs} components centred at 18 keV and 36 keV respectively, to model the fundamental and harmonic CRSF, respectively. 

All the training fits left absorption-like residuals between 5 and 10 keV. An absorptive feature around 8 keV has been reported by \cite{tobrej2023high_1907nustar}. We, therefore, lowered the allowed \textit{TKF} model centre between 6 and 9 keV. None of the best fitting training models could fit the validation data set of Obs. Sn.12. Therefore, single-step fitting was performed skipping training. The best fitting model on Obs. Sn.12 contain \textit{TKF}, centred around 8 keV at a lower centre value than other cases.

The spectral parameters of the best fitting models for Obs. Sn.12 of 4U 1907+09 are given in Tables \ref{tab:cyc1} and \ref{tab:tkf1}. The spectral fit for the time-averaged spectrum of Obs. Sn.12 (having TKF) is shown in Fig.~\ref{fig:1907}.}

\subsection{4U 1538-52}
4U 1538-52 (4U 1538-522) is a bright, persistent eclipsing HMXB pulsar 6.6 kpc away, in which the NS accretes matter from the stellar wind of B-type Super-giant companion. The pulsar has a spin period of $526$ s, and the binary has a relatively short orbital period of 3.73 d (See \citealt{XRPreview_Malacaria_2020}, \citealt{4u1538-52_hemphill_2019ApJ...873...62H}, and references therein.). CRSF has been reported in the source spectrum at $\sim$22 keV and $\sim$49 keV \citep{4u1538-52_hemphill_2019ApJ...873...62H}.

\begin{figure}
    \centering
    \includegraphics[width=\linewidth]{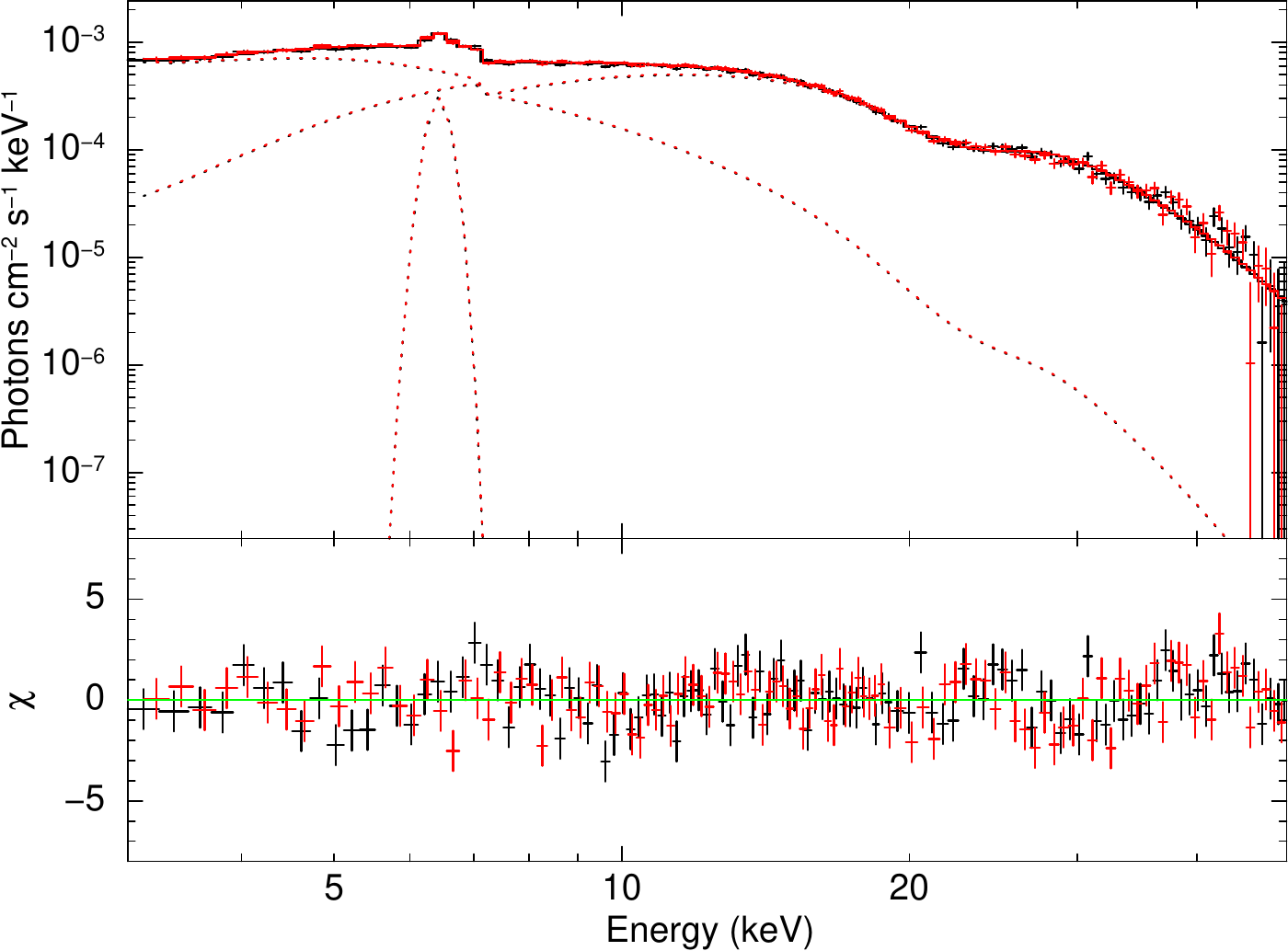}
    \caption{Spectral fit on time average spectrum of 4U 1538-52 Obs. Sn.13 covering the eclipse and out-of-eclipse phases, with \texttt{NPEX} continuum model, having no \textit{TKF}. The top panel shows the best-fit model and the bottom panel shows the residuals to the best-fit model.}
    \label{fig:1538-ecl}
\end{figure}

{We analyzed the spectrum from one observation of the source Obs. Sn.13 (Table~\ref{tab:obs_catalogue}). The light curve indicated that this \textit{NuSTAR} observation spans the ingress, eclipse, and egress phases of the binary. Therefore, we extracted two sets of spectra: i) the time-averaged spectrum of the entire observation, and ii) the spectrum of the out-of-eclipse phase ($\sim$11 ks of data covering ingress and egress phases).

Galactic absorption towards the source could not be constrained by the fit, therefore we fixed it to the galactic value of 7$\times10^{21}$ atoms cm$^{-2}$. Iron line emission was fitted with a \texttt{gaussian} model. CRSF at $\sim$22 keV is fitted a \texttt{gabs} model. 

The best fitting models on both the time-averaged and out-of-eclipse spectra do not contain \textit{TKF}.

The spectral parameters of the best fitting models for Obs. Sn.13 of 4U 1538-52 are given in Tables \ref{tab:cyc1} and \ref{tab:tkf1}. The spectral fit for the time-averaged spectrum of Obs. Sn. 13 (having no TKF) is shown in Fig.~\ref{fig:1538-ecl}.}

\subsection{Cepheus X--4}
Cep X--4 (GS 2138+56) is a transient {HMXB pulsar} with $\rm{P_{spin}}\sim66.2$ s located 3.8 kpc away, {in which} the NS accretes from {a} Be-type stellar companion in a $20$ d binary orbit (See \citealt{cepx4_bonnet_1998A&A...332L...9B}, \citealt{cepx4_vybornov2017luminosity}, and references therein.). CRSF has been found in the source at 30 keV \citep{cepx4_mihara_1991ApJ...379L..61M}, and the asymmetric line profile of CRSF in the \textit{NuSTAR} spectra was described with a combination of two Gaussian absorption model components centered at $\sim30$ keV and $\sim19$ keV respectively \citep{cepx4_furst_2015ApJ...806L..24F}. 

\begin{figure}
     \centering
     \includegraphics[width=\linewidth]{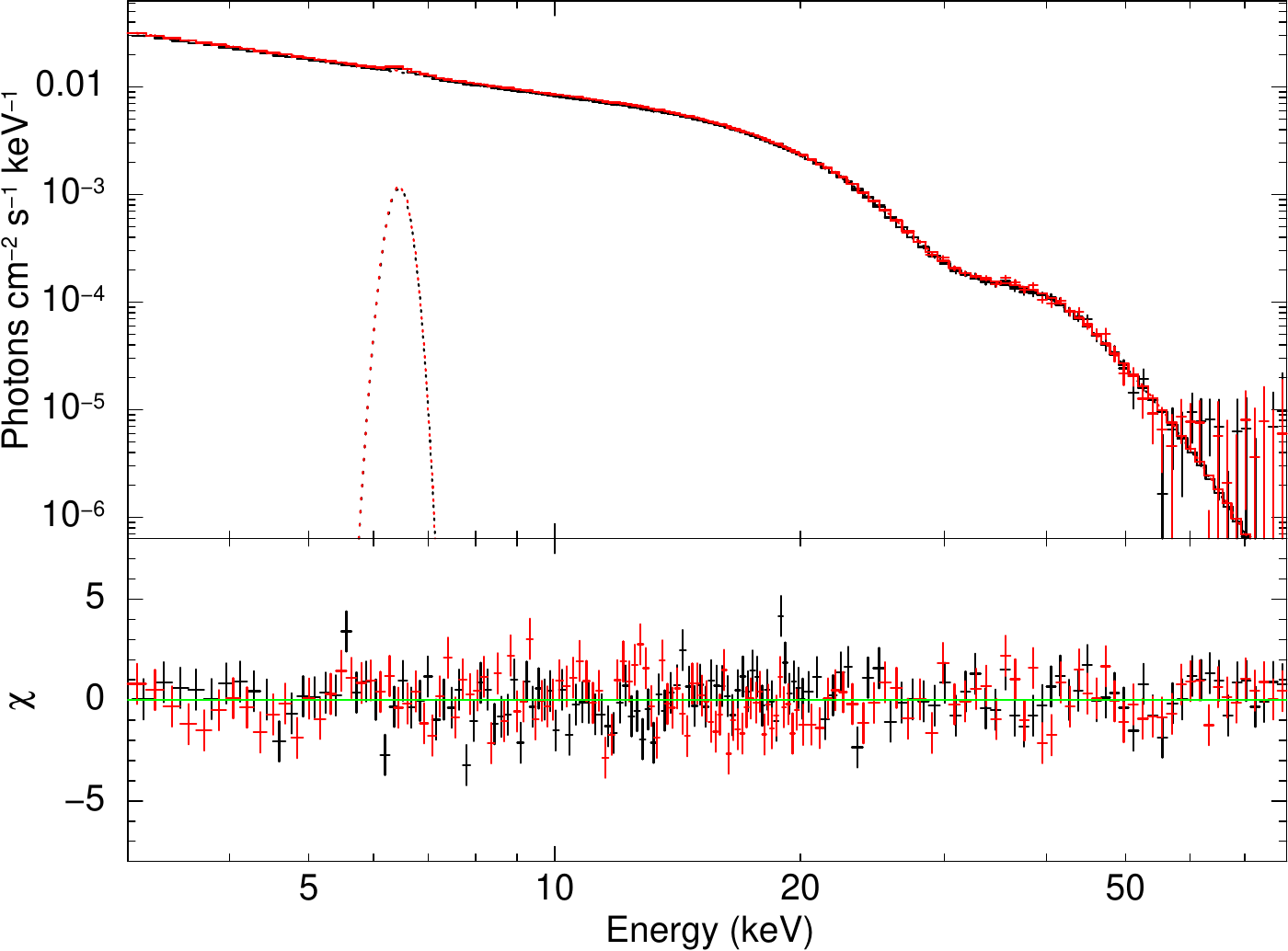}
     \caption{Spectral fit on Cep X--4 Obs. Sn.14 with \texttt{compTT} continuum model, having no \textit{TKF}. The top panel shows the best-fit model and the bottom panel shows the residuals to the best-fit model.}
     \label{fig:cepx4o2}
 \end{figure}

{ We analyzed the 3$-$60 keV spectra from two observations Obs. Sn.14 and 15 (Table~\ref{tab:obs_catalogue}). Obs. Sn.14 probed the peak luminosity phase of the outburst of the source in 2014, while Obs. Sn.15 covered the luminosity decline phase of the same outburst. Iron line emission was fitted with a \texttt{gaussian} model. The CRSF at $\sim$22 keV having an asymmetric profile is fitted with the combination of two \texttt{gabs} models \citep{cepx4_bhargava_2019MNRAS}.

The best fitting models on both observations do not contain \textit{TKF}.

The spectral parameters of the best fitting models for both the observations of Cep X--4 are given in Tables \ref{tab:cyc1} and \ref{tab:tkf1}. The spectral fit for the time-averaged spectrum of Obs. Sn. 14 (having no TKF) is shown in Fig.~\ref{fig:cepx4o2}.
}
 
\subsection{4U 1626-67}
4U 1626-67 is an XRP in a Low mass Ultra-compact XRB (UCXB) {pulsar} hosting the NS having a spin period of $7$ s \citep{1626_mcclintock_1977Natur.270..320M} and $0.04$ M$_\odot$ companion star, with a short binary orbital period of about 42 minutes \citep{1626_middleditch_1981ApJ...244.1001M}. CRSF has been reported in the spectrum at 37 keV \citep{1626_orlandini_1998ApJ...500L.163O}. Another peculiarity of the spectrum is the presence of low energy soft excess, which is usually modeled with a black-body component of kT $\sim0.6$ keV \citep[See][]{1626_kii_1986PASJ...38..751K}. 

\begin{figure}
    \centering
    \includegraphics[width=\linewidth]{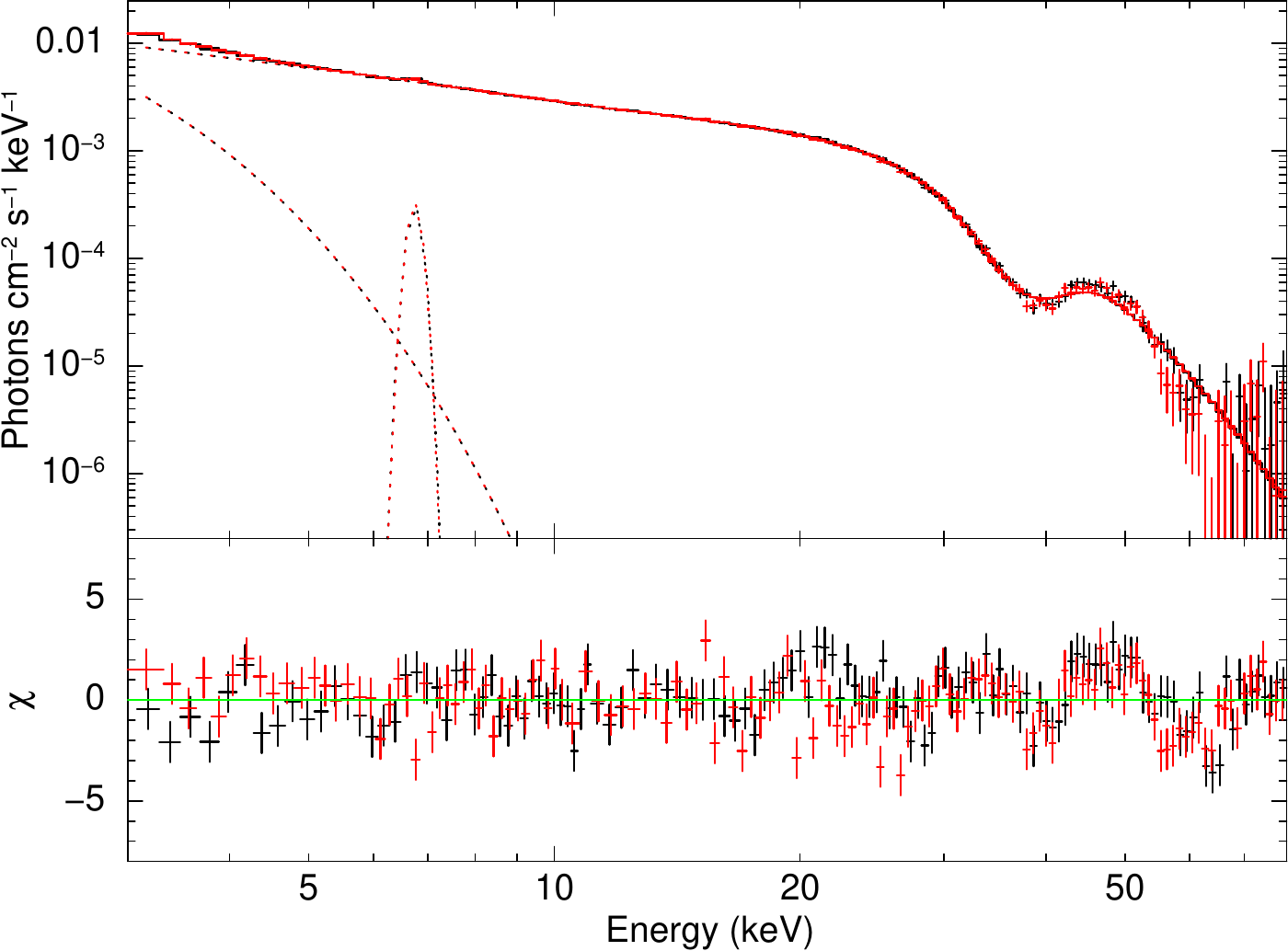}
    \caption{Spectral fit with \texttt{FDcut} model on 4U 1626-67 Obs. Sn.16, having no \textit{TKF}. The top panel shows the best-fit model and the bottom panel shows the residuals to the best-fit model.}
    \label{fig:1626}
\end{figure}

{ We have analyzed the spectrum from the observation Obs. Sn.16 (Table~\ref{tab:obs_catalogue}). All the power-law continuum models having a high energy cutoff could fit the spectrum well. The iron emission line was fitted with a \texttt{gaussian} model. CRSF at around 37 keV was fitted with a \texttt{gabs} model.

None of the best fitting training models could fit the validation data set of Obs. Sn.16. Therefore, single-step fitting was performed on Obs. Sn.16 skipping training. The best fitting model on Obs. Sn.16 does not contain \textit{TKF}.

The spectral parameters of the best fitting models for Obs. Sn.16 of 4U 1626-67 are given in Tables \ref{tab:cyc1} and \ref{tab:tkf1}. The spectral fit for the time-averaged spectrum of Obs. Sn.16 (having no TKF) is shown in Fig.~\ref{fig:1626}.
}

\subsection{SMC X--2}\label{sec:smcx2}
SMC X--2 (2S 0052-739) is a bright transient HMXB XRP in the Small Magellanic Cloud (SMC) dwarf galaxy, located 65 kpc away \citep{smcx2_li_1977IAUC.3125....1L}. The binary hosts the pulsar having spin period $\sim2.37$ s  \citep{smcx2_corbet_2001ApJ...548L..41C} and an O-type stellar companion \citep{smcx2_crampton_1978ApJ...223L..79C} in a 28.6 d orbit \citep{smcx2_schurch_2008ATel.1670....1S}. The X-ray spectrum is generally modelled with an absorbed power-law continuum modified by CRSF centered at $\sim27$ keV \citep{SMCX-2_jaiswal_2016MNRAS.461L..97J}.

\begin{figure}
    \centering
    \includegraphics[width=\linewidth]{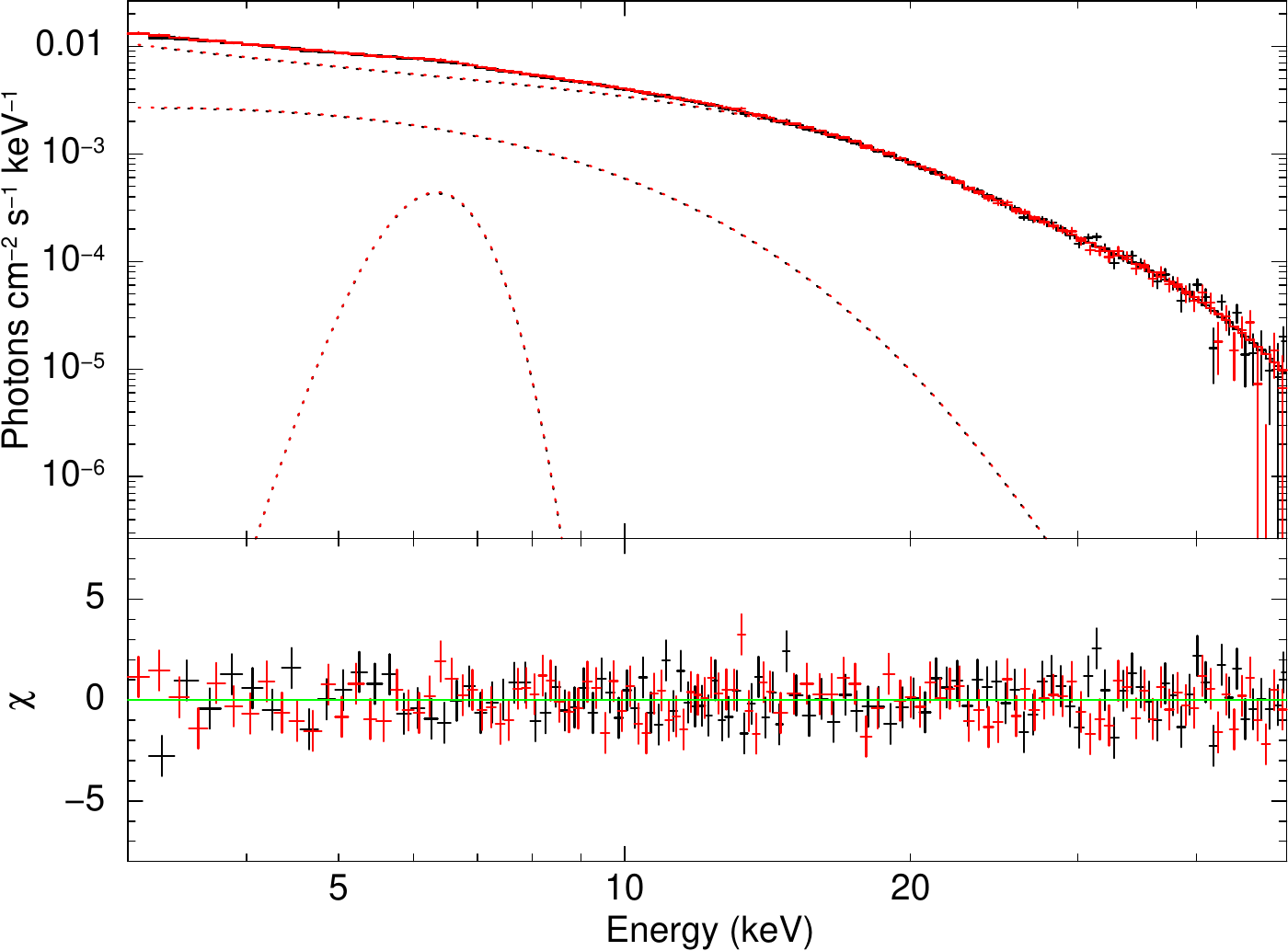}
    \caption{Spectral fit on SMC X--2 Obs Sn.17 with \texttt{compTT} continuum model, having no \textit{TKF}. The top panel shows the best-fit model and the bottom panel shows the residuals to the best-fit model.}
    \label{fig:smcx2}
\end{figure}

{We analysed the spectrum from three observations Obs. Sn. 17, 18 and 19 (Table~\ref{tab:obs_catalogue}) that probed the outburst of the source in 2015. Iron fluorescence line was present, which was fitted with a \texttt{gaussian}. CRSF at $\sim27$ keV was fitted with a  \texttt{gabs}. The width of CRSF was not constrained by the fit in Obs. Sn.19, it was thus fixed to the average width obtained from other observations.

The best fitting models of all three observations did not require \textit{TKF}.

The spectral parameters of the best fitting models for Obs. Sn.17, 18 and 19 of SMC X--2 are given in Tables \ref{tab:cyc1} and \ref{tab:tkf1}. The spectral fit for Obs. Sn.17 (having no TKF) is shown in Fig.~\ref{fig:smcx2}.}

\subsection{IGR J17544--2619}
IGR J17544--2619 is a Super-giant fast X-ray transient located $4$ kpc away \citep[See][and references therein]{17544_rampy_2009ApJ...707..243R} in which, the NS in an HMXB accretes from the wind of an O-type Super-giant companion. The binary has an orbital period of 5 d \citep{17544_clark_2009MNRAS.399L.113C}, and the NS has an (unconfirmed) spin period of 71.5 s (\citealt{17544_drave_2012A&A...539A..21D} and \citealt{17544_drave_2014MNRAS.439.2175D}). The spectrum is usually modelled with a combination of low energy thermal blackbody component and a high-energy non-thermal comptonization component, modified by fundamental CRSF at 17 keV and its harmonic at 30 keV \citep{17544_bhalerao_2015MNRAS.447.2274B}.

\begin{figure}
    \centering
    \includegraphics[width=\linewidth]{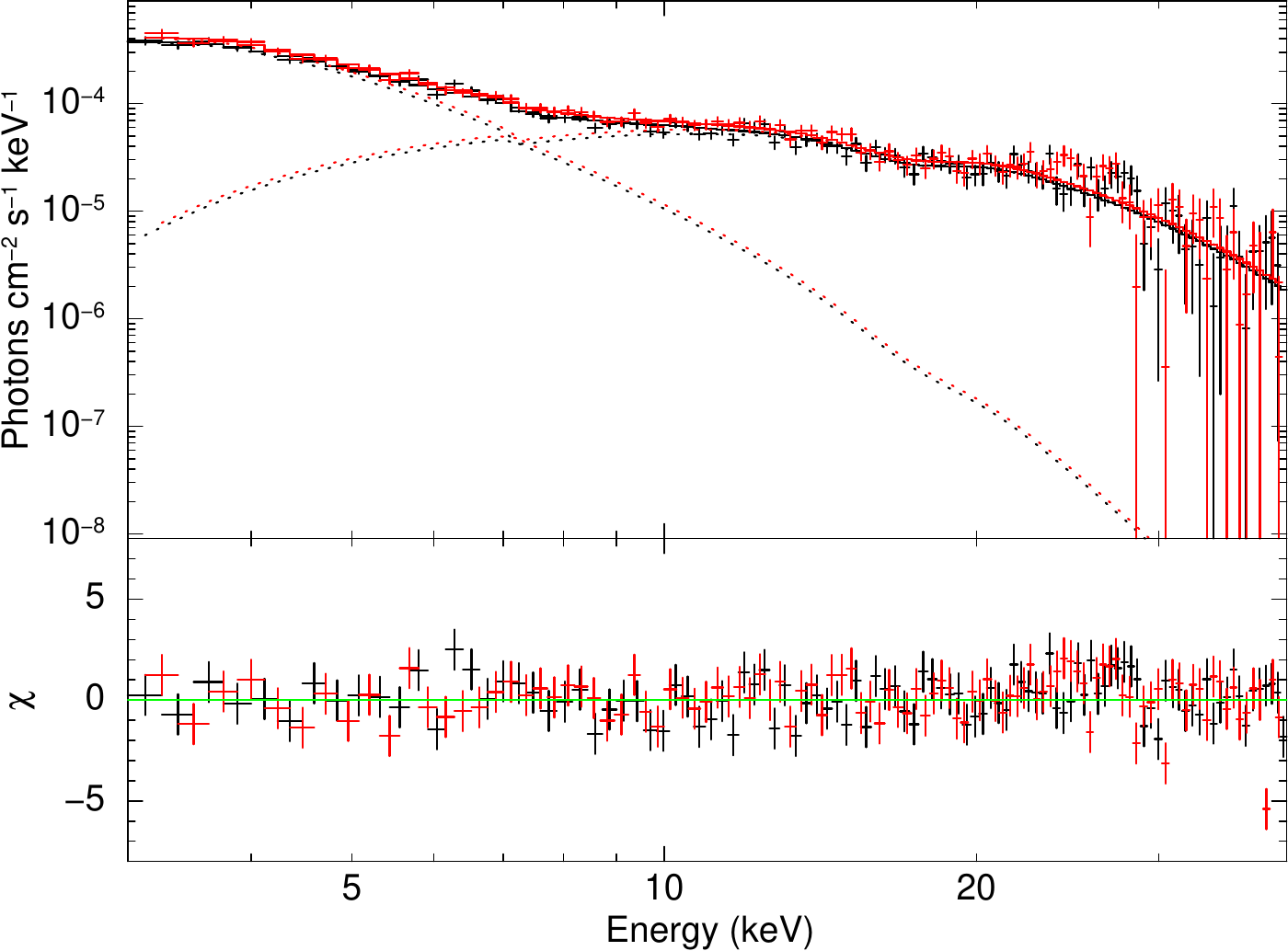}
    \caption{Spectral fit on IGR J17544--2619 Obs Sn.20 with \texttt{NPEX} continuum model, having no \textit{TKF}. The top panel shows the best-fit model and the bottom panel shows the residuals to the best-fit model.}
    \label{fig:igrj17544}
\end{figure}

{We analysed the 3$-$40 keV spectrum from Obs. Sn.20 (Table~\ref{tab:obs_catalogue}). The spectrum was dominated by background photons above 40 keV. CRSF present at $\sim18$ keV was modelled with a \texttt{gabs}.

The best fitting model on Obs. Sn.20 does not contain \textit{TKF}.

The spectral parameters of the best fitting models for Obs. Sn.20 of IGR J17544--2619 are given in Tables \ref{tab:cyc1} and \ref{tab:tkf1}. The spectral fit for Obs. Sn.20 (having no TKF) is shown in Fig.~\ref{fig:igrj17544}.
}

\subsection{IGR J16393--4643}
IGR J16393--4643 is an HMXB {pulsar} in which the NS accretes from the wind of a stellar companion of an unknown spectral class. The pulsar has a spin period of 911 s, and the binary has an orbital period of $\sim$3.7 d \citep[See][and references therein]{16393_thompson_2006ApJ...649..373T}. It shows an unusual partial eclipse \citep{Islam_2014_igrj6393} which has also been interpreted to be the absorption of X-rays in the stellar corona with a grazing line of sight to the source \citep{Kabiraj_2019}. Galactic ISM heavily obscures photons from the source and the broadband spectrum is usually modelled with a heavily absorbed power-law continuum modified by CRSF at 30 keV \citep{16393_bodaghee_2016ApJ...823..146B}.

\begin{figure}
    \centering
    \includegraphics[width=\linewidth]{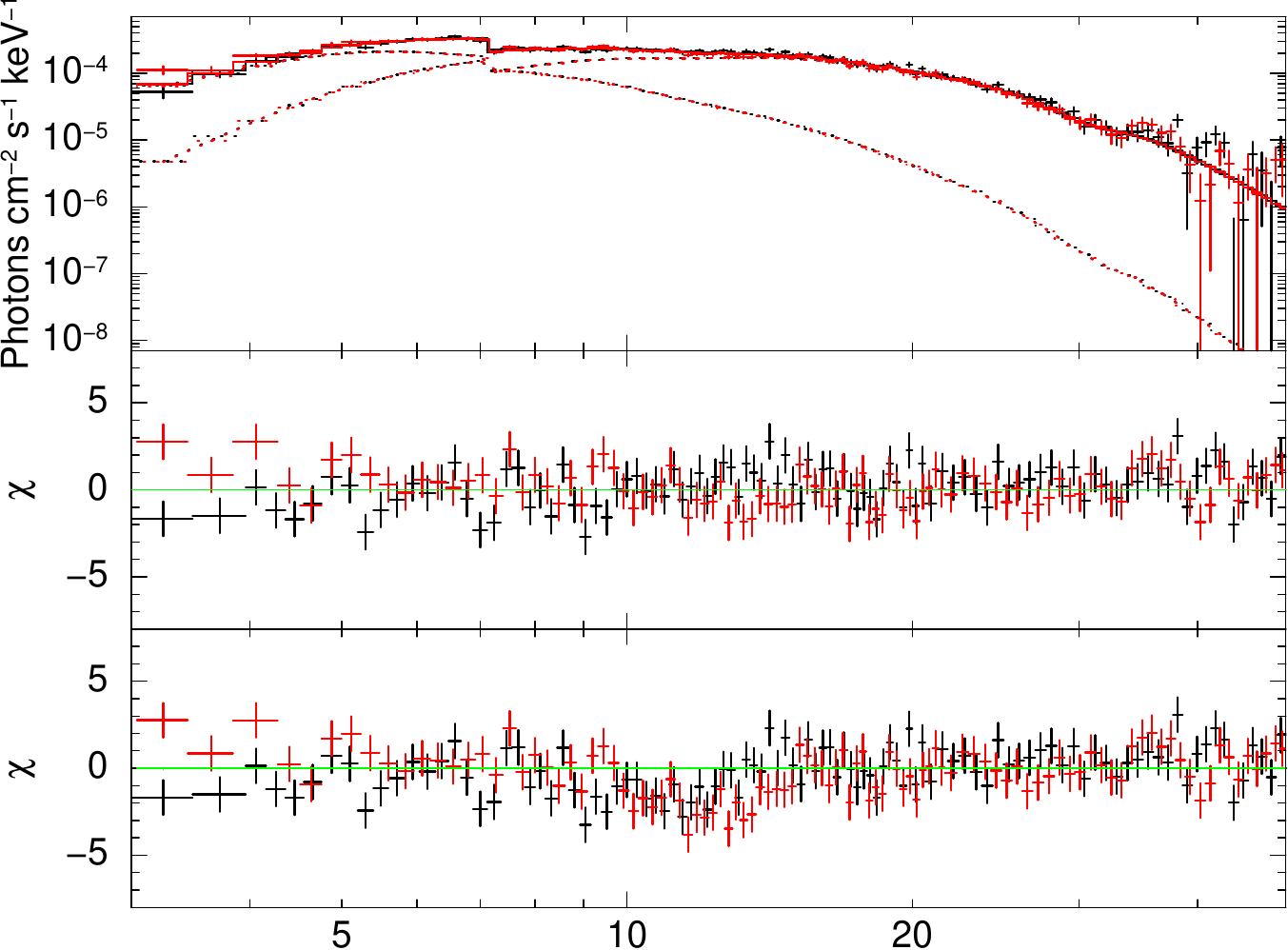}
    \caption{Spectral fit on IGR J16393--4643 Obs Sn.21 with \texttt{NPEX} continuum model, having \textit{TKF}. The top panel shows the best-fit model, the middle panel shows the residuals to the best-fit model, and the bottom panel shows the residuals when the strength of the \texttt{gabs} component modelling \textit{TKF} is set to 0.}
    \label{fig:16393}
\end{figure}

{We analysed the 3--50 keV spectrum from Obs. Sn.21 (Table~\ref{tab:obs_catalogue}). The pointed observation is contaminated by stray light from a nearby source GX 340+0, but we have taken care of this during the selection of the source region. Galactic absorption towards the source could not be constrained by the fit, therefore we fixed it to the galactic value of 2.15$\times10^{21}$ atoms cm$^{-2}$. CRSF present at $\sim30$ keV was modelled with a \texttt{gabs}.

The best fitting model on Obs. Sn.21 contain \textit{TKF}.

The spectral parameters of the best fitting models for Obs. Sn.21 of IGR J16393--4643 are given in Tables \ref{tab:cyc1} and \ref{tab:tkf1}. The spectral fit for Obs. Sn.21 (having TKF) is shown in Fig.~\ref{fig:16393}.
}

\subsection{2S 1553--542}
2S 1553--542 is a transient HMXB {pulsar} located at a distance of $20\pm4$ kpc on the far-side arms of Milkyway (\citealt{1553_tsygankov_2016MNRAS.457..258T} and \citealt{1553_lutovinov_2016MNRAS.462.3823L}). It consists of a NS with a spin period of 9.3 s accreting from a Be-type companion star in a 31 d binary orbit \citep[See][and references therein]{1553_kelley_1983ApJ...274..765K}. CRSF has been reported in the source at $\sim24$ keV \citep{1553_tsygankov_2016MNRAS.457..258T}.

\begin{figure}
    \centering
    \includegraphics[width=\linewidth]{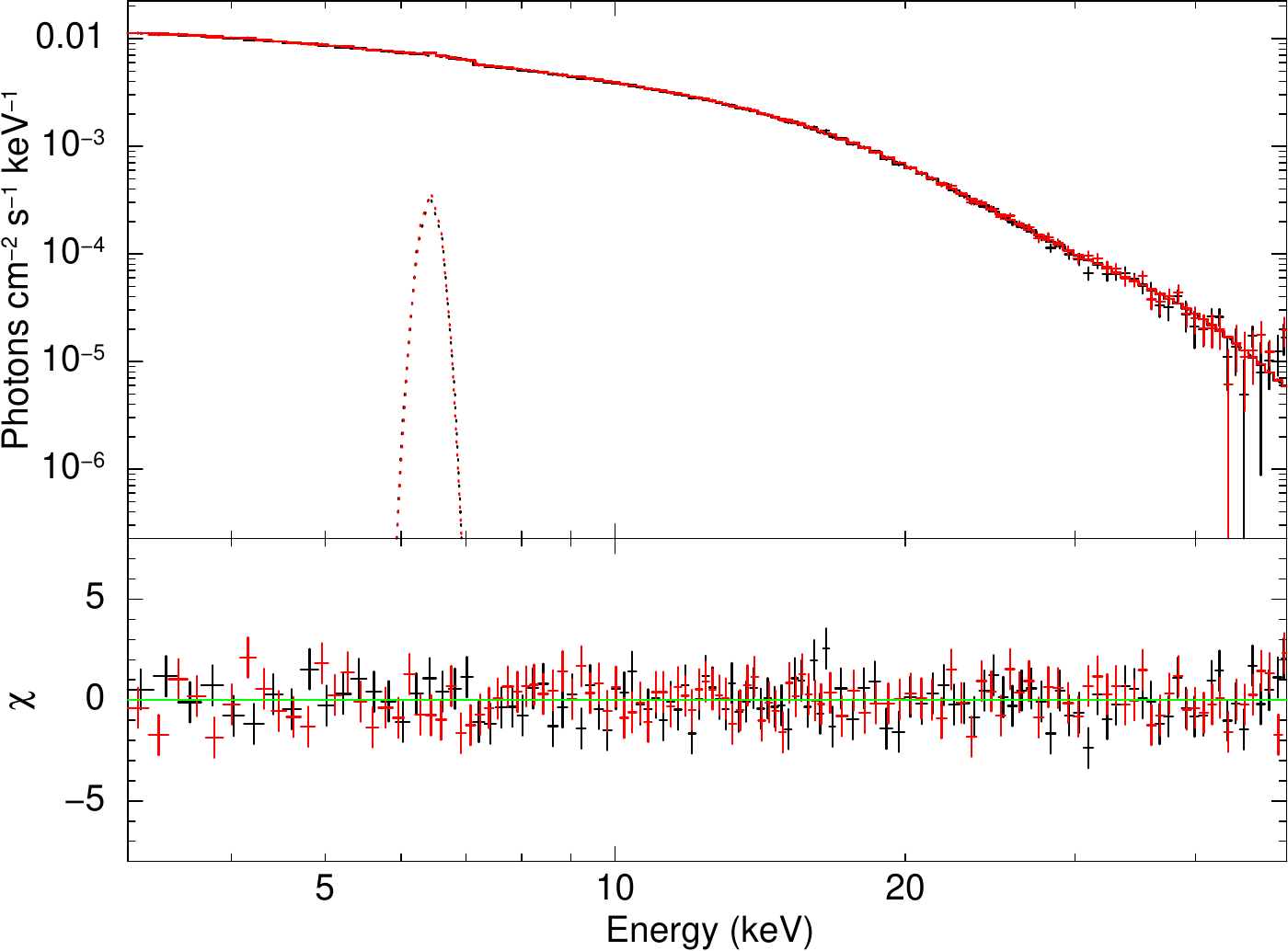}
    \caption{Spectral fit on 2S 1553--542 Obs. Sn.22 with \texttt{FDcut} continuum model, having no \textit{TKF}. The top panel shows the best-fit model and the bottom panel shows the residuals to the best-fit model.}
    \label{fig:1553}
\end{figure}

{We analysed the 3$-$50 keV spectrum from Obs. Sn. 22 (Table~\ref{tab:obs_catalogue}). Iron line emission was fitted with a \texttt{gaussian} model. CRSF at $\sim28$ keV was fitted with \texttt{gabs}.

Since a high energy cutoff was present below 15 keV, the models containing a high energy cutoff were also included while fitting the training data set. The best fitting model on Obs. Sn. 22 does not require \textit{TKF}.

The spectral parameters of the best fitting models for Obs. Sn.22 of 2S 1553--542 are given in Tables \ref{tab:cyc1} and \ref{tab:tkf1}. The spectral fit for Obs. Sn.22 (having no TKF) is shown in Fig.~\ref{fig:1553}.}

\subsection{RX J0520.5--6932}
RX J0520.5--6932 is a transient XRP in the Large Magellanic Cloud \citep{0520_schmidtke_1996PASP..108..668S}. The HMXB hosts the NS with spin period of $\sim8$ s \citep{0520_Vasilopoulos_2013ATel.5673....1V} in a $24.4$ d orbit around a Be-type companion star (See \citealt{0520_coe_2001MNRAS.324..623C}, \citealt{0520_kuehnel_2014ATel.5856....1K}, and references therein). CRSF has been reported in the spectrum at $31$ keV \citep{0520_tendulkar_2014ApJ...795..154T}.

\begin{figure}
    \centering
    \includegraphics[width=\linewidth]{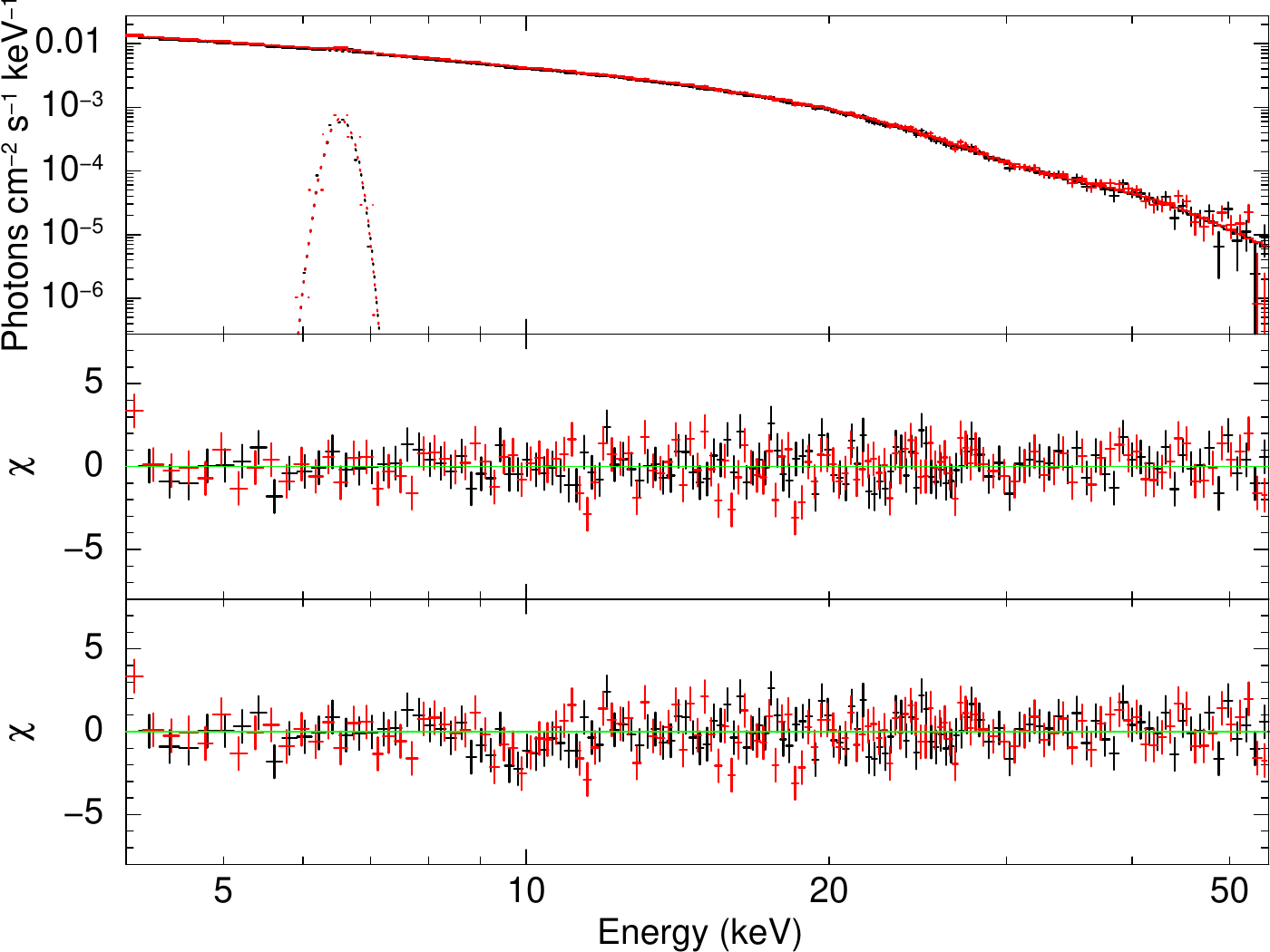}
    \caption{Spectral fit on RX J0520.5--6932 Obs. Sn.23 with \texttt{cutoffpl} continuum model, having \textit{TKF}. The top panel shows the best-fit model, the middle panel shows the residuals to the best-fit model, and the bottom panel shows the residuals when the strength of the \texttt{gabs} component modelling \textit{TKF} is set to 0.}
    \label{fig:rxj0520}
\end{figure}

{ We analyzed the spectrum from two observations Obs. Sn.23 and 24 (Table~\ref{tab:obs_catalogue}). The spectra are background-dominated above 55 keV in both observations. Also, FPMA and FPMB spectra showed cross-calibration anomalies in 3--4 keV for Obs. Sn.23. Hence we analyzed the 3$-$55 keV spectrum of Obs. Sn.23 and 4$-$55 keV spectrum of Obs. Sn.24. Galactic absorption towards the source could not be constrained by the fit, therefore we fixed it to the galactic value of $2\times10^{21}$ atoms cm$^{-2}$. Iron emission line was fitted with a \texttt{gaussian}. CRSF at $\sim30$ keV was fitted with a \texttt{gabs}. 

The best fitting models on both observations contain \textit{TKF}.

The spectral parameters of the best fitting models for Obs. Sn.23 and 24 of RX J0520.5--6932 are given in Tables \ref{tab:cyc1} and \ref{tab:tkf1}. The spectral fit for Obs. Sn.23 (having TKF) is shown in Fig.~\ref{fig:rxj0520}.}

\subsection{Cen X--3}
Cen X--3 is a bright and persistent HMXB {pulsar} that hosts a NS that accretes from the wind of an O-type companion star. The binary has an orbital period of about 2.1 d \citep{cenx3_schreier_1972ApJ...172L..79S}, and the NS has a spin period of $4.8$ s \citep{cenx3_giacconi_1971ApJ...167L..67G}. CRSF is present in the source at 30 keV \citep[See][and references therein]{cenx3_gunjan_2021MNRAS.500.3454T}.

\begin{figure}
    \centering
    \includegraphics[width=\linewidth]{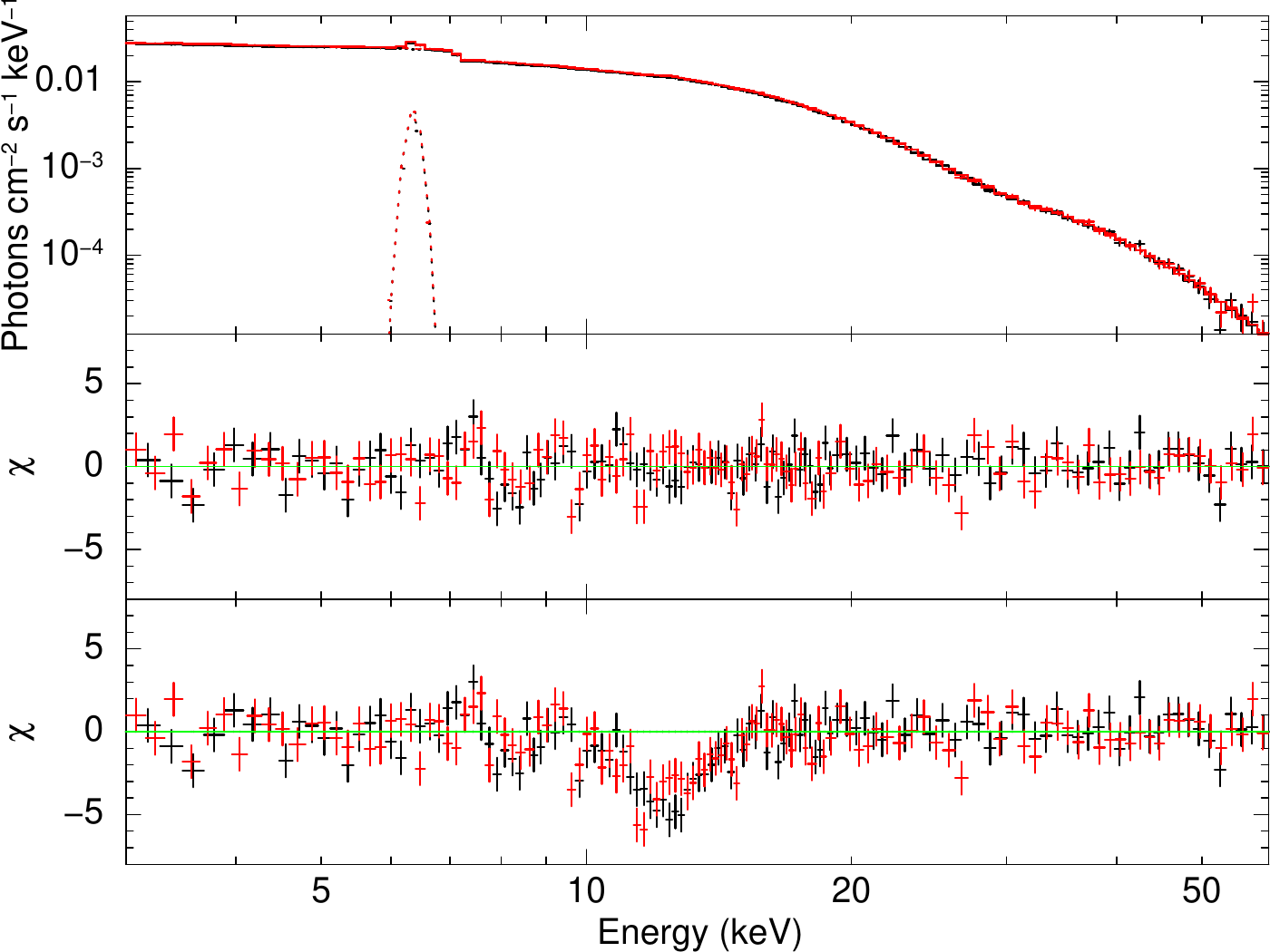}
    \caption{Spectral fit with \texttt{mplccut} model on Cen X--3 Obs. Sn.25, having \textit{TKF}. The top panel shows the best-fit model, the middle panel shows the residuals to the best-fit model, and the bottom panel shows the residuals when the strength of the \texttt{gabs} component modelling \textit{TKF} is set to 0.}
    \label{fig:cenx3}
\end{figure}

{ We analysed the  3$-$60 keV \textit{NuSTAR} spectrum from Obs. Sn.25 (Table~\ref{tab:obs_catalogue}). Iron emission line was fitted with a \texttt{gaussian}. CRSF present at $\sim30$ keV was fitted with a \texttt{gabs}. 

Since a high energy cutoff was present below 15 keV, the models containing a high energy cutoff were also included while fitting the training data set. None of the best fitting training models could fit the validation data set of Obs. Sn.25. Therefore, single-step fitting was performed skipping training. The best fitting model on Obs. Sn.25 contain \textit{TKF}.

The spectral parameters of the best fitting models for Obs. Sn.25 of Cen X--3 are given in Tables \ref{tab:cyc1} and \ref{tab:tkf1}. The spectral fit for Obs. Sn.25 (having TKF) is shown in Fig.~\ref{fig:cenx3}.}

\subsection{GX 301--2}\label{gx301m2}
GX 301--2 is a bright HMXB {pulsar} located $\sim4$ kpc away, hosting a NS accreting from the wind of a hyper-giant 35 M$_\odot$ donor star. The pulsar has a relatively long spin period of $\sim680$ s, and the binary has an eccentric orbit with an orbital period of $41$ d which is manifested as periodic flares during the periastron passage of the NS \citep[See][and references therein]{3012_doroshenko_2010_refId0}. Its broadband spectrum is usually modelled by a power-law continuum modified by two distinct CRSFs at $35$ keV and $55$ keV \citep[See][and references therein]{nabizadeh_2019A&A...629A.101N}.

\begin{figure}
    \centering
    \includegraphics[width=\linewidth]{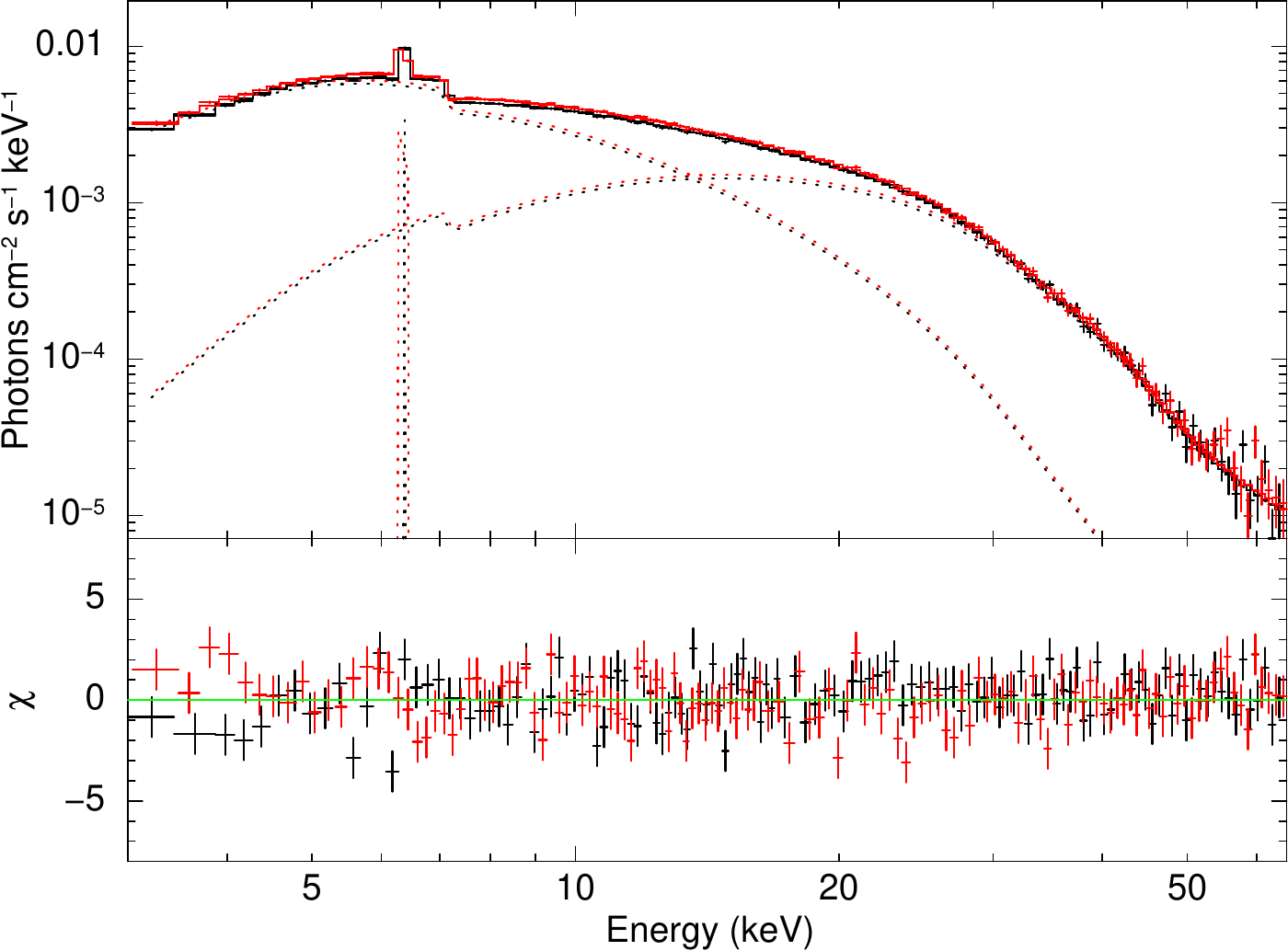}
    \caption{Spectral fit using \texttt{mplcut} model on GX 301--2 Obs. Sn.26, having no \textit{TKF}. The top panel shows the best-fit model and the bottom panel shows the residuals to the best-fit model.}
    \label{fig:gx301m2}
\end{figure}

{ We analysed the 3$-$65 keV spectra from two observations Obs. Sn.26 and 27 respectively in Table~\ref{tab:obs_catalogue}. Since Obs. Sn.27 contains a flare, and therefore we analysed time-averaged, flare and out-of-flare spectra separately. Iron fluorescence line was present in all the spectra, which were fitted with a \texttt{gaussian}. The width of the iron emission line was not constrained by the fit in the out-of-flare-state spectrum of Obs. Sn.38 and we fixed it to 10 eV. Two CRSFs were present at $\sim30$ keV and $\sim50$ keV, and those were fitted with two \texttt{gabs}. The width of CRSF at $50$ keV could be constrained in Obs. Sn.27 out-of-flare state and the flaring state spectrum of Obs. Sn.27. Therefore, it was frozen to the CRSF width obtained from time-averaged spectrum of Obs. Sn.27.

The best fitting models on none of the spectra contain \textit{TKF}.

The spectral parameters of the best fitting models for Obs. Sn.26 and 27 of GX 301--2 are given in Tables \ref{tab:cyc1} and \ref{tab:tkf1}. The spectral fit for Obs. Sn.26 (having no TKF) is shown in Fig.~\ref{fig:gx301m2}.}

\subsection{XTE J1829--098}\label{xtej1829}
XTE J1829--098 is a transient XRP having a spin period of $7.8$ s. Even though the nature of the binary is considered an HMXB based on the observed hard spectrum, the classification of the companion is still uncertain. Assuming an O- or B-type nature to the companion, \cite{xtej1829_dist_halpern2007x} estimated the distance to the binary as 10 kpc. The binary orbital period is considered to be the outburst recurrence interval of  $\sim246$ d. Its broadband spectrum is usually modeled with a power-law continuum, and a CRSF was reported at $\sim15$ keV \citep[See][and references therein]{shtykovsky_2019MNRAS.482L..14S}.

\begin{figure}
    \centering
    \includegraphics[width=\linewidth]{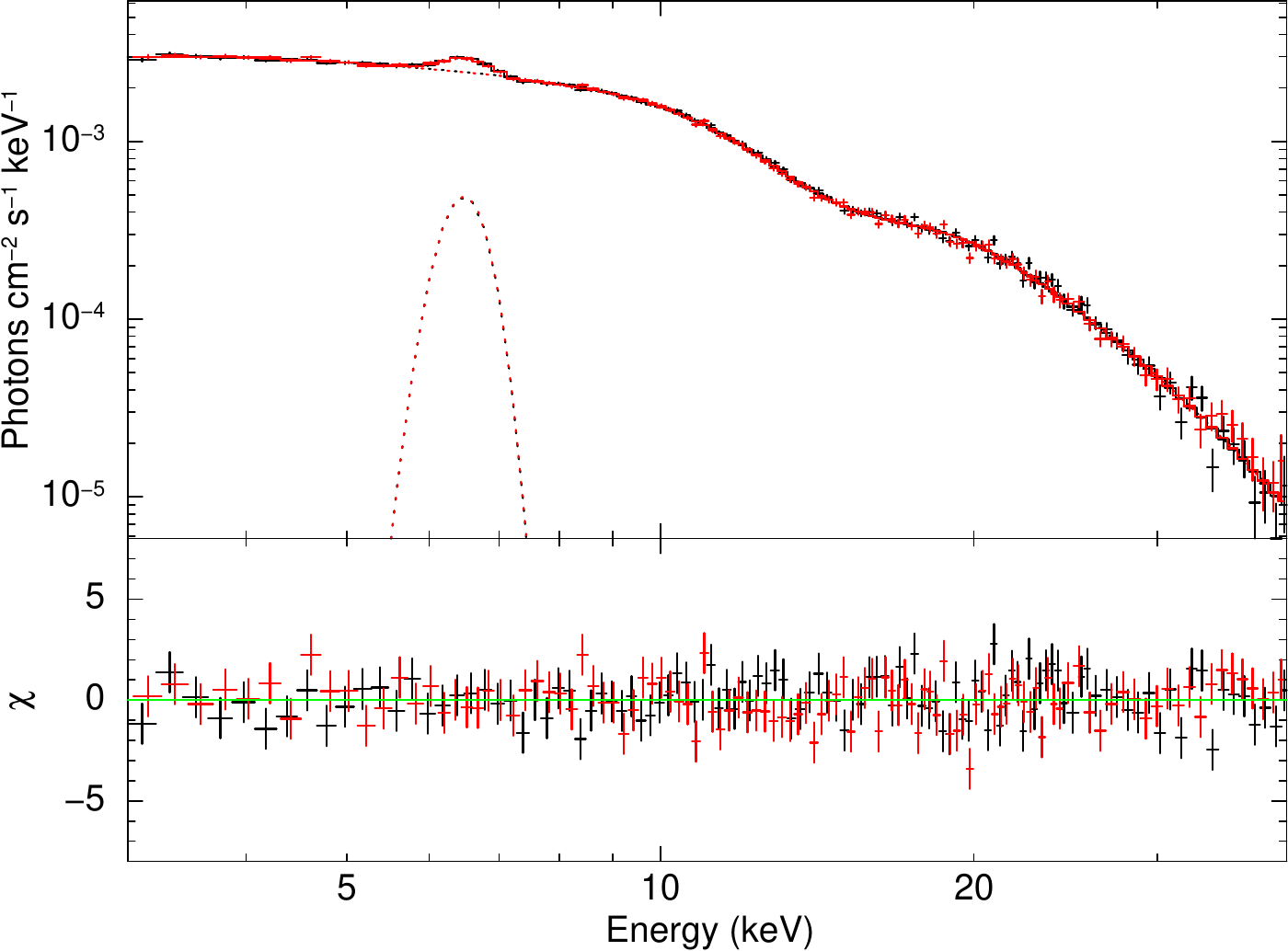}
    \caption{Spectral fit using \texttt{newhcut} model on XTE J1829--098 Obs. Sn. 28, having no \textit{TKF}. The top panel shows the best-fit model and the bottom panel shows the residuals to the best-fit model.}
    \label{fig:xtej1829}
\end{figure}

{We analyzed the 3$-$40 keV \textit{NuSTAR} spectrum of the source (Obs. Sn.28 in Table~\ref{tab:obs_catalogue}) that probed an outburst of the source in 2018. The ISM absorption column density could be constrained by the fit, and the iron fluorescence line was fitted with a \texttt{gaussian}. CRSF present at $\sim15$ keV was fitted with a \texttt{gabs} model. The cutoff energy of the best fit power-law with high energy cutoff models is close to $\sim10$ keV. Moreover, the CRSF present at $\sim15$ keV would impact the \textit{TKF} estimation. Therefore we performed direct validation to find the best-fitting model without \textit{TKF}.

The spectral parameters of the best fitting models for Obs. Sn.28 of XTE J1829--098 are given in Tables \ref{tab:cyc1} and \ref{tab:tkf1}. The spectral fit for Obs. Sn.28 (having no TKF) is shown in Fig.~\ref{fig:xtej1829}.}

\subsection{V 0332+63}
V 0332+53 is a bright transient {HMXB pulsar} located 7 kpc away hosting an O-type stellar companion and the NS having a spin period of $4.4$ s. The binary has an eccentric orbit with an orbital period of 34 d. This is one of the few pulsars that exhibit multiple harmonics of the CRSF in the spectrum. The spectrum is usually modelled by a power-law continuum modified by a CRSF at 28 keV and its two harmonics at 49 keV and 72 keV respectively \citep[See][and references therein]{v0332_tsygankov_10.1111/j.1365-2966.2006.10610.x}.

\begin{figure}
    \centering
    \includegraphics[width=\linewidth]{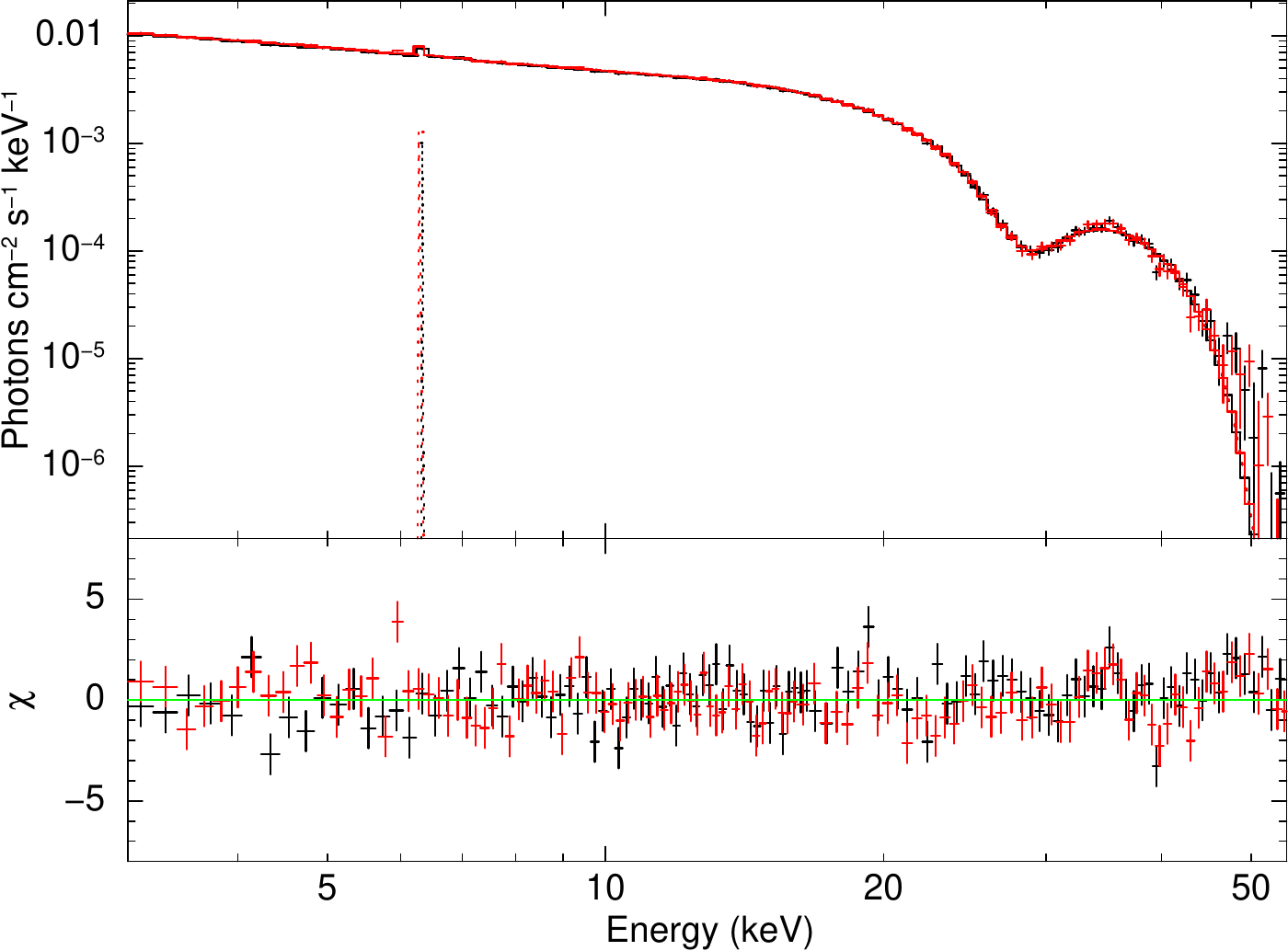}
    \caption{Spectral fit on V 0332+53 Obs. Sn.30 with \texttt{newhcut} continuum model, having no \textit{TKF}. The top panel shows the best-fit model and the bottom panel shows the residuals to the best-fit model.}
    \label{fig:v0332p53_o2}
\end{figure}

{We analysed the 3$-$55 keV spectrum from the four observations Obs. Sn.29, 30, 31 and 32 (Table~\ref{tab:obs_catalogue}). Iron fluorescence line was present in all the spectra, which were fitted with a \texttt{gaussian}. The width of the iron emission line was not constrained by the fits and we fixed it to 10 eV. CRSF at $\sim28$ keV and its first harmonic at $\sim56$ keV fitted with \texttt{cyclabs}. We also tried using \texttt{gabs} to fit the CRSF, but it left residuals at high energies. The width of harmonic CRSF could not be constrained in the best fit models on Obs. Sn.29 and 30 containing \textit{TKF}. Therefore, it was fixed to the CRSF width obtained from the best fit models on Obs. Sn.29 and 30 that does not contain \textit{TKF}.

The best-fitting models on Obs. Sn.29 and 32 do not contain \textit{TKF}. Even though the best fitting model Obs. Sn.30 and 31 contain \textit{TKF}, there also exist similar well-fitting models in which \textit{TKF} is absent.

The spectral parameters of the best fitting models for all four observations of V 0332+63 are given in Tables \ref{tab:cyc1} and \ref{tab:tkf1}. The spectral fit for Obs. Sn.30 (having no TKF) is shown in Fig.~\ref{fig:v0332p53_o2}.}

\subsection{XTE J1858+034}
XTE J1858+034 is a transient Be-XRP with spin period $\sim221$ s located $\sim10$ kpc away. The regular outburst period of $\sim380$ s is considered as the orbital period of the binary. The broadband spectrum is generally modelled by a power-law continuum with a cutoff at high energy, modified by iron fluorescence line at $6.5$ keV and CRSF at $\sim48$ keV \citep[See][and references therein]{malacaria_1858034_2021ApJ...909..153M}.

\begin{figure}
    \centering
    \includegraphics[width=\linewidth]{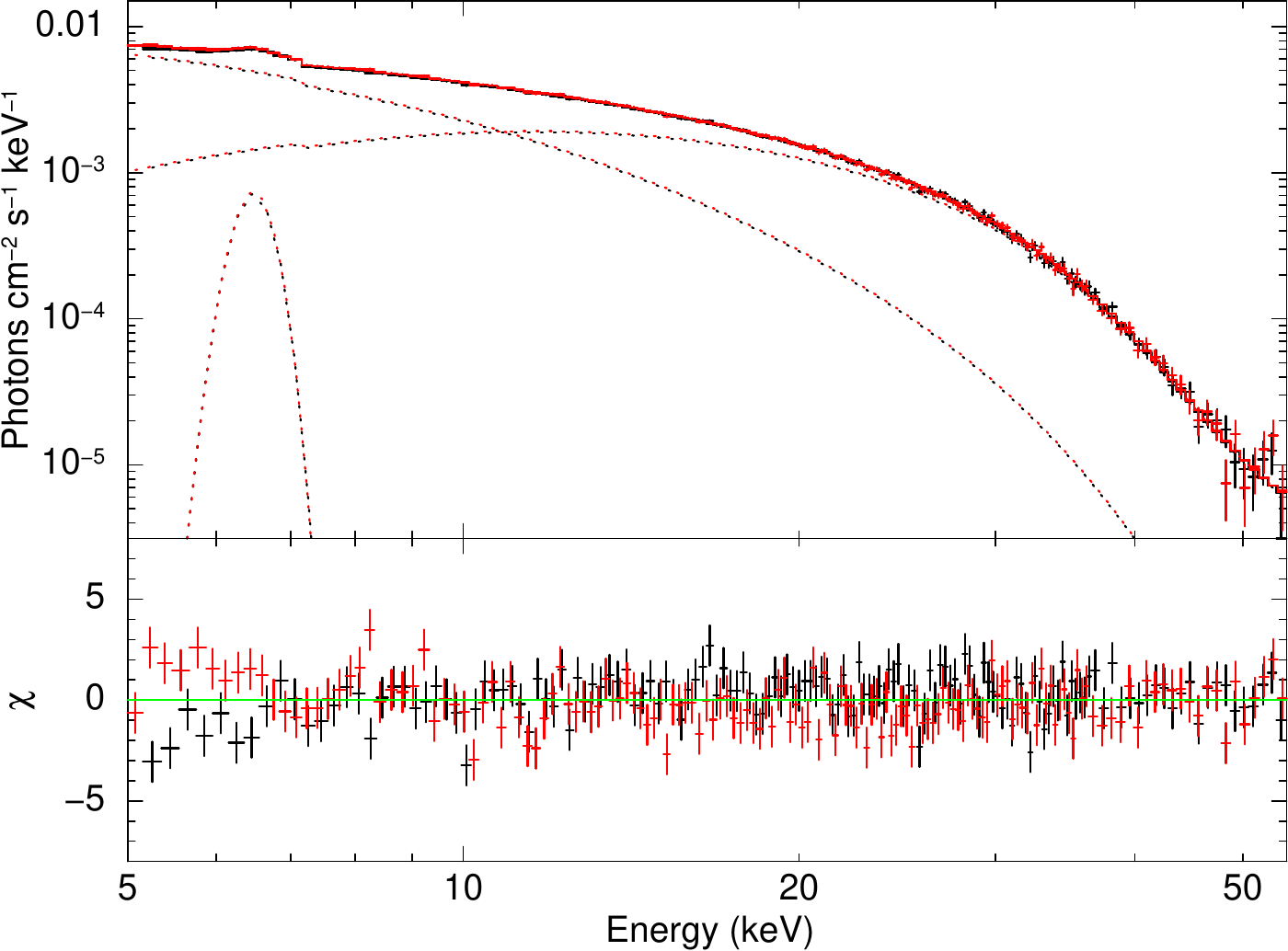}
    \caption{Spectral fit using \texttt{NPEX} continuum model on XTE J1858+034 Obs. Sn.33, having no \textit{TKF}. The top panel shows the best-fit model and the bottom panel shows the residuals to the best-fit model.}
    \label{fig:1858p034}
\end{figure}

{We analysed the 5--55 keV spectrum of the source. The spectrum is background-dominated above 55 keV, and the FPMA/FPMB spectra below 5 keV showed cross-calibration anomalies. Iron fluorescence line was present, which was fitted with a \texttt{gaussian}. CRSF centred around 50 keV was fitted with \texttt{gabs}.

The best-fitting models on Obs. Sn.33 does not contain \textit{TKF}.

The spectral parameters of the best fitting models for Obs. Sn.33 of XTE J1858+034 are given in Tables \ref{tab:cyc2} and \ref{tab:tkf2}. The spectral fit for Obs. Sn.33 (having no TKF) is shown in Fig.~\ref{fig:1858p034}.}

\subsection{4U 1700--37}\label{subsec:4u1700}
4U 1700--37 is a transient HMXB in which a NS accretes from the wind of a supergiant O-type star in a $3.5$ d binary orbit. Even though coherent pulsations were reported at $67$ s from Tenma observation during a flare, none of the other observatories has detected pulsations in the subsequent observations. The spectrum of 4U 1700--37 is usually modelled with power-law continuum models, and CRSF has been reported at two different energies of $16$ keV \citep{4U1700_118_2020MNRAS.493.3045B} and $37$ keV \citep{4u1700_150_reynolds1999bepposax}.

Obs. Sn.34 (Table~\ref{tab:obs_catalogue}) contains a flare as evident from the light curve. We, therefore, analysed the 3$-$79 keV \textit{NuSTAR} time-averaged, flare and out-of-flare spectra from Obs. Sn.34. Galactic absorption towards the source could not be constrained by the fit, therefore we fixed it to the galactic value of $5\times10^{21}$ atoms cm\textsuperscript{-2}. Iron fluorescence line was present, which is fitted with a \texttt{gaussian}. The width of the iron emission line was not constrained by the fit and we fixed it to 10 eV. CRSF at $\sim16$ keV was fitted with a \texttt{gabs} model. The width of CRSF was not constrained by the fit in the flaring state spectrum, it was thus fixed to the width obtained from the time-averaged spectrum.

The best fitting models on the time-averaged, flaring state and out-of-flare state spectrum of Obs. Sn.34 contains \textit{TKF}.

\begin{figure}
     \centering
     \includegraphics[width=\linewidth]{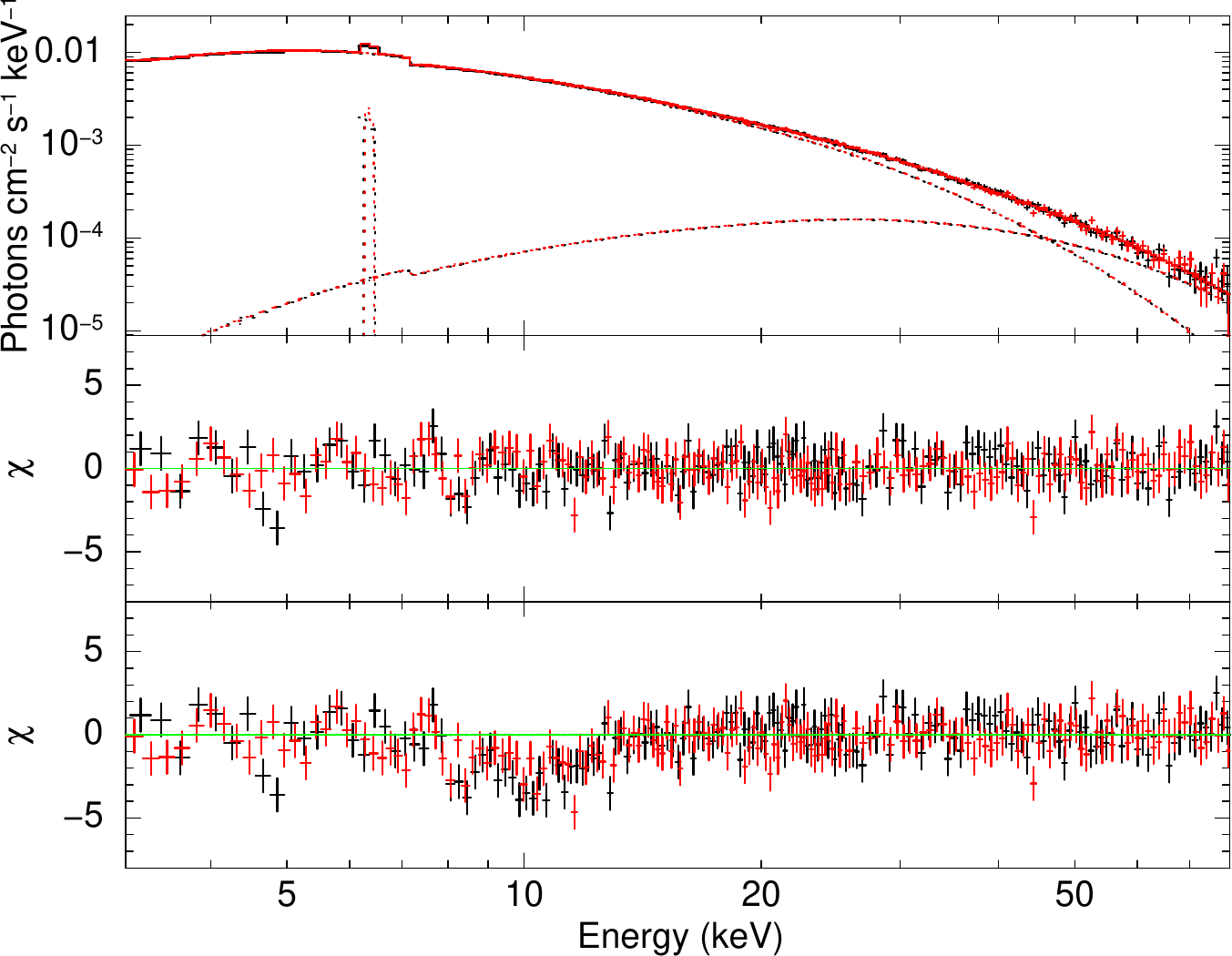}
     \caption{Spectral fit using \texttt{NPEX} continuum model on 4U 1700--37 Obs. Sn.34 time-averaged spectrum, having \textit{TKF}. The top panel shows the best-fit model, the middle panel shows the residuals to the best-fit model, and the bottom panel shows the residuals when the strength of the \texttt{gabs} component modelling \textit{TKF} is set to 0.}
     \label{fig:4u1700}
 \end{figure}

The spectral parameters of the best fitting models for Obs. Sn.34 of 4U 1700--37 are given in Tables \ref{tab:cyc2} and \ref{tab:tkf2}. The spectral fit for Obs. Sn.34 (having TKF) is shown in Fig.~\ref{fig:4u1700}.
 
 \subsection{LMC X--4}\label{subsec:lmcx4}
 LMC X--4 is a persistent flaring {HMXB pulsar} (P\textsubscript{spin}$\sim$13.5 s) located $\sim50$ kpc away in the Large Magellanic Cloud (LMC) satellite galaxy, hosting a NS pulsar that accretes from the disk-fed matter of an O-type companion star in a $1.4$ d orbit. The spectrum of LMC X--4 is usually fitted with power-law continuum models with fluorescence lines of iron \citep{lmcx4_120_levine1991lmc}.
 
\begin{figure}
    \centering
    \includegraphics[width=\linewidth]{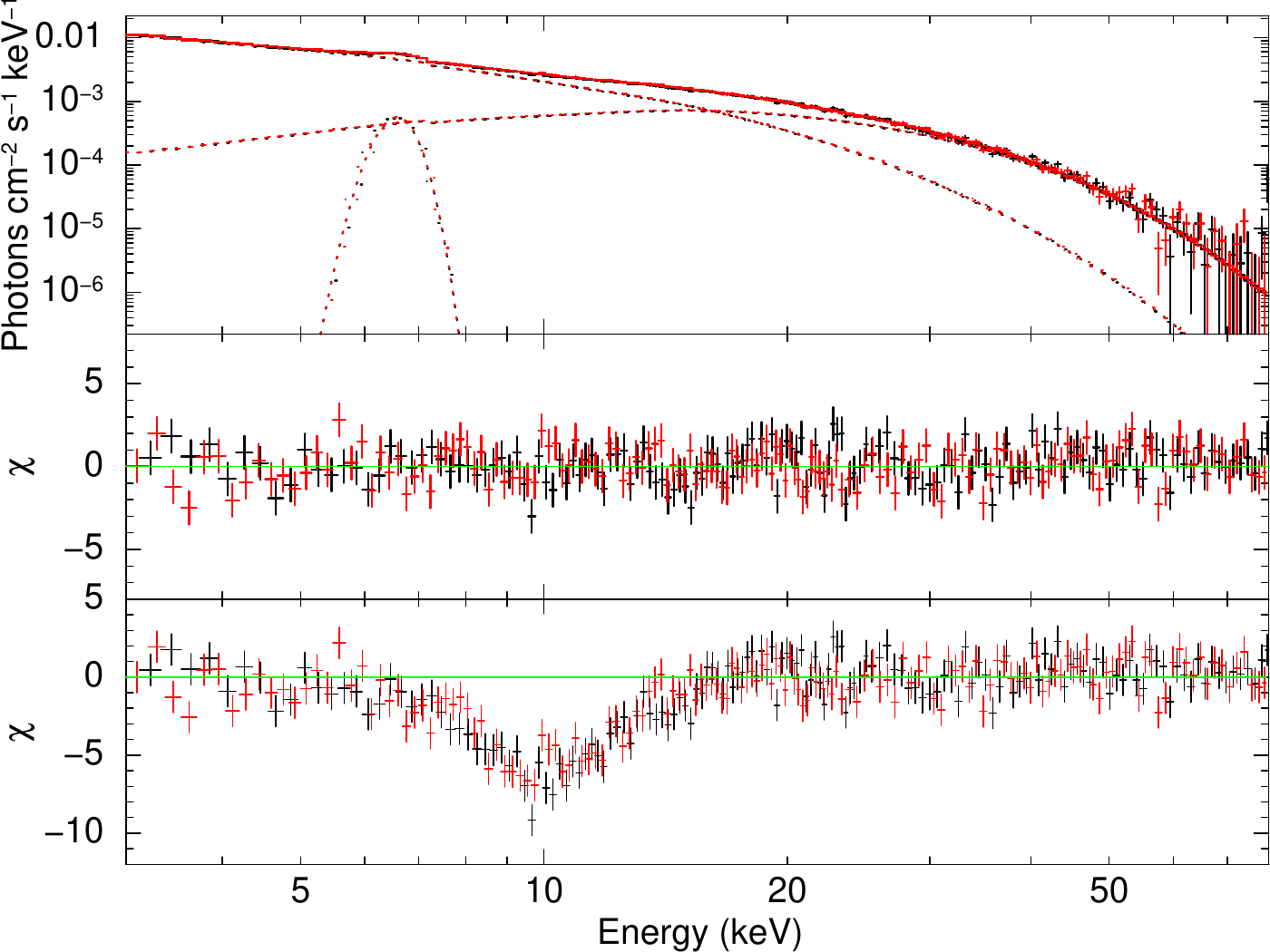}
    \includegraphics[width=\linewidth]{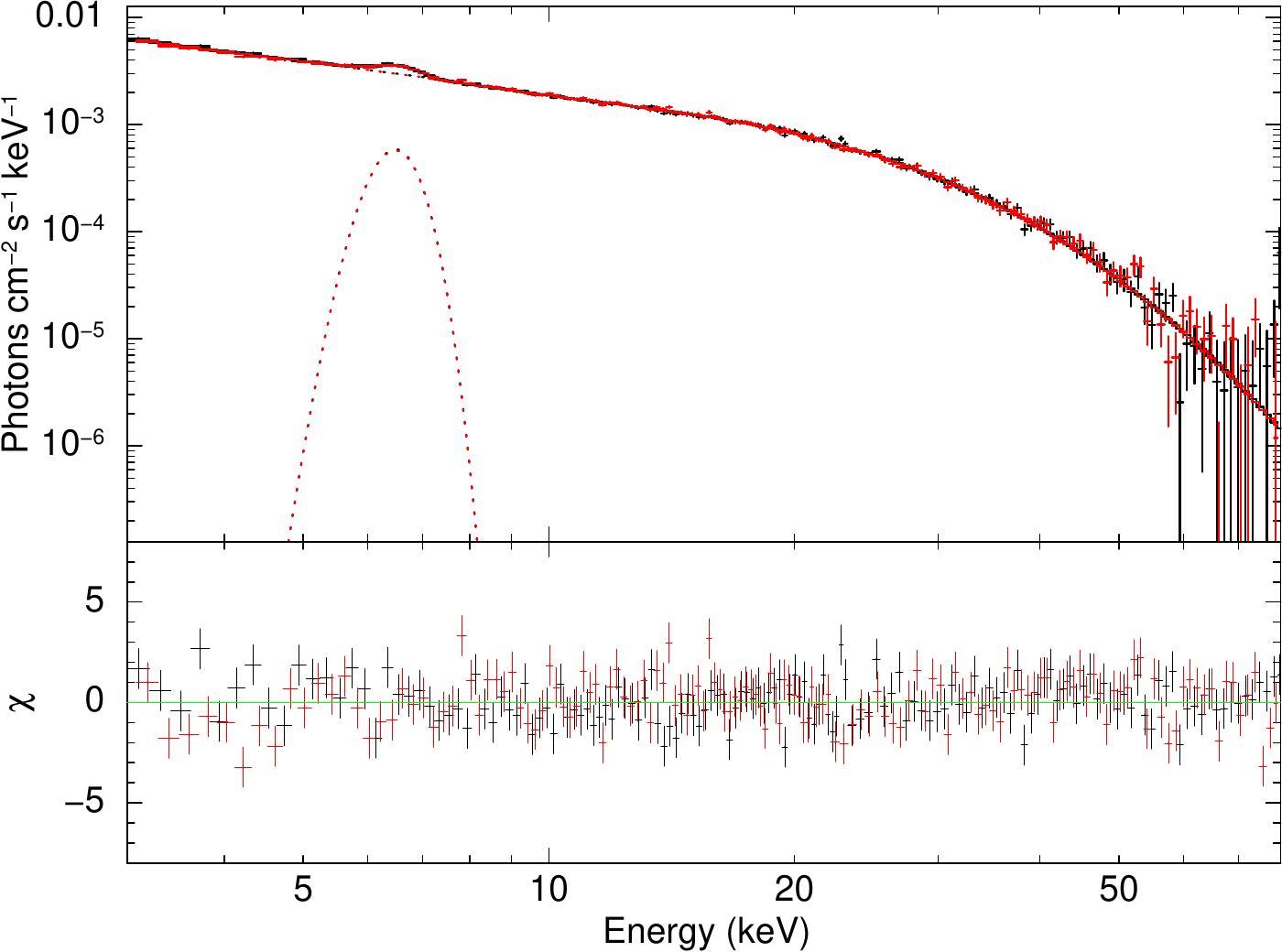}
    \caption{Top: Spectral fit on LMC X--4 Obs. Sn.35 time-averaged spectrum. with \texttt{NPEX} continuum model, having \textit{TKF}. The top panel shows the best-fit model, the middle panel shows the residuals to the best-fit model, and the bottom panel shows the residuals when the strength of the \texttt{gabs} component modelling \textit{TKF} is set to 0. Bottom: Spectral fit on LMC X--4 Obs. Sn.35 out-of-flare state spectrum with \texttt{FDcut} continuum model, having no \textit{TKF}. The top panel shows the best-fit model and the bottom panel shows the residuals to the best-fit model.}
    \label{fig:lmcx4}
\end{figure}

{ We analysed six spectra from four \textit{NuSTAR} observations Obs. Sn.35, 36, 37 and 38 (Table~\ref{tab:obs_catalogue}). Obs. Sn.35 and 38 have flares in the light curves, and therefore the separated flare and out-of-flare spectra were independently analysed. Galactic absorption towards the source could not be constrained by the fit, therefore we fixed it to the galactic value of $8\times10^{20}$ atoms cm\textsuperscript{-2}. Iron fluorescence line was present in all the spectra, which were fitted with a \texttt{gaussian}. The width of the iron emission line was not constrained by the fit in the out-of-flare-state spectrum of Obs. Sn.38 and we fixed it to 10 eV.

The best fitting models on the time-averaged and flaring-state spectrum of Obs. Sn.35, and time-averaged spectrum of Obs. Sn.38 contain \textit{TKF}. The best fitting model on the out-of-flare-state spectrum of Obs. Sn.38 does not contain \textit{TKF}. Even though the best fitting models on the out-of-flare-state spectrum of Obs. Sn.35, the flaring-state spectrum of Obs. Sn.38 and time-averaged spectra of Obs. Sn.36 and 37 contain \textit{TKF}, there also exist similar well-fitting models in which \textit{TKF} is absent.
 
The spectral parameters of the best fitting models for all four observations of LMC X--4 are given in Tables \ref{tab:cyc2} and \ref{tab:tkf2}. The spectral fit for the time-averaged spectrum of Obs. Sn.35 (having TKF) and out-of-flare-state spectrum of Obs. Sn.35 (having no TKF) are shown in Fig.~\ref{fig:lmcx4}.}
 
 \subsection{IGR J17329-2731}\label{subsec:igr17329}
 
IGR J17329-2731 is a transient XRP in a Symbiotic X-ray Binary (SyXB) hosting the slow spinning $\sim6680$ s pulsar accreting from the wind of a late M-type giant stellar companion. The source is estimated to be at a distance of $2.7$ kpc. The broadband spectrum is modelled with a power-law continuum modified by CRSF at $\sim21$ keV and iron fluorescence lines \citep{igrj17329_119_bozzo2018igr}.
 
\begin{figure}
     \centering
     \includegraphics[width=\linewidth]{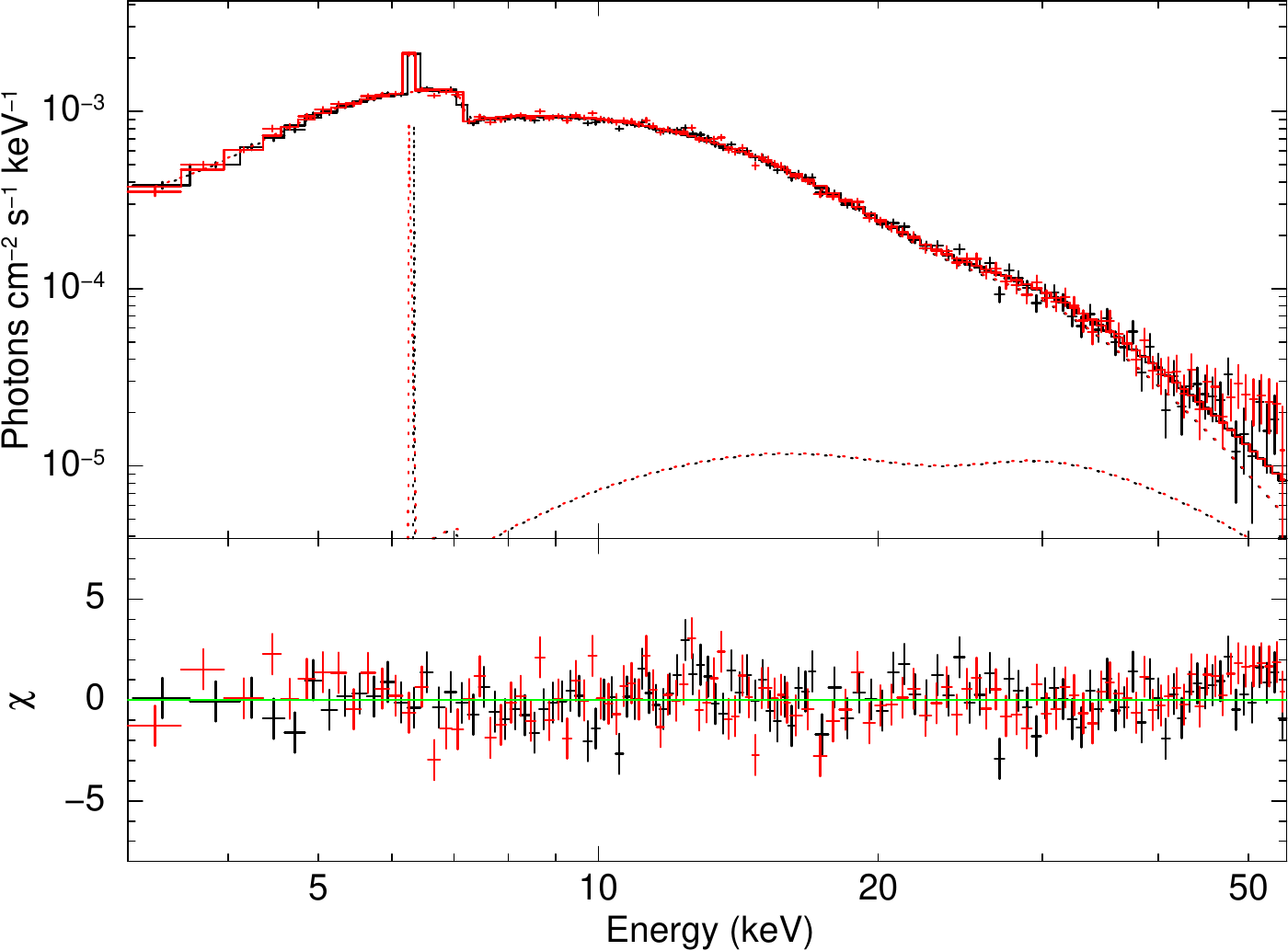}
     \caption{Spectral fit on IGR J17329-2731 Obs. Sn.39 using \texttt{NPEX} continuum model, having no \textit{TKF}. The top panel shows the best-fit model and the bottom panel shows the residuals to the best-fit model.}
     \label{fig:17329}
 \end{figure}

{We analysed the 3$-$55 keV spectrum from the observation Obs. Sn.39 (Table~\ref{tab:obs_catalogue}). Galactic absorption towards the source could not be constrained by the fit, therefore we fixed it to the galactic value of $3\times10^{21}$ atoms cm\textsuperscript{-2}. Iron fluorescence line was present, which is fitted with a \texttt{gaussian}. The width of the iron emission line was not constrained by the fit and we fixed it to 10 eV. CRSF at $\sim22$ keV was fitted with a \texttt{gabs} model.

The best fitting models on Obs. Sn.39 does not contain \textit{TKF}.

The spectral parameters of the best fitting models for Obs. Sn.39 of IGR J17329-2731 are given in Tables \ref{tab:cyc2} and \ref{tab:tkf2}. The spectral fit for Obs. Sn.39 (having no TKF) is shown in Fig.~\ref{fig:17329}.
 }
 
 \subsection{SMC X--1}\label{subsec:smcx1}
 
SMC X--1 is an HMXB pulsar in the Small Magellanic Cloud located $\sim61$ kpc away. This eclipsing binary consists of a $\sim0.7$ s NS pulsar in $\sim3.89$ d orbit around a B-type Super-giant companion star and is one of the few HMXBs that accrete via Roche lobe overflow. The broadband spectrum is usually modelled by absorbed power-law models and iron fluorescence line \citep[See][and references therein]{smcx1_130_pike2019observing}.
 
\begin{figure}
     \centering
     \includegraphics[width=\linewidth]{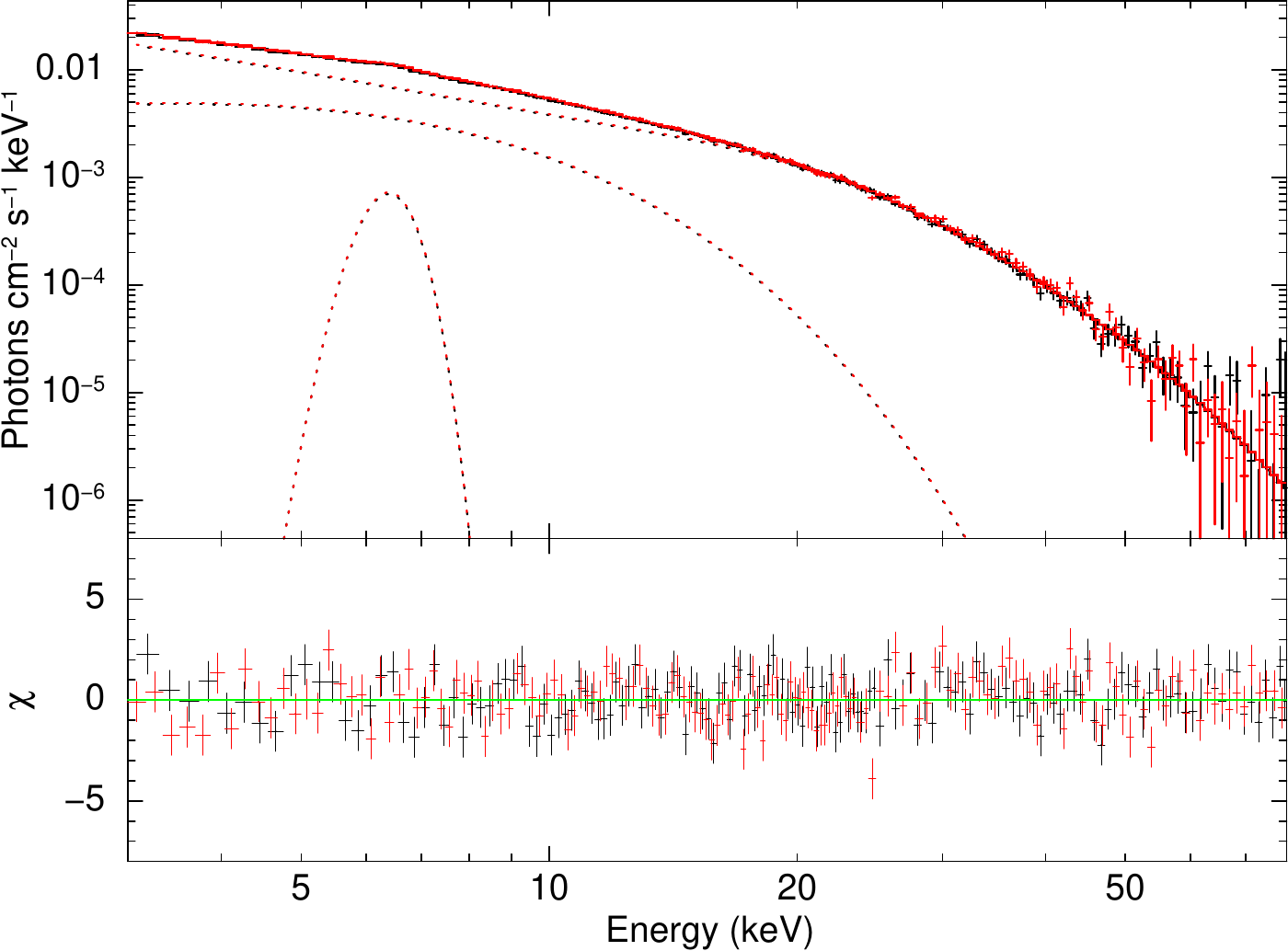}
     \caption{Spectral fit performed on SMC X--1 Obs. Sn.40 using \texttt{mplcut} continuum model, having no \textit{TKF}. The top panel shows the best-fit model and the bottom panel shows the residuals to the best-fit model.}
     \label{fig:smcx1}
 \end{figure}

{We analysed the spectra of SMC X--1 from four \textit{NuSTAR} observations Obs. Sn.40, 41, 42 and 43 (Table~\ref{tab:obs_catalogue}). Galactic absorption towards the source could not be constrained by the fit, therefore we fixed it to the galactic value of $5\times10^{21}$ atoms cm\textsuperscript{-2}. Iron fluorescence line was present, which is fitted with a \texttt{gaussian}.
 
The best fitting models on Obs. Sn.40, 41 and 42 do not contain \textit{TKF}. Even though the best fitting models on Obs. Sn.43 could contain \textit{TKF} there also exists a similar well-fitting model in which \textit{TKF} is absent.

The spectral parameters of the best fitting models for all four observations of SMC X--1 are given in Tables \ref{tab:cyc2} and \ref{tab:tkf2}. The spectral fit for the time-averaged spectrum of Obs. Sn.40 (having no TKF) is shown in Fig.~\ref{fig:smcx1}.}
 
 \subsection{GRO J1008--57}\label{subsec:groj1008}

GRO J1008--57 is a transient {HMXB pulsar} located $\sim$5 kpc away, hosting a $\sim94$ s pulsar and a Be-type stellar companion. The source shows regular outbursts at a period of $\sim249$ d, interpreted as the binary orbital period. The spectrum is usually modelled with power-law continuum modified by CRSF at $\sim78$ keV (See \citealt{groj1008_138_kuhnel2013gro}, \citealt{groj1008_137_bellm2014confirmation} and references therein).

\begin{figure}
    \centering
    \includegraphics[width=\linewidth]{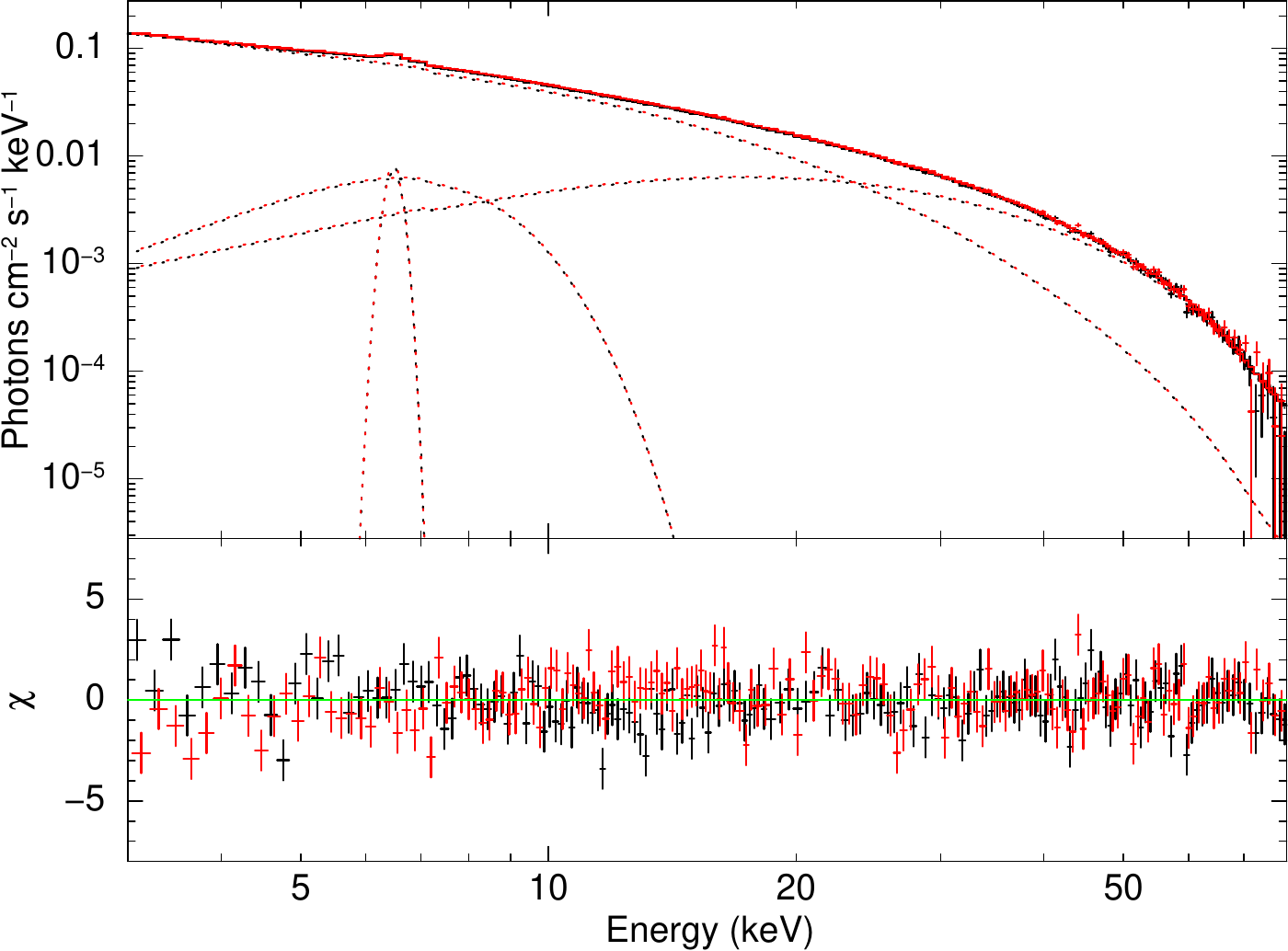}
    \caption{Figure showing spectral fit using \texttt{NPEX} continuum model on GRO J1008--57 Obs. Sn.44, having no \textit{TKF}. The top panel shows the best-fit model and the bottom panel shows the residuals to the best-fit model.}
    \label{fig:groj1008}
\end{figure}

{ We analysed the 3$-$79 keV \textit{NuSTAR} spectrum from Obs. Sn.44. Galactic absorption towards the source could be constrained by the fit. Iron fluorescence line was present, but fitting it with a \texttt{gaussian} line left residuals in the iron region. This could be attributed to improper modelling of the complex iron fluorescence emission region, as the presence of lines at 6.4 keV, 6.6 keV and 7 keV have been reported in the \textit{Suzaku} observations in \cite{groj1008_135_yamamoto2014firm}. Therefore, we used a combination of two \texttt{gaussian} models to fit the complex iron region (similar to \ref{subsec:herx1}). The cyclotron line parameters could not be constrained by the fit, therefore we froze the CRSF line energy to the best fit value obtained by \cite{groj1008_137_bellm2014confirmation} from simultaneous \textit{Suzaku}-HXD (20-100 keV) and \textit{NuSTAR} (3$-$79 keV) spectra.

None of the best fitting training models could fit the validation data sets in Obs. Sn.44. Therefore, single-step fitting was performed on Obs. Sn.44, skipping training. The best fitting model on Obs. Sn.44 does not contain \textit{TKF}.

The spectral parameters of the best fitting models for Obs. Sn.44 of GRO J1008--57 are given in Tables \ref{tab:cyc2} and \ref{tab:tkf2}. The spectral fit for Obs. Sn.44 (having no TKF) is shown in Fig.~\ref{fig:groj1008}.}

 \subsection{GX 304--1}\label{subsec:gx304m1}
GX 304--1 is a transient Be-XRP located $\sim2$ kpc away that hosts the $\sim272$ s pulsar and Be-type companion star in a long $\sim133$ d orbit. It exhibits regular Type-I outbursts during its periastron passage. The luminosity of the source varies over a wide range spanning several orders of magnitude like other Be-XRPs. CRSF has been detected in the high luminosity outburst-state spectrum at $\sim54$ keV, which was modelled with a power-law continuum model \citep{gx304_141_yamamoto2011discovery}. The low luminosity spectrum is modelled by a two-component comptonization model \citep{gx304_144_tsygankov2019dramatic}.

\begin{figure}
    \centering
    \includegraphics[width=\linewidth]{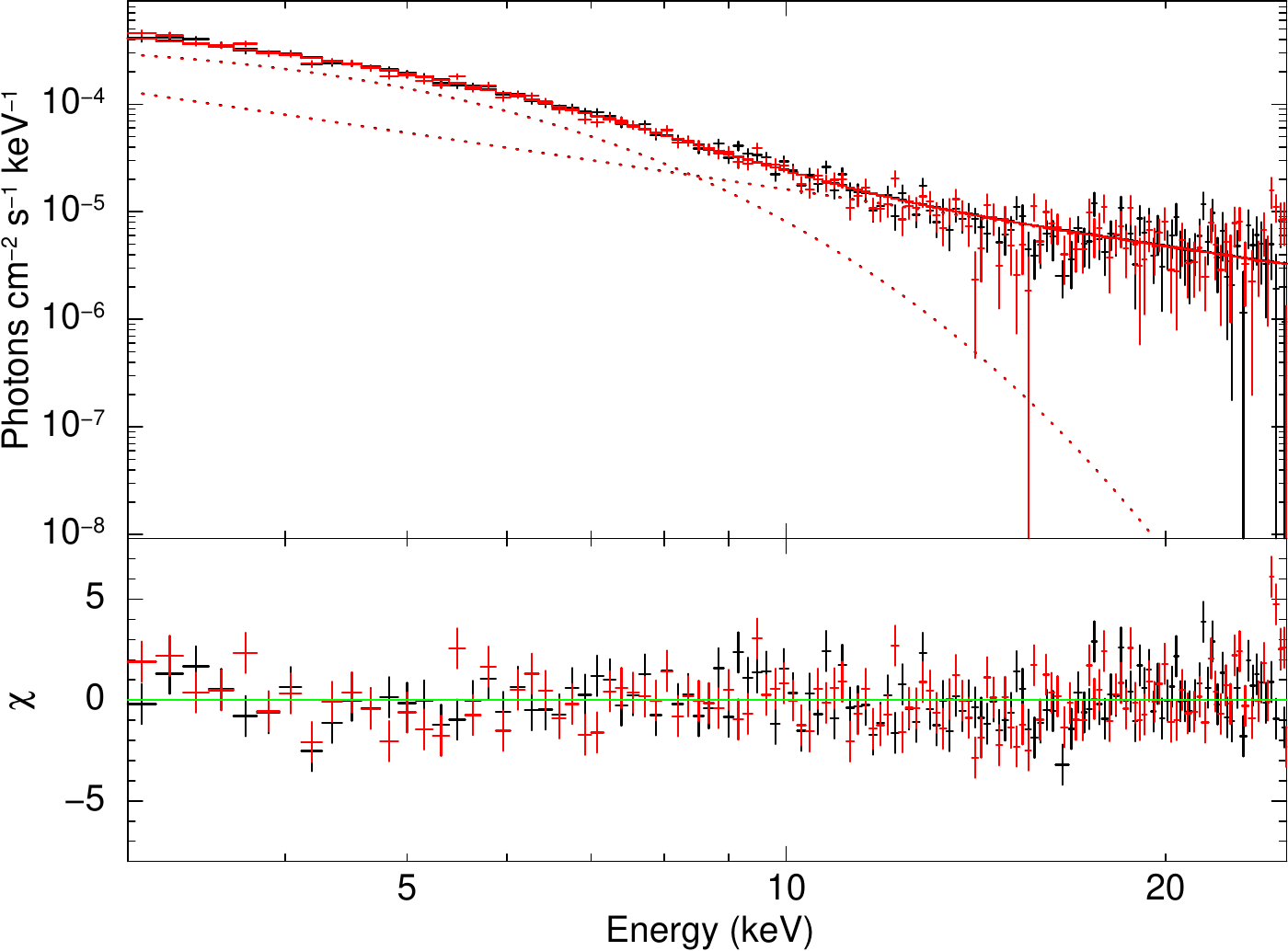}
    \caption{Spectral fit using \texttt{NPEX} continuum model performed on GX 304--1 Obs. Sn.45, having no \textit{TKF}. The top panel shows the best-fit model and the bottom panel shows the residuals to the best-fit model.}
    \label{fig:gx304m1}
\end{figure}

{ We analysed the 3$-$25 keV \textit{NuSTAR} spectrum from Obs. Sn.45, during which the source was in a low-luminosity state. The spectrum is dominated by background photons above 25 keV. This observation has poor photon statistics, with an average count rate of $\sim0.21$ counts s$^{-1}$ and a total exposure of $\sim$58 ks (See Table~\ref{tab:obs_catalogue}). Therefore we re-binned the spectrum such that it has a minimum of 1 count per energy bin and have used Cash-statistic (cstat in XSPEC) as the fit statistic for analysing this \textit{NuSTAR} observation.

We found that some of the model combinations given in Section~\ref{ref:subsec:analysis-method} could fit the training and validation sets. However, as discussed in \cite{gx304_144_tsygankov2019dramatic}, the spectrum of GX 304--1 shows a transition from single power-law to double power-law, when the source luminosity drops by $\sim2$ orders of magnitude. We found that double power-law models could also fit the spectrum. The best fitting model on Obs. Sn.45 does not contain \textit{TKF}.

The spectral parameters of the best fitting models for Obs. Sn.45 of GX 304--1 are given in Tables \ref{tab:cyc2} and \ref{tab:tkf2}. The spectral fit for Obs. Sn.45 (having no TKF) is shown in Fig.~\ref{fig:gx304m1}.}  

\subsection{1A 0535+26}\label{sec:1a0535p26}
1A 0535+26 is a transient {HMXB pulsar} located 2 kpc away that hosts the pulsar spinning with a period of 104 s in an eccentric and long 110 d orbit around the Be-type stellar companion. It exhibits regular Type-I outbursts during the periastron passage and Type-II outbursts due to companion stellar activity. It is known to exhibit long out-of-flare states as well (See \citealt{1a0535_ballhausen_2017}, and references therein), and the source flux, therefore, spans several orders of magnitude. The source spectrum shows fundamental CRSF at about 45 keV and first harmonic at about 100 keV \citep[See][and references therein]{1a0535_Sartore_2015}. The high luminosity level spectra are usually modelled with power-law continuum models \citep{1a0535_Mandal_2022} while the low luminosity spectra are modelled with a two-component comptonization model \citep{1a0535_tsygankov2019MNRAS.487L..30T}.

\begin{figure}
    \centering
    \includegraphics[width=\linewidth]{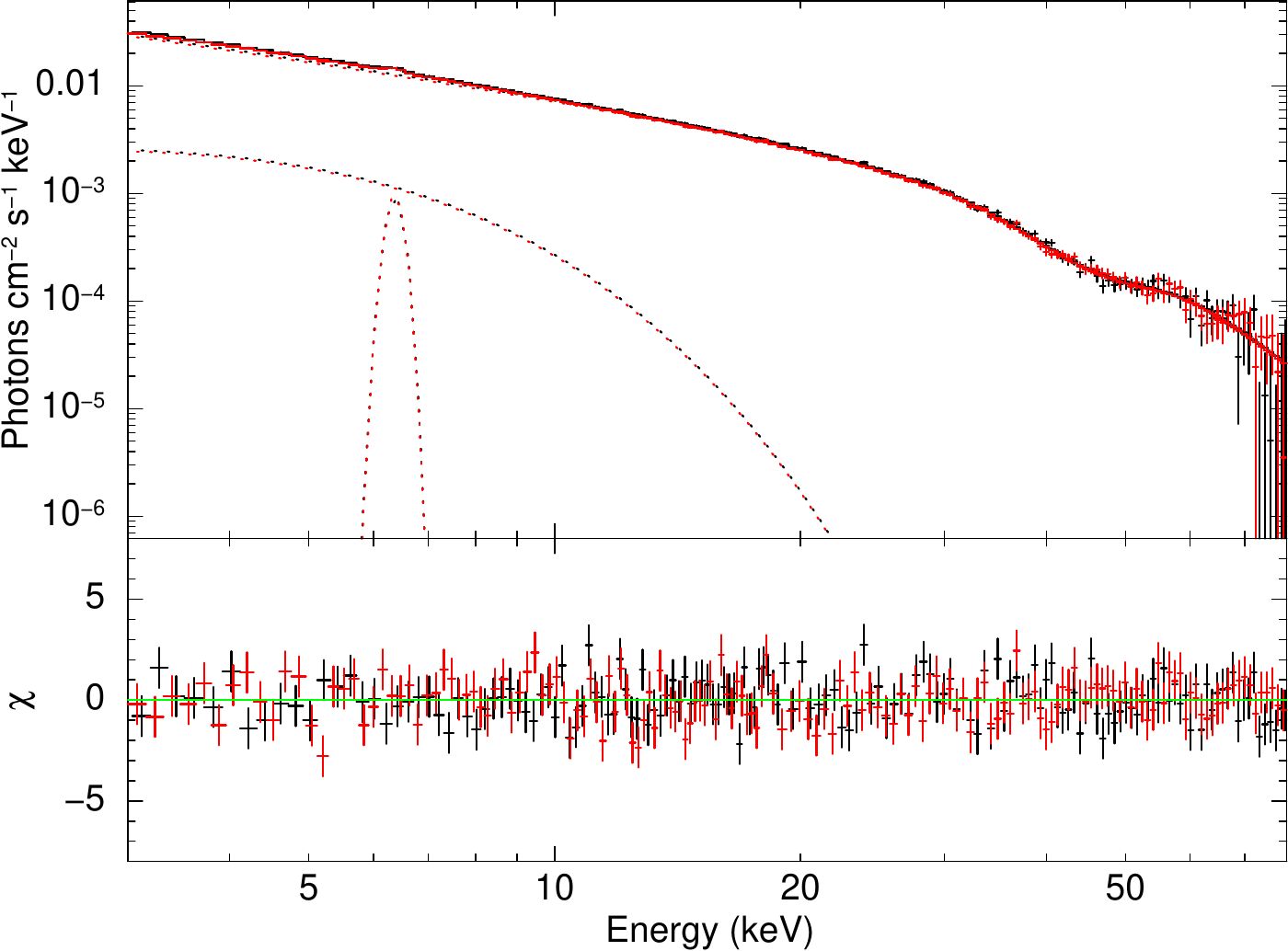}
    \caption{Spectral fit using \texttt{FDcut} continuum model performed on 1A 0535+26 Obs. Sn.46, having no \textit{TKF}. The top panel shows the best-fit model and the bottom panel shows the residuals to the best-fit model.}
    \label{fig:1a0535p26}
\end{figure}

{ We analysed spectra from three \textit{NuSTAR} observations of the source, Obs. Sn.46, 47 and 48 (Table~\ref{tab:obs_catalogue}). Obs. Sn. 47 and 48 were dominated by background photons above 50 keV. Galactic absorption towards the source could not be constrained by the fit in Obs. Sn.48, therefore we fixed it to the galactic value of $4\times10^{21}$ atoms cm\textsuperscript{-2}. Iron fluorescence line was present in Obs. Sn.46 and 47, which were fitted with a \texttt{gaussian}. CRSF at $\sim45$ keV was fitted with a \texttt{gabs} model in Obs. 46 and 48, and it was not required in Obs. Sn.47. The width of CRSf could not be constrained by the fit in Obs. Sn.48, therefore it was fixed to the best fit value obtained from Obs. Sn.46.

The best fitting models on Obs. Sn.49 and 50 contain \textit{TKF}. All the models that could fit Obs. Sn.48 by Training/Validation gave systematic wavy residuals in 3--20 keV, regardless of the best fitting models having no \textit{TKF}. Therefore, Obs. Sn.48 was fitted manually with a two-component \texttt{CompTT} model (See \citealt{1a0535_tsygankov2019MNRAS.487L..30T}), and it does not contain \textit{TKF}. 

The spectral parameters of the best fitting models for all three observations of 1A 0535+26 are given in Tables \ref{tab:cyc2} and \ref{tab:tkf2}. The spectral fit for the time-averaged spectrum of Obs. Sn.46 (having no TKF) is shown in Fig.~\ref{fig:1a0535p26}.
}

\subsection{GRO J2058+42}\label{subsec:groj2058}
GRO J2058+42 is a bright transient {HMXB pulsar} with P\textsubscript{spin}$\sim$198 s located 9 kpc away, hosting the pulsar and a Be-type companion star \citep{groj2058_wilson_intro}. The binary orbital period of this system is ambiguous due to the ambiguity in the perceived intensity levels of successive outbursts. An outburst interval of 55 d is interpreted as the orbital period when each outburst is assumed to be occurring due to the periastron passage, or 110 d if the successive outbursts are distinguishable, with one from the apastron and the next from the periastron. \cite{groj2058_nustar_molkov} reported the presence of CRSF at $\sim10$ keV and two of its harmonics at $\sim20$ keV and $\sim30$ keV in the \textit{NuSTAR} spectrum in a narrow spin phase interval. The presence of spin phase-dependent CRSF and two of its harmonics were later reported by \cite{groj2058_astrosat_kallol} in the \textit{Astrosat} spectrum as well. However, analysis of the spin-phase averaged \textit{NuSTAR} spectrum by \cite{groj2058_nustar_kabiraj2020broad} didn't require these CRSF lines. The presence of the possible CRSF near 10 keV makes it almost impossible to detect \textit{TKF} even if it exists.

\begin{figure}
    \centering
    \includegraphics[width=\linewidth]{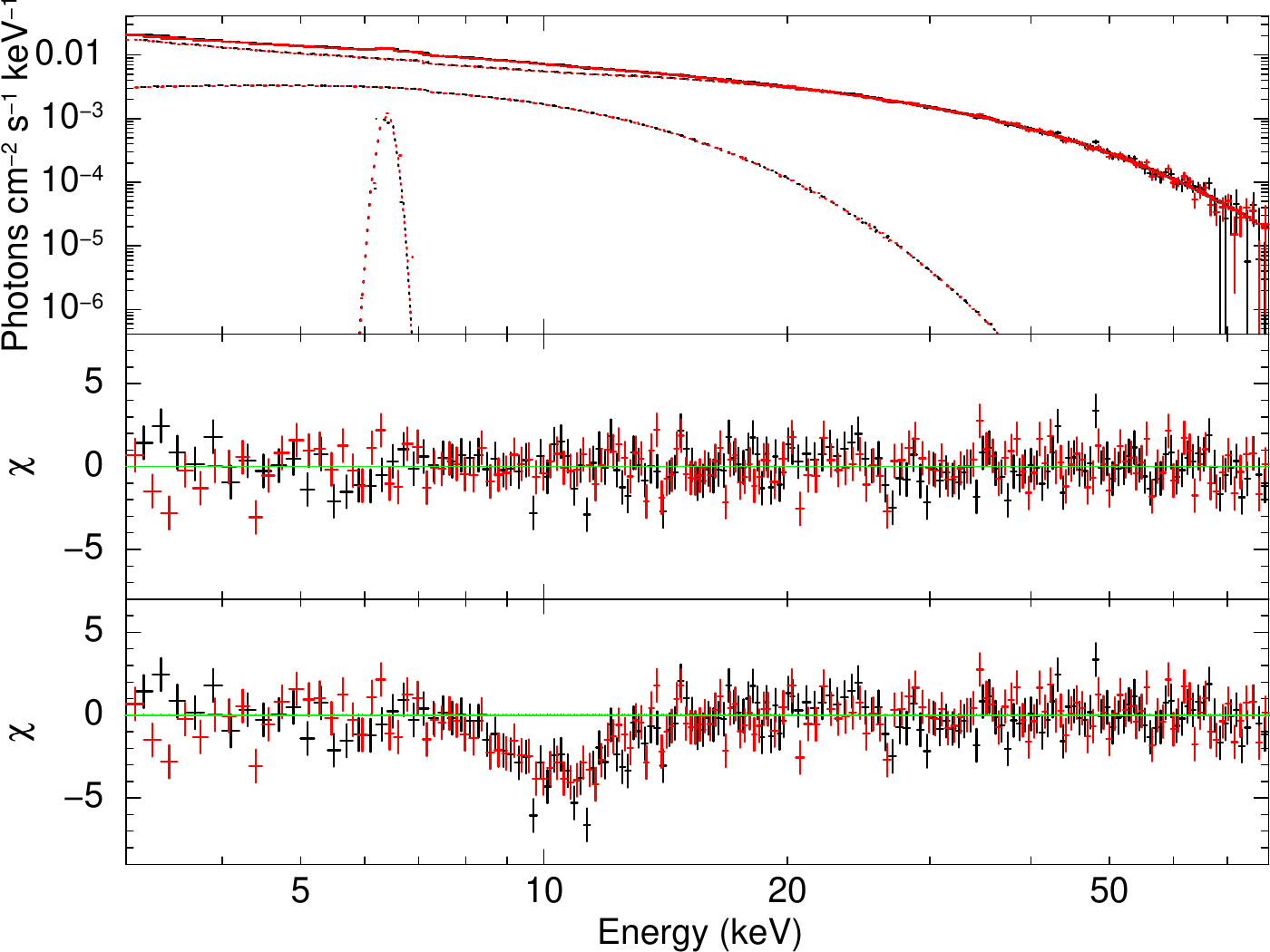}
    \includegraphics[width=\linewidth]{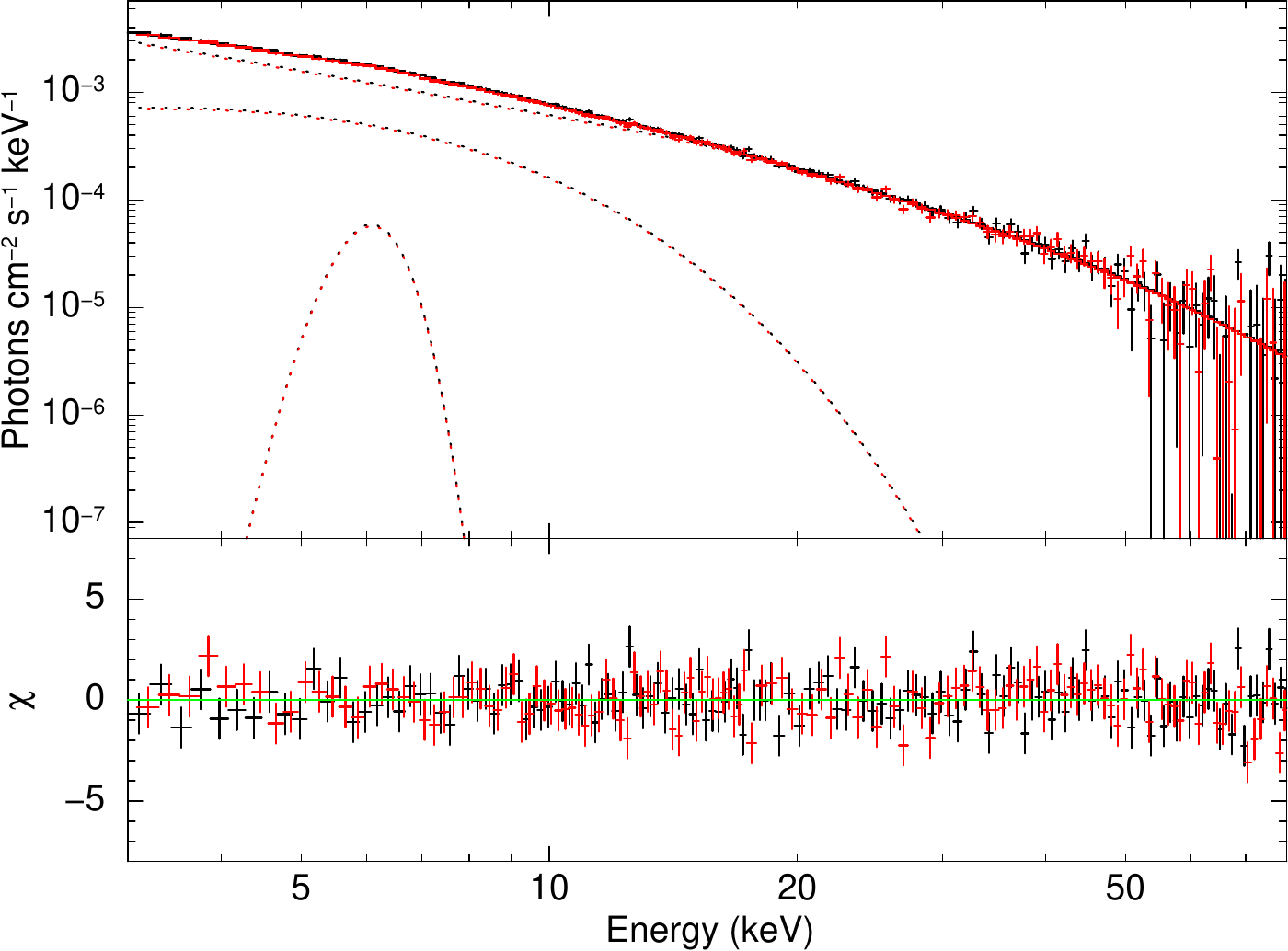}
    \caption{Top: Spectral fit on GRO J2058+42 Obs. Sn.49 with \texttt{compTT} continuum model, having \textit{TKF}. The top panel shows the best-fit model, the middle panel shows the residuals to the best-fit model, and the bottom panel shows the residuals when the strength of the \texttt{gabs} component modelling \textit{TKF} is set to 0. Bottom: Spectral fit on GRO J2058+42 Obs. Sn.51 with \texttt{newhcut} continuum model, having no \textit{TKF}. The top panel shows the best-fit model and the bottom panel shows the residuals to the best-fit model.}
    \label{fig:groj2058}
\end{figure}

{ We analysed the spectra from three \textit{NuSTAR} observations of the source Obs. Sn.49 (3$-$79 keV), 50 (3$-$79 keV) and 51 (3$-$50 keV) (Table~\ref{tab:obs_catalogue}). Galactic absorption towards the source could not be constrained by the fit, therefore we fixed it to the galactic value of $6\times10^{21}$ atoms cm\textsuperscript{-2}. Iron fluorescence line was present, which is fitted with a \texttt{gaussian}. No significant residuals resembling the reported CRSF line were visible during the fitting, probably because it becomes evident only if one performs spin-phase resolved spectroscopy (See \citealt{groj2058_nustar_molkov}).

The best fitting models on Obs. Sn.49 and 50 contain \textit{TKF}, while the best fitting model on Obs. Sn.51 does not contain \textit{TKF}.

The spectral parameters of the best fitting models for all three observations of GRO J2058+42 are given in Tables \ref{tab:cyc2} and \ref{tab:tkf2}. The spectral fits for Obs. Sn.49 (having TKF) and Obs. Sn.51 (having no TKF) are shown in Fig.~\ref{fig:groj2058}.}

\subsection{1E 1145.1--6141}\label{subsec:1145.1--6141}
1E 1145.1--6141 is a persistent {HMXB pulsar} located $\sim8.5$ kpc away, hosting the 297 s pulsar that accretes from the wind of B-type Supergiant companion star in an eccentric ($e\sim0.2$), 14.4 d orbit (See \citealt{1e1145_rxte_Ray_2002}, \citealt{1e1145_nustar_bcp}), and references therein.

\begin{figure}
    \centering
    \includegraphics[width=\linewidth]{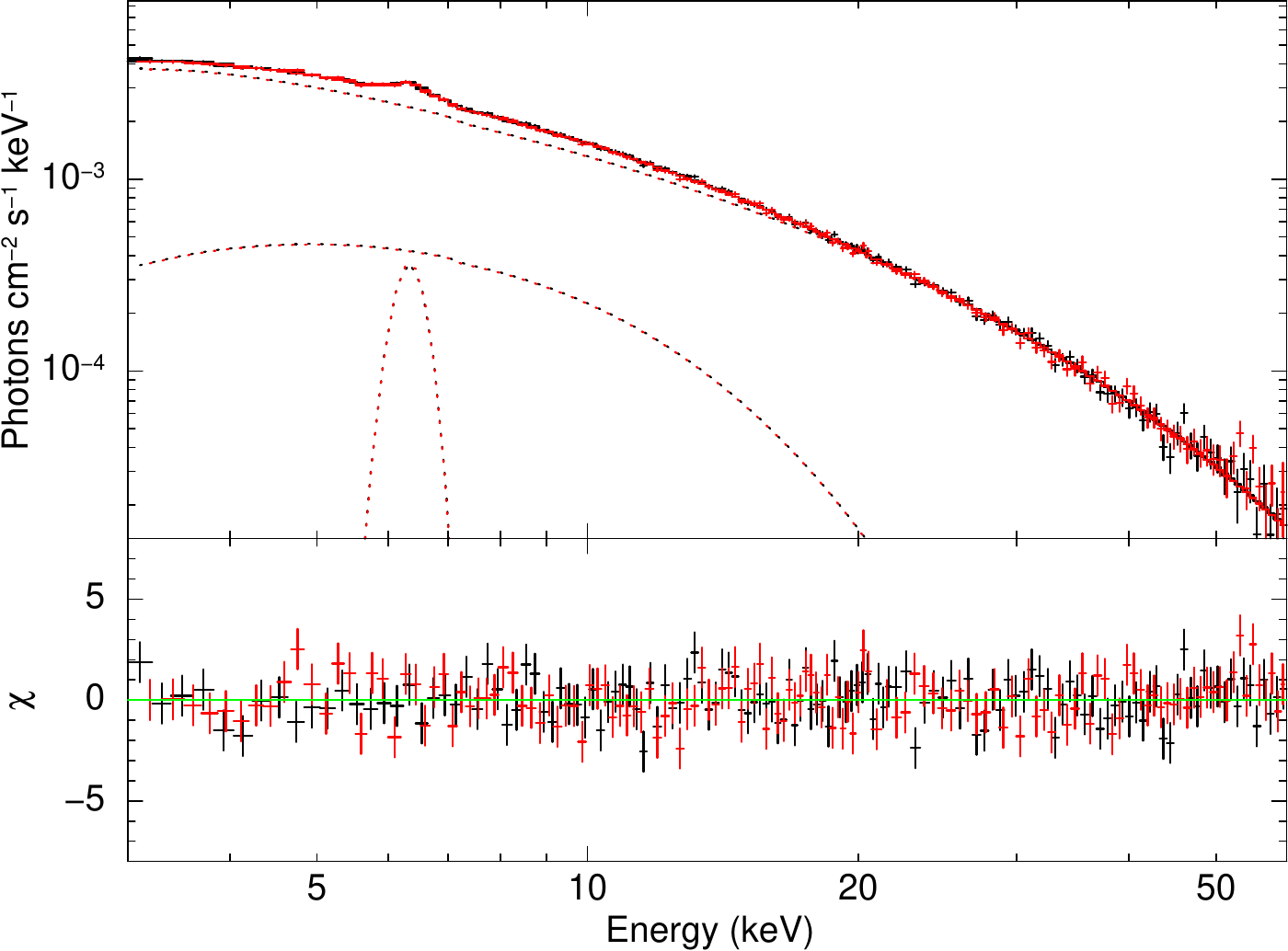}
    \caption{Spectral fit using \texttt{newhcut} continuum model on 1E 1145.1--6141 Obs. Sn.52, having no \textit{TKF}. The top panel shows the best-fit model and the bottom panel shows the residuals to the best-fit model.}
    \label{fig:1e1145}
\end{figure}

{ We analysed the 3$-$60 keV spectrum from one \textit{NuSTAR} observation of the source, Obs. Sn.52 in Table~\ref{tab:obs_catalogue}. Iron fluorescence line was present, which is fitted with a \texttt{gaussian}.

The best fitting model on Obs. Sn.52 does not contain \textit{TKF}.

The spectral parameters of the best fitting models for Obs. Sn.52 of 1E 1145.1--6141 are given in Tables \ref{tab:cyc2} and \ref{tab:tkf2}. The spectral fit for the time-averaged spectrum of Obs. Sn.52 (having no TKF) is shown in Fig.~\ref{fig:1e1145}.
}

\subsection{OAO 1657--415}\label{subsec:oao}
OAO 1657--415 is an {HMXB pulsar} located $\sim2.2$ kpc away, hosting the 38 s pulsar in an eccentric orbit ($e\sim0.1$) around the early type Ofpe/WN9 Supergiant. The binary has an orbital period of about 10.4 d \citep[See][and references therein]{oao1657_nustar_enzo}. The spectrum of OAO 1657--415 is generally characterized by heavily absorbed power-law models modified by CRSF at $\sim36$ keV (\citealt{oao_1657_bepposax}, \citealt{oao1657_suzaku_pragati}).

\begin{figure}
    \centering
    \includegraphics[width=\linewidth]{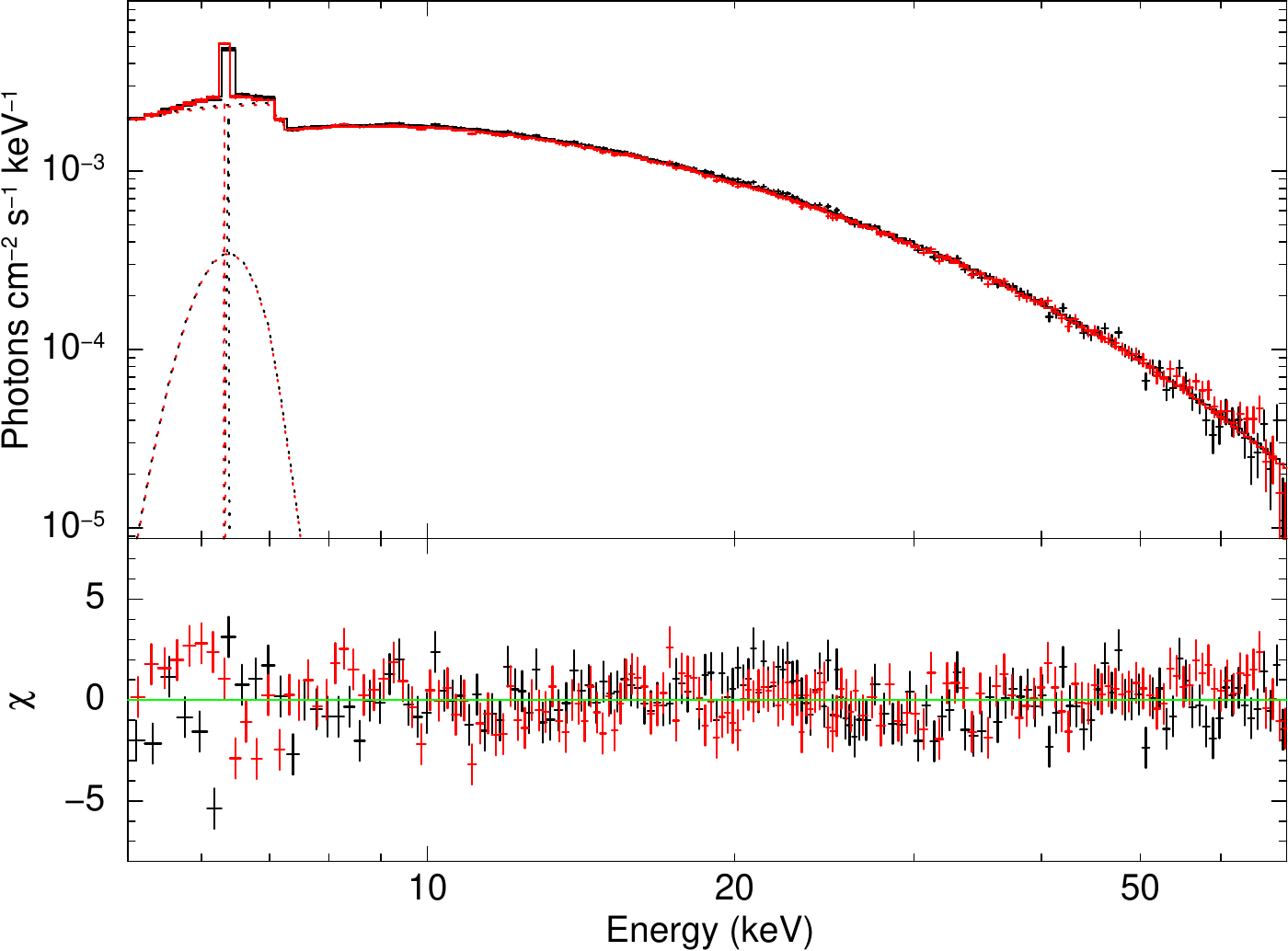}
    \caption{Spectral fit using \texttt{cutoffpl} continuum model on OAO 1657--415 Obs. Sn.53. Top panel shows the best fit model and the bottom panel shows the residuals to the best fit model.}
    \label{fig:oao1657}
\end{figure}

{ We analysed the 5--70 keV spectrum from one \textit{NuSTAR} observation of the source, Obs. Sn.53 in Table~\ref{tab:obs_catalogue}. The FPMA module was contaminated by stray light from a nearby source, because of which there was a significant deviation in spectra from FPMA and FPMB below $\sim5$ keV. We tried selecting an annular background region around the source region in FPMA so that the background will include any streak of stray light contaminating the source region. Even though the residuals improved, residuals still deviated in FPMA and FPMB below 5 keV. Galactic absorption towards the source could not be constrained by the fit, therefore we fixed it to the galactic value of $1.81\times10^{22}$ atoms cm\textsuperscript{-2}. Iron fluorescence line was present, which is fitted with a \texttt{gaussian}. The width of the iron emission line was not constrained by the fit and we fixed it to 10 eV. CRSF was not present in the spectrum.

The best fitting model on Obs. Sn.53 does not contain \textit{TKF}.

The spectral parameters of the best fitting models for Obs. Sn.53 of OAO 1657--415 are given in Tables \ref{tab:cyc2} and \ref{tab:tkf2}. The spectral fit for the time-averaged spectrum of Obs. Sn.53 (having no TKF) is shown in Fig.~\ref{fig:oao1657}.}

\subsection{EXO 2030+375}\label{subsec:exo2030p375}

EXO 2030+375 is a transient {HMXB pulsar} with P\textsubscript{spin}$\sim42$ s located about 7 kpc away, that hosts the NS and a Be-companion star in an eccentric (e$\sim$0.36) 46 d orbit (See \citealt{exo2030_intro_parmar} and \citealt{exo2030_wilson_ditance}). EXO 2030+375 is a well-studied source, and it exhibits regular Type-I outbursts during every periastron passage. Its outburst spectrum is generally modelled with an absorbed power-law model having an exponential cutoff at high energy, with emission lines of iron (See \citealt{exo2030_rxte_reig} and \citealt{exo2030_suzaku_Naik_2013}).

\begin{figure}
     \centering
     \includegraphics[width=\linewidth]{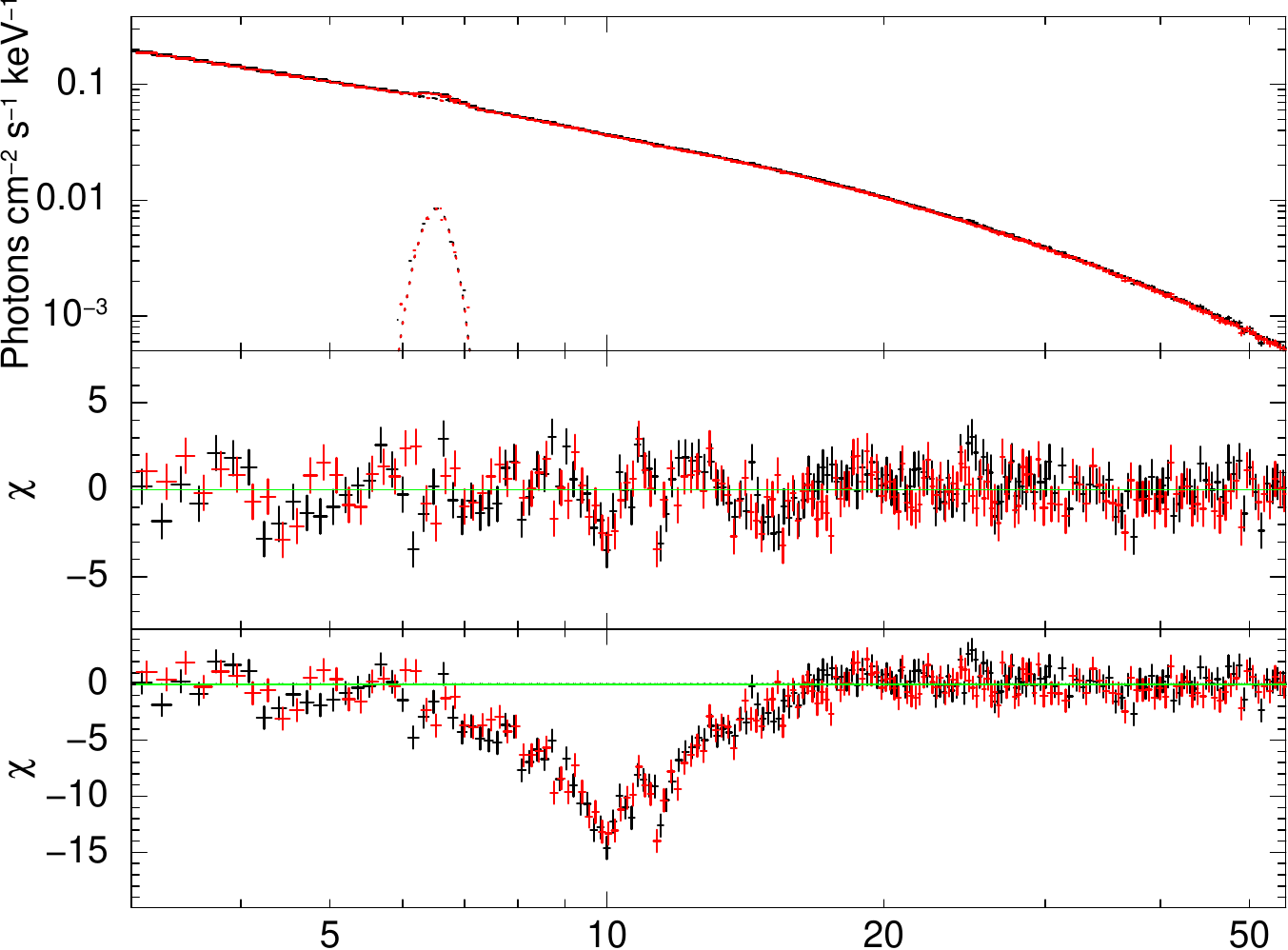}
     \includegraphics[width=\linewidth]{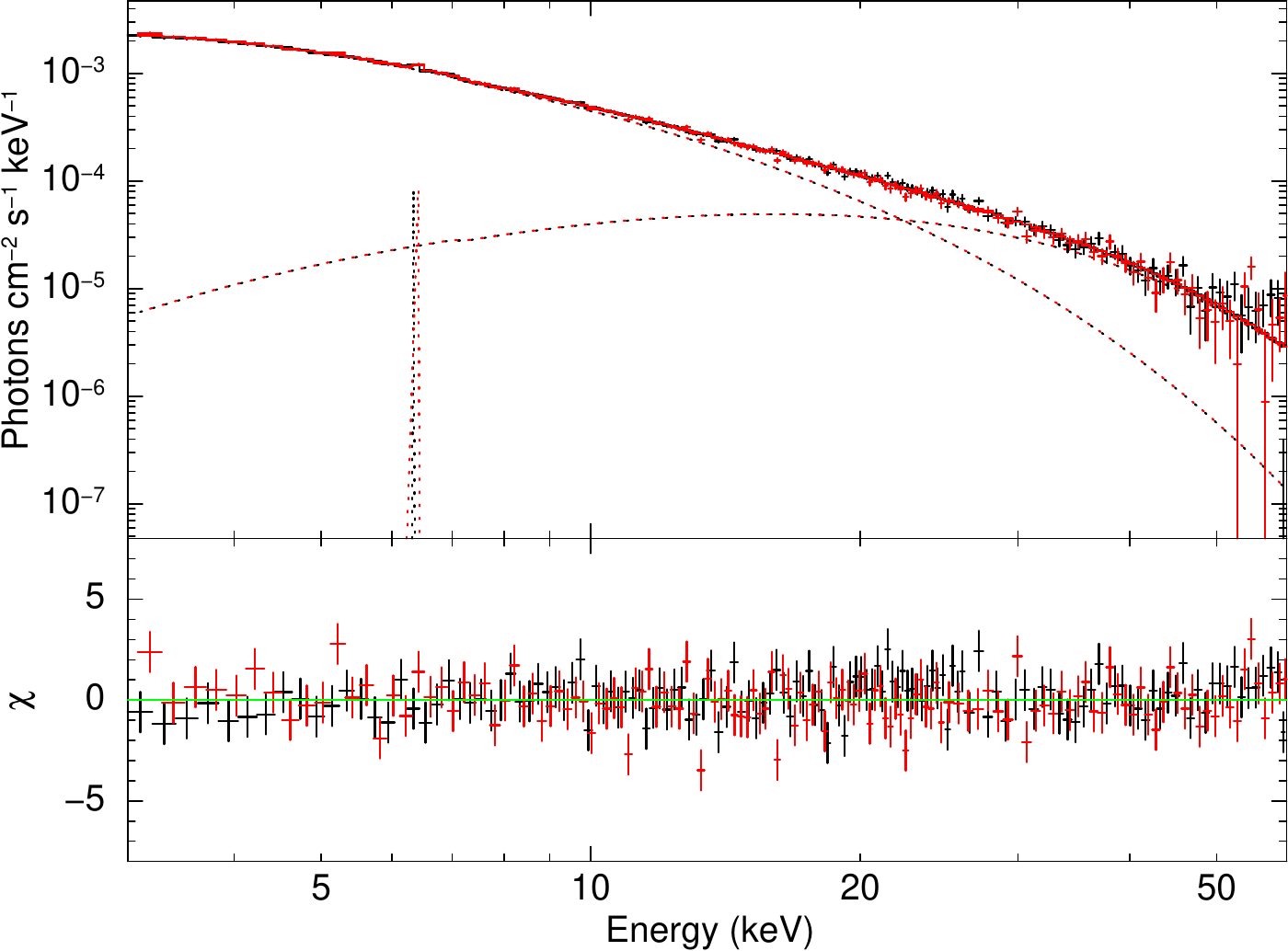}
     \caption{Top: Spectral fit on EXO 2030+375 Obs. Sn.54 with \texttt{cutoffpl} continuum model, having \textit{TKF}. The top panel shows the best-fit model, the middle panel shows the residuals to the best-fit model, and the bottom panel shows the residuals when the strength of the \texttt{gabs} component modelling \textit{TKF} is set to 0. Bottom: Spectral fit on EXO 2030+375 Obs. Sn.55 with \texttt{NPEX} continuum model, having no \textit{TKF}. The top panel shows the best-fit model and the bottom panel shows the residuals to the best-fit model.}
     \label{fig:exo2030}
 \end{figure}

{We analysed the spectra from three \textit{NuSTAR} observations, Obs. Sn.54 (3$-$55 keV), Obs. Sn.55 (3$-$79 keV) and Obs. Sn.56 (3$-$79 keV) (Table~\ref{tab:obs_catalogue}). Galactic absorption towards the source could not be constrained by the fit, therefore we fixed it to the galactic value of $8\times10^{21}$ atoms cm\textsuperscript{-2}. Iron fluorescence line was present in all the observations, which were fitted with a \texttt{gaussian}. The width of the iron emission line was not constrained by the fit in Obs. Sn.55 and we fixed it to 10 eV.

The best fitting models on Obs. Sn.54 and 56 contain \textit{TKF}, while the best fitting model on Obs. Sn.55 does not contain \textit{TKF}. 

The spectral parameters of the best fitting models for all three observations of EXO 2030+375 are given in Tables \ref{tab:cyc2} and \ref{tab:tkf2}. The spectral fit for the time-averaged spectrum of Obs. Sn.54 (having \textit{TKF}) and Obs. Sn.55 (having no TKF) are shown in Fig.~\ref{fig:exo2030}.}

\subsection{IGR J19294+1816}\label{subsec:igrj19294}
IGR J19294+1816 is a transient {HMXB pulsar} located $\sim11$ kpc away in the Perseus spiral arm of Milkyway, hosting the NS and a Be-type companion. The NS has a spin period of 12.4 s, and the binary has an orbital period of about 117 d (See \citealt{igrj19294_rodriguez}, and references therein). The source spectrum is characterized by an absorbed power-law. Discovery of CRSF was claimed in the RXTE spectrum at $\sim35$ keV by \cite{igrj19294_roy_jayashree}. However, a reanalysis of the same data by \cite{igtj19294_nustar_tsygankov} could not detect this CRSF. They detected the CRSF at $\sim42$ keV in the \textit{NuSTAR} data, and the same line was detected by \cite{igrj19294_astrosat_raman} in the \textit{Astrosat} data.

\begin{figure}
    \centering
    \includegraphics[width=\linewidth]{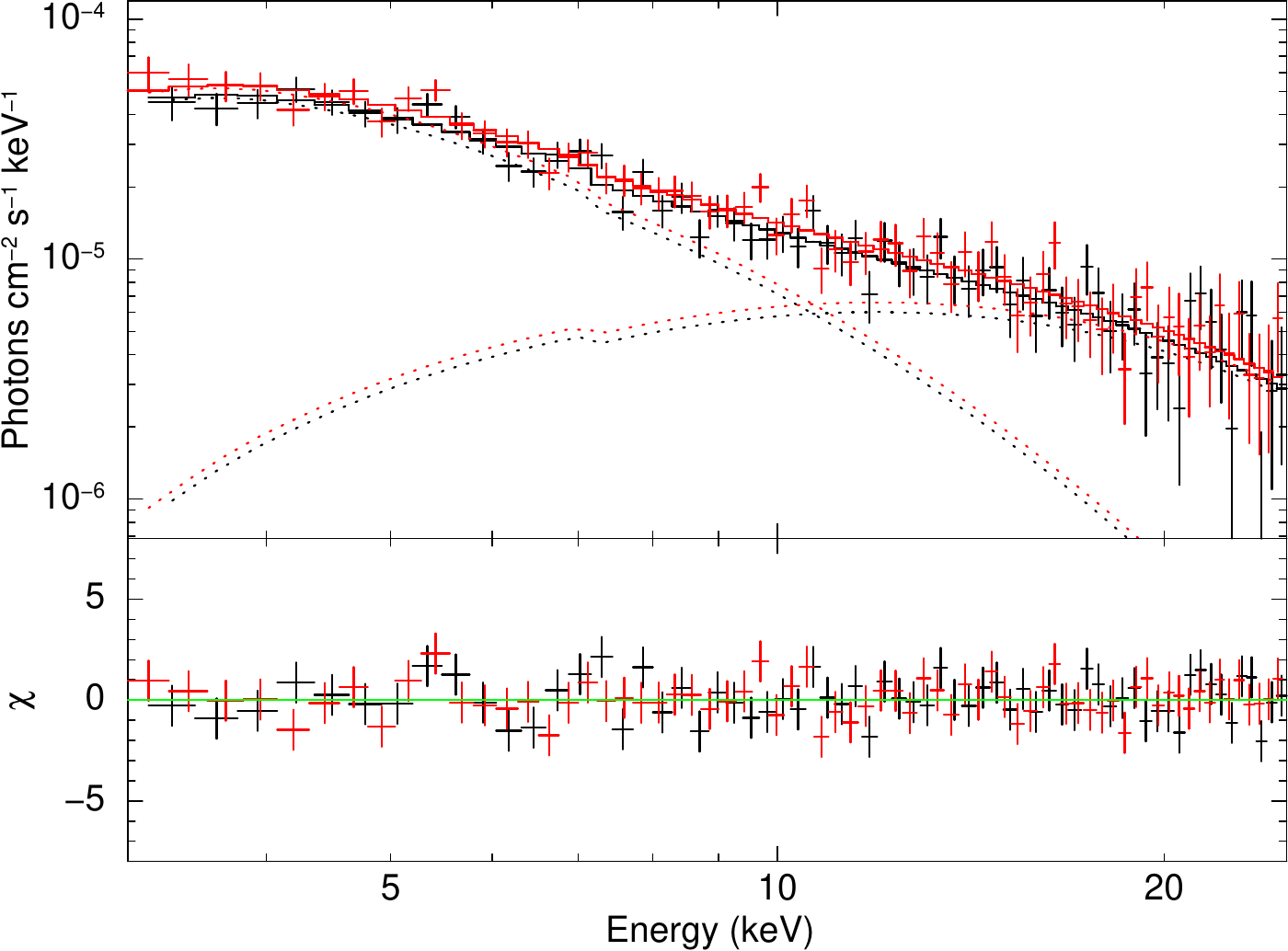}
    \caption{Spectral fit using \texttt{NPEX} continuum model on IGR J19294+1816 Obs. Sn.57, having no \textit{TKF}. The top panel shows the best-fit model and the bottom panel shows the residuals to the best-fit model.}
    \label{fig:igrj19294}
\end{figure}

{ We analysed the spectra from two \textit{NuSTAR} observations, Obs. Sn.57 and Obs. Sn.58 (Table~\ref{tab:obs_catalogue}). Obs. Sn.57 and 58 were dominated by background photons above 20 keV and 50 keV respectively. Therefore 3--20 keV spectrum of Obs. Sn.57 and 3$-$50 keV spectrum of Obs. Sn.58 were analysed. Iron fluorescence line was present only in Obs. Sn.57, which was fitted with a \texttt{gaussian}. The spectrum of Obs. Sn.58 could be fitted without the requirement of the CRSF reported in \cite{igtj19294_nustar_tsygankov}. 

The best models for validation were based on the top 10\% AIC on the training data set of Obs. Sn.57 because only one model was in the top 5\% AIC. The best fitting models on Obs. Sn.56 and 57 do not contain \textit{TKF}.

The spectral parameters of the best fitting models for both observations of IGR J19294+1816 are given in Tables \ref{tab:cyc2} and \ref{tab:tkf2}. The spectral fit for Obs. Sn.57 (having no \textit{TKF}) is shown in Fig.~\ref{fig:igrj19294}.}

\section{Discussion} \label{disc}

We (re)analysed the \textit{NuSTAR} broadband spectra of 30 XRPs (Table~\ref{tab:obs_catalogue}), and 9 of them viz., Vela X--1, Her X--1, XTE J1946+274, 4U 1907+09, 4U 1538-52, Cep X--4, SMC X--1, GX 304--1 and EXO 2030+375 (See Table~\ref{tab:survey}) have previous report(s) of \textit{TKF}. After separating the flare and out-of-flare spectra in observations containing flares, separating the eclipse and out-of-eclipse spectra in observations containing eclipses, {and by performing a systematic two-step spectral analysis of each observation given in Table~\ref{tab:obs_catalogue} that gives importance to data in the 3$-$15 keV (where \textit{TKF} is present) in the model selection, we detected \textit{TKF} in 11 sources viz., Her X--1, Vela X--1, XTE J1946+274, 4U 1907+09, IGR J16393--4643, RX J0520.5--6932, Cen X--3, 4U 1700--37, LMC X--4, GRO J2058+42 and EXO 2030+375. This is contrary to previous reports of \textit{TKF} in 4U 1538-52, Cep X--4, SMC X--1 and GX 304--1, and no previous reports of \textit{TKF} in IGR J16393--4643, RX J0520.5--6932, Cen X--3, 4U 1700--37, LMC X--4 and GRO J2058+42. The centre of the \texttt{gabs} used to model the \textit{TKF} varies from 9$-$11 keV, while the width varies between 1$-$2 keV. The significance of this feature varies, which is very low in some observations (See Figs.~\ref{fig:16393} and \ref{fig:rxj0520}), moderately significant in some (See Figs.~\ref{fig:herx1}, \ref{fig:xtej1946p274}, \ref{fig:cenx3}) and prominent in others (See Figs.~\ref{fig:velax1}, \ref{fig:lmcx4} and \ref{fig:exo2030}).

The best fitting composite model(s) on the broadband (3$-$79 keV) data were identified from a set of composite models screened based on how well they fit the 3$-$15 keV narrower passband (10 keV range) data. The best fitting model was selected based on AIC score that considers the fit statistic but also penalizes for the model's complexity (Refer section~\ref{ref:subsec:analysis-method}). The presence of \textit{TKF} was assessed from the presence of \texttt{gabs} component modelling \textit{TKF} in the composite model. As a consequence of this model selection methodology, different observations of the same source were sometimes fitted with different continuum models. This is however justified as the literature shows that the observations acquired at different times from the same source have often been fitted with different continuum models (See Table~\ref{tab:survey}). Also, there are instances where the best fitting models from this work have \textit{TKF} but individual spectral analysis of some of these observations reported by other authors do not report this feature (See for eg. \citealt{cenx3_gunjan_2021MNRAS.500.3454T}, \citealt{17544_bhalerao_2015MNRAS.447.2274B} and \citealt{16393_bodaghee_2016ApJ...823..146B}). However, it is inevitable when the selection of best fitting model on a large set of observations happens in a systematic fashion.

Of the 58 different observations, 14 observations from 11 sources have shown the presence of this feature. Due to the presence of a CRSF at $\sim15$ keV, this feature could not be checked on one observation of XTE J1829--098 (Obs. Sn.28). Because none of the mentioned models could give good fits to the spectrum, Obs. Sn.48 of 1A 0535+26 could not be checked for this feature. Therefore, the presence of \textit{TKF} could be checked in 56 observations out of 58. Single-step validation fitting process had to be performed on 8 observations as the best-fit training models does not fit the validation data set. The two step fitting process as outline before could be performed on 48 observations out of 58. 

We screened the best fitting models on the validation set within the top 5\% AIC score and have reported the best two models of two different kinds amongst them, in which one contains \textit{TKF} and the other doesn't. In some cases, the best-fitting models on the validation set within the top 5\% AIC score only contain one kind of model, in which case only one is reported. 36 out of 58 observations only had the best fitting models that do not contain \textit{TKF}. Out of the rest, 6 out of 22 could be well fitted with two different kinds of composite models in which one contains \textit{TKF} while the other doesn't contain \textit{TKF}. Of the rest, 14 out of 16 include only the composite model containing \textit{TKF} in the top 5\% AIC. The two left out observations (Obs. Sn.35 and 38) showed something peculiar; these are the two observations of LMC X--4 and have a flare in each of them. The best fitting models on the time-averaged spectra of Obs. Sn.38 contain \textit{TKF}, while the best fitting models on the separated flaring state and out-of-flare state spectra of Obs. Sn.38 do not contain \textit{TKF}. This is not, however, entirely true for Obs. Sn. 35. While the time-averaged spectra of 35 could only be fitted with models containing \textit{TKF} and the out-of-flare state spectrum does not need the feature, the flaring state spectrum shows the presence of it. This indicates that a combination of the flare and out-of-flare state spectrum could also mimic \textit{TKF} like residuals. However, this was not observed in 4U 1700--37 (Obs. Sn.34), where the best fitting models to time-averaged as well as separated flare and out-of-flare state spectrum contains \textit{TKF}. {A bar chart representation of the \textit{TKF} detected with different continuum models in this work is shown in Fig.\ref{fig:tkf-results}. Of the 67 spectra analysed in this work from 58 observations, \texttt{NPEX} was the continuum model that fitted the majority of the spectra (28 out of 67), and \textit{TKF} was detected in 19 spectra. However, the 19 \textit{TKF} detections also include the 6 spectral sets of LMC X--4, where the presence of flare is possibly causing the presence of \textit{TKF}. Excluding those 6, there are 13 spectral sets with a \textit{TKF} detection.}

\begin{figure}
    \centering
    \includegraphics[width=\columnwidth]{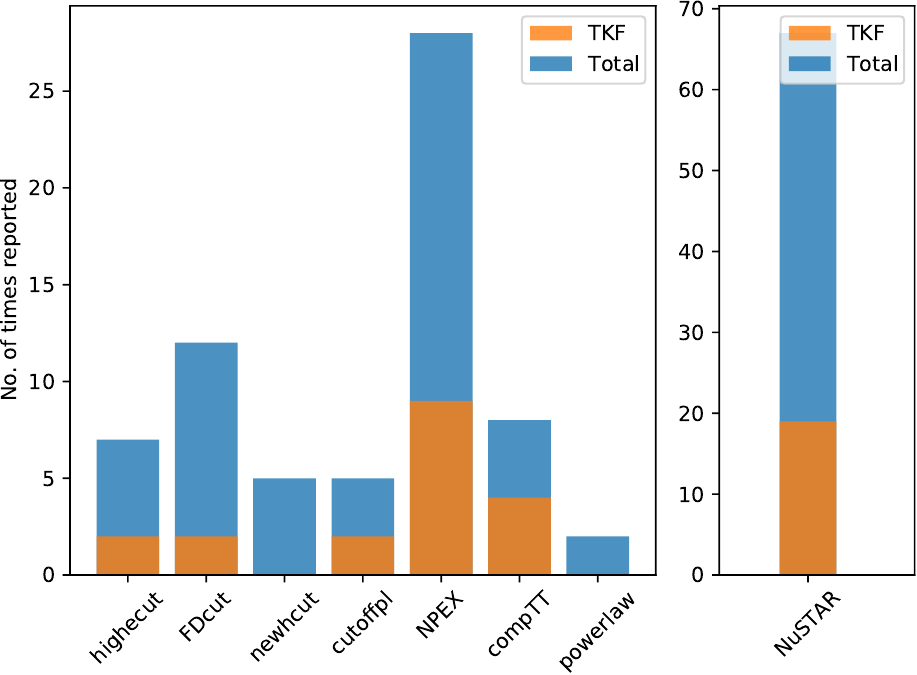}
    \caption{The bar chart shows the results of our analysis of the \textit{NuSTAR} spectral data. The distributions of \textit{TKF} found for different continuum models are shown.}
    \label{fig:tkf-results}
\end{figure}

In 4U 1907+09 (Obs. Sn.12), there were systematic absorption residuals left by almost every fit near 8 keV, and hence we re-performed the fitting by allowing the centre of the \texttt{gabs} accounting \textit{TKF} to vary down to a lower limit of 7 keV. This is an exception to other cases of detection where the centre of \textit{TKF} stayed within 9-11 keV. We also found that the best fitting models on none of the \textit{NuSTAR} observations of KS 1947+300 have an absorption feature near 12 keV (Section~\ref{subsec:ks1947p300}, Fig.~\ref{fig:1947p300}), whose identification as CRSF \citep{ks1947+300_furst_2014ApJ...784L..40F} or an artefact due to two-component comptonization model in low luminosity XRPs \citep{xtej1946_doroshenko_2017_refId0} is debated.}

The accretion-powered X-ray pulsars show flares and dips in their light curves, and sometimes there are spectral changes associated with the change in intensity \citep{pradhan2014variations}. Analysis of time-averaged spectrum, including persistent intensity, dips, flares, eclipses etc., is equivalent to fitting spectra of the sum of different power-law components with a single power-law. We have taken care of this in the current analysis. XRPs also show a significant variation in absorption column density with time \citep{pradhan2014variations}. We have used a partial covering absorption model to account for the column density variations at short intervals which may happen due to the clumpiness of the stellar wind of the companion. In particular, \textit{TKF} was not observed in broadband X-ray spectra of XRPs when a partial covering absorption component was used in \cite{Maitra_2013-xtej1946} and \cite{maitra2012timing}. As mentioned in Section. \ref{sec:intro}, analysis of the same data by different groups also {have} obtained different results regarding the \textit{TKF}. In a systematic spectral study of the CRSF in XRPs using \textit{BeppoSAX} by \cite{intro_doroshenko_2017ASS}, \textit{TKF} was not required to model the spectrum in any of the sources as well. This feature being consistently at or near 10 keV, is unlikely to be related to the magnetic field of the neutron star. It was noted earlier that instead of some physical reason, the \textit{TKF} appears from modelling the continuum below the cutoff energy with a simple model, and the \texttt{NPEX} model was moderately successful in removing this feature from the residuals in some cases \citep{into_coburn_2002}.

It is also worth noting that features that perfectly resemble \textit{TKF} have been reported in other classes of sources like non-pulsating neutron star X-ray binaries as well, for instance, the \textit{Ginga} observation of GX 5-1 reported in \citealt{intro_asai_1994PASJ...46..479A}. However, this additional feature is neither reported nor has any mention in the subsequent observations of the same source by \textit{INTEGRAL} \citep{gx5-1_integral_2005A&A...443..599P} and \textit{NuSTAR} \citep{gx5-1_Homan_2018}. In the recent \textit{NuSTAR} observations, the authors have used multi-component continuum models that have transition near 10 keV (See Fig.~2 of \citealt{gx5-1_Homan_2018}). \textit{TKF} seen in XRPs could also be similar in nature.

\section{Conclusion}
`\textit{10 keV feature}' has been frequently reported in the spectrum of XRPs since the \textit{RXTE} era. \textit{NuSTAR} is the most ideal X-ray observatory as of now which can study `\textit{10 keV feature}', and we performed a systematic search for the `\textit{10 keV feature}' in archival \textit{NuSTAR} observations of XRPs. {We found the `\textit{10 keV feature}' in 16 out of 58 \textit{NuSTAR} observations of XRPs. It could be fitted with a gaussian absorption model centred around 10 keV in 15 out of 16 cases (exception 4U 1907+09). Of these 16 reports, two of the observations could have an appearance of this feature due to flare in the observation. We however also report that individual analyses of some of these observations in the literature have reported models without the `\textit{10 keV feature}'.}

\section*{Acknowledgements}

{The authors thank the anonymous referee for the constructive comments, especially for pointing out the bias in the model selection in the original manuscript.} This research has made use of data and/or software provided by the High Energy Astrophysics Science Archive Research Center (HEASARC), which is a service of the Astrophysics Science Division at NASA/GSFC. This research has made use of the NuSTAR Data Analysis Software (NuSTARDAS) jointly developed by the ASI Science Data Center (Italy) and the California Institute of Technology (USA).

\section*{Data Availability}

All the data underlying this research work is publicly available in the High Energy Astrophysics Science Archive Research Center (HEASARC Data Archive).



\bibliographystyle{mnras}
\bibliography{example} 



\appendix
\section{Table: Best fitting spectral parameters} \label{Tables}
The best fit spectral parameters for all the observations analyzed in this work are { given in four tables}.


\newpage

\begin{table*}
    \centering
    \tiny
    \rotatebox{90}{
    \begin{tabular}{clc|ccccccccccccccccc}
        \hline
        \hline
         Source &\multicolumn{2}{c}{Observation details} &\multicolumn{6}{c}{Continuum parameters} &\multicolumn{3}{c}{Cyclotron Resonance Scattering Feature}\\
         &Obs. Sn. &Energy range & &$\Gamma_1$ &norm.\textsubscript{1} &norm.\textsubscript{2}$^\dagger$ / T$_0^\ddagger$ (keV)  &E\textsubscript{cut} (keV) / kT$_e^\ddagger$ (keV) &E\textsubscript{fold} (keV) / $\tau^\ddagger$ &E\textsubscript{cyc} (keV) &$\sigma$\textsubscript{cyc} (keV) &Strength$^\S$\\
         \hline
         Her X--1 &1 (ta) &3$-$79 &NPEX &$1.11_{-0.07}^{+0.08}$ &$0.18_{-0.02}^{+0.03}$ &$(3\pm0.1)\times10^{-4}$ &$5.75_{-0.06}^{+0.06}$ &-  &$38.20_{-0.33}^{+0.34}$ &$7.12_{-0.29}^{+0.30}$ &$15.76_{-1.22}^{+1.32}$\\
         &1 (oe) &3$-$79 &HEC &$0.95_{-0.01}^{+0.01}$ &$0.108_{-0.001}^{+0.001}$ &- &$20.51_{-0.34}^{+0.40}$ &$10.13_{-0.23}^{+0.23}$ &$38.19_{-0.53}^{+0.56}$ &$6.18_{-0.53}^{+0.59}$ &$8.86_{-1.23}^{+1.45}$\\
         &2 &3$-$79 &FDC &$0.82_{-0.05}^{+0.05}$ &$0.20_{-0.02}^{+0.02}$ &- &$21.07_{-3.34}^{+4.37}$ &$8.27_{-0.39}^{+0.22}$ &$37.31_{-0.22}^{+0.23}$ &$6.67_{-0.34}^{+0.46}$ &$13.91_{-1.58}^{+2.71}$\\
         &3 &3$-$79 &FDC &$0.77_{-0.05}^{+0.04}$ &$0.24_{-0.01}^{+0.01}$ &- &$15.75_{-2.68}^{+2.27}$ &$8.53_{-0.11}^{+0.10}$ &$37.16_{-0.20}^{+0.20}$ &$6.75_{-0.28}^{+0.25}$ &$13.13_{-1.03}^{+1.14}$\\
         
         Vela X--1 &4 &3$-$79 &compTT &- &$0.26_{-0.03}^{+1.12}$ &$0.47_{-0.47}^{+0.12}$ &$6.56_{-0.08}^{+0.09}$ &$12.83_{-0.20}^{+0.21}$ &$24.79_{-1.13}^{+0.97}$ &$4.60^\star$ &$0.54_{-0.12}^{+0.15}$\\
         &&&&&&&&&$54.48_{-0.67}^{+0.77}$ &$7.50_{-0.65}^{+0.79}$ &$15.93_{-2.09}^{+2.73}$\\
         &5 &3$-$79 &CPL &$1.05_{-0.06}^{+0.05}$ &$0.27_{-0.03}^{+0.03}$ &- &$20.91_{-2.02}^{+1.03}$ &- &$26.37_{-1.26}^{+1.24}$ &$3.83_{-1.32}^{+1.15}$ &$0.82_{-0.42}^{+0.49}$\\
         &&&&&&&&&$57.18_{-1.24}^{+0.83}$ &$9.59_{-1.23}^{+-9.59}$ &$34.56_{-7.16}^{+4.80}$\\
         &6 &3$-$79 &compTT &- &$(4.2\pm0.2)\times10^{-3}$ &$1.01_{-0.02}^{+0.02}$ &$8.41_{-0.43}^{+0.51}$ &$9.72_{-0.10}^{+0.26}$ &$27.05_{-0.53}^{+0.58}$ &$5.83_{-0.53}^{+0.57}$ &$3.31_{-0.79}^{+0.80}$\\
         &&&&&&&&&$52.59_{-0.97}^{+1.07}$ &$8.95_{-1.16}^{+-8.95}$ &$25.67_{-6.93}^{+7.96}$\\
         &" &3$-$79 &compTT &- &$0.10_{-0.01}^{+0.40}$ &$0.29_{-0.29}^{+0.11}$ &$7.29_{-0.22}^{+0.19}$ &$10.32_{-0.31}^{+0.33}$ &$24.85_{-0.74}^{+0.78}$ &$5.83^\star$ &$2.34_{-0.29}^{+0.42}$\\
         &&&&&&&&&$50.79_{-0.66}^{+0.75}$ &$6.07_{-0.84}^{+1.24}$ &$10.45_{-2.26}^{+4.19}$\\
         &7 (f) &3$-$79 &HEC &$1.16_{-0.06}^{+0.04}$ &$0.27_{-0.04}^{+0.03}$ &- &$21.80_{-0.70}^{+0.68}$ &$15.04_{-1.97}^{+1.53}$ &$27.71_{-2.51}^{+1.26}$ &$3.40_{-1.49}^{+2.12}$ &$1.10_{-0.59}^{+1.29}$\\
         &&&&&&&&&$56.72_{-2.34}^{+1.93}$ &$9.16_{-2.27}^{+-9.16}$ &$34.91_{-14.07}^{+10.50}$\\
         
         XTE J1946+274 &8 &3$-$60 &HEC &$0.99_{-0.04}^{+0.04}$ &$0.07_{-0.01}^{+0.01}$ &- &$18.10_{-0.19}^{+0.19}$ &$9.41_{-0.24}^{+0.26}$ &$38.55_{-0.81}^{+0.87}$ &$5.03_{-0.64}^{+0.71}$ &$3.63_{-0.84}^{+1.06}$\\
         KS 1947+300 &9 &3$-$79 &HEC &$1.17_{-0.07}^{+0.07}$ &$0.23_{-0.04}^{+0.05}$ &- &$7.92_{-0.31}^{+0.27}$ &$25.71_{-1.18}^{+1.26}$ &- &- &-\\
         &10 &3$-$79 &NPEX &$0.80_{-0.04}^{+0.03}$ &$0.32_{-0.04}^{+0.03}$ &$(2.1\pm0.3)\times10^{-5}$ &$10.70_{-0.22}^{+0.25}$ &- &- &- &-\\
         &11 &3$-$79 &NPEX &$0.69_{-0.04}^{+0.03}$ &$0.15_{-0.02}^{+0.02}$ &$(1.0\pm0.2)\times10^{-5}$ &$11.31_{-0.31}^{+0.38}$ &- &- &- &-\\
         
         4U 1907+09 &12 &3$-$45 &CompTT &- &$0.021_{-0.002}^{+0.013}$ &$0.36_{-0.36}^{+0.11}$ &$4.20_{-0.03}^{+0.03}$ &$19.44_{-0.36}^{+0.39}$ &$17.98_{-0.08}^{+0.08}$ &$2.69_{-0.11}^{+0.12}$ &$3.02_{-0.17}^{+0.21}$\\
         &&&&&&&&&$36.70_{-0.37}^{+0.42}$ &$3.28_{-0.31}^{+0.35}$ &$7.09_{-0.83}^{+0.93}$\\
         
         4U 1538-52 &13 (ta) &3$-$50 &NPEX &$1.45_{-0.11}^{+0.11}$ &$0.05_{-0.01}^{+0.01}$ &$(5.0\pm0.3)\times10^{-5}$ &$4.77_{-0.05}^{+0.05}$ &- &$22.25_{-0.17}^{+0.17}$ &$3.49_{-0.17}^{+0.18}$ &$6.37_{-0.44}^{+0.47}$\\
         & 13 (ooe) &3$-$50 &NPEX &$1.11_{-0.11}^{+0.11}$ &$0.07_{-0.02}^{+0.01}$ &$(1.01\pm0.07)\times10^{-4}$ &$4.81_{-0.06}^{+0.06}$ &- &$22.35_{-0.20}^{+0.21}$ &$3.39_{-0.20}^{+0.22}$ &$6.18_{-0.51}^{+0.55}$\\
         
         Cep X--4 &14 &3$-$79 &CompTT &- &$0.13_{-0.01}^{+1.06}$ &$0.36_{-0.36}^{+0.08}$ &$4.75_{-0.06}^{+0.05}$ &$14.30_{-0.19}^{+0.21}$ &$19.34_{-0.59}^{+-19.34}$ &$2.57_{-0.55}^{+0.84}$ &$0.37_{-0.15}^{+0.37}$\\ &&&&&&&&&$30.47_{-0.14}^{+0.14}$ &$5.17_{-0.20}^{+0.19}$ &$14.08_{-0.71}^{+0.80}$\\
         &15 &3$-$79 &NPEX &- &$1.96_{-0.92}^{+1.70}$ &$0.06_{-0.03}^{+0.13}$ &$(1.53\pm0.09)\times10^{-4}$ &$4.55_{-0.06}^{+0.06}$ &$18.06_{-0.67}^{+0.79}$ &$2.36_{-0.56}^{+0.72}$ &$0.40_{-0.17}^{+0.26}$\\
         &&&&&&&&&$29.35_{-0.22}^{+0.22}$ &$4.60_{-0.26}^{+0.26}$ &$12.95_{-1.05}^{+1.13}$\\
         
         4U 1626-67 &16 &3$-$79 &FDC &$0.93_{-0.05}^{+0.04}$ &$0.042_{-0.003}^{+0.003}$ &- &$23.99_{-2.29}^{+2.25}$ &$7.61_{-0.23}^{+0.20}$ &$37.30_{-0.13}^{+0.13}$ &$4.76_{-0.17}^{+0.19}$ &$17.28_{-0.97}^{+1.17}$\\
         
         SMC X--2 &17 &3$-$50 &CompTT &- &$0.03_{-0.01}^{+0.38}$ &$0.31_{-0.31}^{+1.18}$ &$4.73_{-0.14}^{+0.29}$ &$16.39_{-2.43}^{+1.89}$ &$26.94_{-0.71}^{+0.86}$ &$6.96_{-1.39}^{+1.32}$ &$5.39_{-1.97}^{+0.70}$\\
         &18 &3$-$50 &NPEX &$0.67_{-0.11}^{+0.13}$ &$0.023_{-0.004}^{+0.006}$ &$(1.2\pm0.1)\times10^{-4}$ &$4.52_{-0.17}^{+0.26}$ &- &$28.92_{-1.68}^{+2.43}$ &$7.20_{-1.51}^{+2.37}$ &$9.23_{-4.05}^{+10.22}$\\
         &19 &3$-$50 &NPEX &$0.96_{-0.16}^{+0.17}$ &$0.02_{-0.00}^{+0.01}$ &$(9.3\pm0.8)\times10^{-5}$ &$4.51_{-0.10}^{+0.10}$ &- &$28.85_{-0.69}^{+0.65}$ &$7.1^\star$ &$9.42_{-1.49}^{+1.49}$\\
         
         IGR J17544--2619 &20 &3$-$40 &NPEX &$2.97_{-0.34}^{+0.37}$ &$0.09_{-0.04}^{+0.09}$ &$4.75_{-0.61}^{+0.73}\times10^{-6}$ &$4.77_{-0.14}^{+0.15}$ &- &$17.13_{-0.58}^{+0.60}$ &$1.89_{-0.61}^{+0.73}$ &$1.47_{-0.55}^{+0.70}$\\
         
         IGR J16393--4643 &21 &3$-$50 &NPEX &$1.19_{-0.31}^{+0.28}$ &$0.014_{-0.007}^{+0.010}$ &$(2.4\pm0.4)\times10^{-5}$ &$4.48_{-0.14}^{+0.13}$ &- &$30.75_{-1.21}^{+1.04}$ &$2.44_{-1.04}^{+0.95}$ &$1.87_{-1.00}^{+0.98}$\\
         
         2S 1553--542 &22 &3$-$50 &FDC &$1.07_{-0.14}^{+0.07}$ &$0.07_{-0.01}^{+0.01}$ &- &$17.02_{-5.85}^{+6.62}$ &$6.21_{-1.15}^{+0.52}$ &$27.13_{-0.70}^{+0.80}$ &$7.75_{-1.16}^{+1.28}$ &$14.92_{-6.80}^{+14.61}$\\
         RX J0520.5--6932 &23 &4$-$55 &CPL &$0.42_{-0.04}^{+0.04}$ &$0.054\pm0.003$ &- &$7.35_{-0.26}^{+0.28}$ &- &$31.32_{-0.70}^{+0.79}$ &$5.53_{-0.67}^{+0.82}$ &$6.78_{-1.38}^{+1.84}$\\
         &24  &4$-$55 &NPEX &$0.23_{-0.03}^{+0.03}$ &$0.036\pm0.002$ &- &$4.93_{-0.10}^{+0.10}$ &- &$31.92_{-0.61}^{+0.68}$ &$7.43_{-0.65}^{+0.76}$ &$13.45_{-2.30}^{+3.02}$\\
         
         Cen X--3 &25 &3$-$60 &HEC &$1.27_{-0.04}^{+0.03}$ &$0.38_{-0.03}^{+0.03}$ &- &$12.48_{-0.13}^{+0.13}$ &$8.92_{-0.15}^{+0.15}$ &$29.02_{-0.29}^{+0.30}$ &$4.93_{-0.35}^{+0.41}$ &$5.22_{-0.56}^{+0.68}$\\
         
         GX 301--2 &26 (ta) &3$-$65 &NPEX &$0.78_{-0.03}^{+0.21}$ &$0.09_{-0.01}^{+0.01}$ &$(6.7\pm1.5)\times10^{-5}$ &$6.59_{-0.31}^{+5.20}$ &- &$34.31_{-1.04}^{+1.50}$ &$4.51_{-1.23}^{+1.24}$ &$2.63_{-1.43}^{+2.68}$\\
         &&&&&&&&&$51.22_{-1.87}^{+8.63}$ &$10.30_{-2.54}^{+9.07}$ &$25.95_{-9.60}^{+107.26}$\\
         &27 (ta) &3$-$65 &NPEX &$0.85_{-0.04}^{+0.14}$ &$0.15_{-0.01}^{+0.01}$ &$1.05_{-0.51}^{+0.33}\times10^{-4}$ &$6.84_{-0.63}^{+4.09}$ &- &$33.88_{-0.84}^{+1.68}$ &$4.83_{-1.16}^{+1.52}$ &$2.77_{-1.31}^{+4.78}$\\
         &&&&&&&&&$53.21_{-2.37}^{+6.38}$ &$13.26_{-4.14}^{+6.09}$ &$54.60_{-27.02}^{+143.92}$\\
         &27 (f) &3$-$65 & NPEX &$0.84_{-0.05}^{+0.04}$ &$0.25_{-0.02}^{+0.03}$ &$3.45_{-0.33}^{+0.26}\times10^{-4}$ &$5.66_{-0.09}^{+0.15}$ &- &$32.64_{-1.32}^{+1.26}$ &$4.04_{-0.90}^{+0.81}$ &$1.82_{-0.74}^{+1.00}$\\
         &&&&&&&&&$53.2^\star$ &$9.88_{-0.84}^{+0.79}$ &$42.89_{-4.54}^{+6.07}$\\
         &27 (q) &3$-$65 &NPEX &$0.89_{-0.04}^{+0.04}$ &$0.13_{-0.01}^{+0.01}$ &$6.96\pm1.15)\times10^{-5}$ &$6.85_{-0.31}^{+0.42}$ &- &$34.30_{-0.78}^{+0.95}$ &$4.74_{-0.77}^{+0.77}$ &$4.20_{-1.52}^{+2.44}$\\
         &&&&&&&&&$53.2^\star$ &$11.40_{-2.02}^{+1.75}$ &$38.99_{-9.70}^{+13.52}$\\
         
         XTE J1829--098 &28 &3$-$40 &NHC &$0.64_{-0.05}^{+0.05}$ &$0.008_{-0.001}^{+0.001}$ &- &$10.31_{-0.47}^{+0.40}$ &$6.53_{-0.28}^{+0.25}$ &$14.61_{-0.09}^{+0.10}$ &$2.66_{-0.20}^{+0.19}$ &$3.53_{-0.57}^{+0.64}$\\

         V 0332+53 &29 &3$-$55 &HEC &$0.67_{-0.03}^{+0.03}$ &$0.054\pm0.004$ &- &$13.85_{-0.43}^{+0.41}$ &$19.23_{-3.73}^{+7.20}$ &$27.82_{-0.07}^{+0.06}$ &$4.98_{-0.24}^{+0.27}$ &$2.21_{-0.08}^{+0.09}$\\
         &"&&&&&&&&&$7.49_{-0.87}^{+1.43}$ &$10.47_{-1.94}^{+-10.47}$\\
         &30 &3$-$55 &NHC &$0.75_{-0.03}^{+0.03}$ &$0.030_{-0.002}^{+0.003}$ &$15.50_{-0.60}^{+0.38}$ &$16.22_{-2.00}^{+2.87}$ &- &$27.89_{-0.07}^{+0.07}$ &$4.20_{-0.26}^{+0.29}$ &$2.08_{-0.08}^{+0.08}$ &\\
         &"&&&&&&&&&$7.2^\star$ &$12.56_{-1.77}^{+1.88}$\\
         &30 &3$-$55 &CompTT &- &$0.031_{-0.003}^{+0.080}$ &$0.45_{-0.45}^{+0.19}$ &$5.87_{-0.31}^{+0.35}$ &$18.68_{-0.67}^{+0.78}$ &$27.67_{-0.07}^{+0.08}$ &$4.44_{-0.25}^{+0.26}$ &$2.06_{-0.07}^{+0.07}$\\
         &"&&&&&&&&&$7.16_{-4.09}^{+2.77}$ &$7.26_{-2.44}^{+21.63}$\\
         &31 &3$-$55 &NHC &$0.71_{-0.05}^{+0.04}$ &$0.015_{-0.002}^{+0.002}$ &- &$16.49_{-0.38}^{+0.35}$ &$17.85_{-2.63}^{+3.92}$ &$29.43_{-0.09}^{+0.09}$ &$4.04_{-0.26}^{+0.28}$ &$2.32_{-0.09}^{+0.09}$\\
         &"&&&&&&&&&$8.95_{-3.69}^{+2.99}$ &$9.88_{-2.69}^{+11.20}$\\
         &31 &3$-$55 &NPEX &$0.37_{-0.07}^{+0.23}$ &$0.020_{-0.001}^{+0.006}$ &$(8.4\pm1.9)\times10^{-5}$ &$6.14_{-0.33}^{+0.29}$ &- &$29.19_{-0.08}^{+0.08}$ &$4.40_{-0.24}^{+0.26}$ &$2.29_{-0.04}^{+0.07}$\\
         &"&&&&&&&&&$8.95^\star$ &$5.66_{-1.06}^{+1.06}$\\
         &32 &3$-$55 &NHC &$0.66_{-0.14}^{+0.07}$ &$0.011_{-0.003}^{+0.002}$ &- &$16.56_{-1.30}^{+0.57}$ &$19.75_{-4.10}^{+9.31}$ &$29.21_{-0.11}^{+0.10}$ &$4.46_{-0.36}^{+0.46}$ &$2.29_{-0.12}^{+0.15}$\\
         &"&&&&&&&&&$8.16_{-4.65}^{+3.69}$ &$11.56_{-3.52}^{+24.98}$\\

         \hline
        \hline
    \end{tabular}
    }
    \begin{tablenotes}
            \item $^\dagger$ Normalization of \texttt{cutoffpl} model with power-law index -2 in \texttt{NPEX} continuum model.
            \item $^\ddagger$ Parameters for the \texttt{CompTT} continuum model.
            \item $^\S$ Optical depth related to strength (line depth) by $\tau_{\rm{cyc}}=\frac{\rm{Strength}_{\rm{cyc}}}{\sqrt{2\pi}\sigma_{\rm{cyc}}}$.
            \item $^\star$ Frozen parameter.
    \end{tablenotes}
    \caption{The best fitting continuum model and Cyclotron resonance scattering feature model parameters from spectral fitting for Obs. Sn.1 to 32.}
    \label{tab:cyc1}
\end{table*}

\begin{table*}
    \centering
    \tiny
    \rotatebox{90}{
    \begin{tabular}{l l||cccccccccccccccc}
       \hline
	 \hline
    &tbabs &\multicolumn{2}{c}{tbpcf} &\multicolumn{2}{c}{bbody} &\multicolumn{3}{c}{Atomic line} &\multicolumn{3}{c}{\textit{10 keV feature}} &C\textsubscript{FPMB} &Fit &AIC Score\\
    Obs. Sn.&nH$_1$ &$f$ &nH$_2$ &kT\textsubscript{BB} &norm.\textsubscript{BB} &E (Fe K$\alpha$) &$\sigma$ (Fe K$\alpha$) &norm. &E\textsubscript{gabs} &$\sigma$\textsubscript{gabs} &strength\textsubscript{gabs} & &$\chi^2$/dof &$\chi^2+2n$\textsubscript{par}\\
	 \hline
    1 (ta) &0.015$^\star$ &$0.45_{-0.04}^{+0.03}$ &$44.45_{-5.70}^{+5.84}$ &- &- &$6.41_{-0.02}^{+0.02}$ &$0.15_{-0.04}^{+0.04}$ &$(1\pm0.1)\times10^{-3}$ &- &- &- &$1.035\pm0.002$ &615.5/455 &655.5\\
    &&&&&&$6.41_{-0.02}^{+0.02}$ &$1.58_{-0.14}^{+0.13}$ &$(6\pm0.1)\times10^{-3}$\\
    1 (oe) &0.015$^\star$ &- &- &- &- &$6.43_{-0.02}^{+0.02}$ &$0.13_{-0.09}^{+0.07}$ &$(1\pm0.3)\times10^{-3}$ &- &- &- &$1.033\pm0.003$ &500.9/433 &536.9\\
    &&&&&&$6.43_{-0.02}^{+0.02}$ &$0.92_{-0.13}^{+0.17}$ &$(4\pm0.4)\times10^{-3}$\\
    2 &0.015$^\star$ &$0.26_{-0.09}^{+0.07}$ &$512.63_{-83.96}^{+47.06}$ &$0.34_{-0.08}^{+0.09}$ &$0.01_{-0.01}^{+0.04}$ &$6.52_{-6.52}^{+-6.52}$ &$0.22_{-0.05}^{+0.04}$ &$(3\pm1)\times10^{-3}$ &$11.05_{-0.22}^{+0.23}$ &$0.88_{-0.22}^{+0.32}$ &$0.05_{-0.02}^{+0.03}$ &$1.033\pm0.001$ &538.3/474 &595.3\\
    &&&&&&$6.52_{-0.01}^{+0.01}$ &$0.67_{-0.11}^{+0.14}$ &$(4\pm0.8)\times10^{-3}$\\
    3 &0.015$^\star$ &$0.42_{-0.03}^{+0.03}$ &$590.47_{-18.52}^{+19.10}$ &$0.32_{-0.05}^{+0.07}$ &$0.02_{-0.01}^{+0.04}$ &$6.49_{-0.01}^{+0.01}$ &$0.18_{-0.03}^{+0.03}$ &$(3\pm0.6)\times10^{-3}$ &$10.21_{-0.12}^{+0.12}$ &$0.23_{-0.10}^{+0.13}$ &$0.012\pm0.004$ &$1.016\pm0.001$ &696.9/488 &742.9\\
    &&&&&&$6.49_{-0.01}^{+0.01}$ &$0.61_{-0.06}^{+0.07}$ &$(6\pm0.6)\times10^{-6}$\\
    4 &0.37$^\star$ &$0.87_{-0.01}^{+0.01}$ &$33.20_{-1.40}^{+1.63}$ &- &- &$6.40_{-0.01}^{+0.01}$ &$0.06_{-0.06}^{+0.04}$ &$(4\pm0.3)\times10^{-3}$ &$10.39_{-0.21}^{+0.24}$ &$1.34_{-0.30}^{+0.36}$ &$0.10_{-0.03}^{+0.04}$ &$1.044\pm0.002$ &552.4/23 &598.4\\
    5 &0.37$^\star$ &$0.84_{-0.02}^{+0.02}$ &$23.49_{-2.06}^{+1.97}$ &- &- &$6.37_{-0.03}^{+0.03}$ &$0.09_{-0.09}^{+0.06}$ &$(1.3\pm0.2)\times 10^{-3}$ &- &- &- &$1.027\pm0.004$ &531.9/435 &565.9\\
    6 &0.37$^\star$ &$0.71_{-0.02}^{+0.02}$ &$26.10_{-1.96}^{+1.86}$ &- &- &$6.37_{-0.02}^{+0.02}$ &$0.05_{-0.05}^{+0.06}$ &$(7.6\pm0.8)\times10^{-4}$ &- &- &- &$1.013\pm0.002$ &503.9/465 &543.9\\
    " &0.37$^\star$ &$0.86_{-0.01}^{+0.01}$ &$25.59_{-0.63}^{+1.17}$ &- &- &$6.36_{-0.02}^{+0.04}$ &$0.00_{-0.00}^{+0.06}$ &$6.97_{-0.62}^{+0.37}\times10^{-4}$ &$11.97_{-0.31}^{+0.32}$ &$1.96_{-0.37}^{+0.39}$ &$0.20_{-0.07}^{+0.10}$ &$1.013\pm0.002$ &540.9/463 &586.9\\
    7 (f) &0.37$^\star$ &$0.88_{-0.03}^{+0.05}$ &$23.45_{-4.53}^{+3.63}$ &- &- &$6.35_{-0.06}^{+0.05}$ &$0.22_{-0.08}^{+0.10}$ &$(2.7\pm0.8)\times10^{-3}$ &- &- &- &$1.02\pm0.01$ &454.1/398 &496.1\\
    8 &$0.10_{-0.10}^{+0.21}$ &$0.21_{-0.05}^{+0.04}$ &$79.11_{-23.90}^{+21.83}$ &$2.09_{-0.07}^{+0.10}$ &$(3.8\pm0.9)\times10^{-3}$ &$6.45_{-0.03}^{+0.03}$ &$0.20_{-0.08}^{+0.08}$ &$(5.2\pm0.1)\times 10^{-3}$ &$10.25_{-0.26}^{+0.26}$ &$1.11_{-0.29}^{+0.40}$ &$0.07_{-0.03}^{+0.05}$ &$0.998\pm0.002$ &506.8/23 &552.8\\
    9 &$1.50_{-1.50}^{+2.65}$ &$0.15_{-0.05}^{+0.03}$ &$463.43_{-88.47}^{+83.43}$ &$0.59_{-0.08}^{+0.17}$ &$(5.4^{+7.6}_{-2.7})\times10^{-3}$ &$6.56_{-0.03}^{+0.03}$ &$0.25_{-0.06}^{+0.06}$ &$(1.5\pm0.3)\times10^{-3}$ &- &- &- &$1.017\pm0.002$ &546.8/486 &580.8\\
    10 &$3.39_{-1.48}^{+1.40}$ &$0.26_{-0.02}^{+0.02}$ &$511.99_{-51.09}^{+52.01}$ &$0.53_{-0.03}^{+0.05}$ &$0.02_{-0.01}^{+0.01}$ &$6.51_{-0.03}^{+0.03}$ &$0.23_{-0.05}^{+0.05}$ &$(1.68\pm0.3)\times10^{-3}$ &- &- &- &$1.015\pm0.002$ &527.4/493 &559.4\\
    11 &$4.00_{-1.38}^{+1.27}$ &$0.12_{-0.03}^{+0.02}$ &$564.48_{-110.40}^{+123.68}$ &$0.55_{-0.03}^{+0.04}$ &$0.014\pm0.002$ &$6.49_{-0.03}^{+0.03}$ &$0.19_{-0.05}^{+0.05}$ &$(9\pm1)\times10^{-4}$ &- &- &- &$1.021\pm0.002$ &583.8/503 &615.8\\
    12 &$6.27_{-0.47}^{+0.49}$ &- &- &- &- &$6.38_{-0.04}^{+0.03}$ &$0.22_{-0.07}^{+0.06}$ &$2.31_{-0.44}^{+0.49}$ &$7.96_{-0.14}^{+0.12}$ &$1.44_{-0.17}^{+0.21}$ &$0.40_{-0.07}^{+0.11}$ &$1.014\pm0.003$ &416.5/299 &456.5\\
    13 (ta) &$0.7^\star$ &$0.87_{-0.02}^{+0.02}$ &$41.77_{-5.17}^{+4.92}$ &- &- &$6.43_{-0.03}^{+0.03}$ &$0.16_{-0.08}^{+0.08}$ &$(2.0\pm0.4)\times10^{-4}$ &- &-&- &$1.02_{-0.01}^{+0.01}$ &363.0/306 &397.0\\
    13 (oe) &$0.7^\star$ &$0.86_{-0.03}^{+0.03}$ &$29.75_{-4.80}^{+4.31}$ &- &- &$6.42_{-0.07}^{+0.07}$ &$0.08_{-0.08}^{+0.15}$ &$(1.8\pm0.7)\times10^{-4}$ &- &- &- &$1.01_{-0.01}^{+0.01}$ &274.3/292 &308.3\\
    14 &$1.53_{-0.24}^{+0.20}$ &$0.33_{-0.02}^{+0.04}$ &$281.64_{-30.67}^{+40.12}$ &- &- &$6.45_{-0.03}^{+0.03}$ &$0.17_{-0.05}^{+0.04}$ &$(8.1\pm1.3)\times10^{-4}$ &- &- &- &$1.034\pm0.002$ &530.7/458 &570.7\\
    15 &$0.10_{-0.10}^{+2.65}$ &- &- &$1.41_{-0.03}^{+0.08}$ &$(2.0\pm0.6)\times10^{-3}$ &$6.39_{-0.05}^{+0.05}$ &$0.23_{-0.10}^{+0.08}$ &$(1.9\pm0.5)\times10^{-4}$ &- &- &- &$1.021\pm0.004$ &440.8/421 &478.8\\
    16 &$0.05_{-0.05}^{+0.41}$ &$0.32_{-0.06}^{+0.05}$ &$620.03_{-71.85}^{+70.62}$ &$0.49_{-0.03}^{+0.03}$ &$(2.07\pm0.27)\times10^{-3}$ &$6.74_{-0.05}^{+0.05}$ &$0.12_{-0.08}^{+0.07}$ &$(1.6\pm0.5)\times10^{-4}$ & & & &$1.002_{-0.003}^{+0.003}$ &691.4/444 &725.4\\
    17 &$0.65_{-0.65}^{+0.96}$ &- &- &$2.19_{-0.28}^{+0.55}$ &$(2.1\pm0.3)\times10^{-3}$ &$6.33_{-0.11}^{+0.10}$ &$0.62_{-0.15}^{+0.19}$ &$(6.8\pm2.0)\times10^{-4}$ &- &- &- &$1.014\pm0.004$ &304.5/317 &336.5\\
    18 &$0.85_{-0.85}^{+0.97}$ &- &- &- &- &$6.30_{-0.08}^{+0.08}$ &$0.34_{-0.09}^{+0.11}$ &$(2.4\pm0.6)\times10^{-4}$ &- &- &- &$1.03_{-0.01}^{+0.01}$ &249.4/304 &277.4\\
    19 &$1.86_{-1.25}^{+1.27}$ &- &- &- &- &$6.31_{-0.10}^{+0.09}$ &$0.28_{-0.15}^{+0.19}$ &$11.6_{-3.8}^{+5.2}\times10^{-5}$ &- &- &- &$1.03_{-0.01}^{+0.01}$ &324.3/300 &352.3\\
    20 &$16.36_{-3.71}^{+3.95}$ &- &- &- &- &- &- &- &- &- &- &$1.09_{-0.03}^{+0.03}$ &236.1/200 &256.1\\
    21 &$2.15^\star$ &$0.97_{-0.02}^{+0.02}$ &$51.65_{-6.29}^{+6.17}$ &- &- &- &- &- &$11.76_{-0.45}^{+0.41}$ &$1.46_{-0.47}^{+0.69}$ &$0.41_{-0.16}^{+0.30}$ &$0.99_{-0.02}^{+0.02}$ &371.4/267 &405.4\\
    22 &$3.77_{-0.84}^{+0.50}$ &$0.16_{-0.05}^{+0.05}$ &$83.94_{-38.63}^{+58.43}$ &- &- &$6.43_{-0.09}^{+0.09}$ &$0.12_{-0.12}^{+0.12}$ &$(1.4\pm0.7)\times10^{-4}$ &- &- &- &$1.012\pm0.003$ &268.6/314 &298.6\\
    23 &$0.2^\star$ &$0.26_{-0.03}^{+0.03}$ &$266.55_{-62.83}^{+58.67}$ &- &- &$6.54_{-0.05}^{+0.05}$ &$0.14_{-0.08}^{+0.07}$ &$(3.7\pm0.9)\times10^{-4}$ &$9.87_{-0.24}^{+0.39}$ &$0.39_{-0.38}^{+0.58}$ &$0.03_{-0.02}^{+0.04}$ &$1.043\pm0.004$ &332.8/312 &366.8\\
    24 &$0.2^\star$ &- &- &- &- &$6.59_{-0.05}^{+0.05}$ &$0.40_{-0.08}^{+0.08}$ &$(23.3\pm0.1)\times10^{-3}$ &$11.69_{-0.27}^{+0.27}$ &$0.37_{-0.30}^{+0.39}$ &$0.02_{-0.02}^{+0.02}$ &$1.022\pm0.003$ &381.8/339 &415.8\\
    25 &$5.27_{-0.63}^{+0.58}$ &$0.52_{-0.02}^{+0.02}$ &$72.38_{-5.64}^{+5.75}$ &- &- &$6.36_{-0.02}^{+0.02}$ &$0.10_{-0.04}^{+0.04}$ &$(2.03\pm0.23)\times10^{-3}$ &$11.22_{-0.23}^{+0.22}$ &$2.52_{-0.16}^{+0.18}$ &$0.98_{-0.14}^{+0.17}$ &$1.027\pm0.002$ &422.9/381 &464.9\\
    26 (ta) &$1.62_{-1.62}^{+7.26}$ &$0.90_{-0.10}^{+0.02}$ &$39.30_{-2.92}^{+2.07}$ &- &- &$6.36_{-0.01}^{+0.01}$ &$0.01^\star$ &$(10.0\pm0.6)\times10^{-4}$ &- &- &- &$1.054_{-0.004}^{+0.004}$ &479.5/387 &519.5\\
    27 (ta) &$5.41_{-5.41}^{+3.56}$ &$0.68_{-0.14}^{+0.14}$ &$31.88_{-2.61}^{+5.63}$ &- &- &$6.35_{-0.01}^{+0.01}$ &$0.01^\star$ &$1.76_{-0.07}^{+0.26}\times10^{-3}$ &- &- &- &$1.065\pm0.003$ &486.8/395 &526.8\\
    27 (f) &$0.10_{-0.10}^{+5.75}$ &$0.82_{-0.15}^{+0.01}$ &$31.62_{-2.64}^{+2.16}$ &- &- &$6.33_{-0.01}^{+0.02}$ &$0.01^\star$ &$(3.04\pm0.15)\times10^{-4}$ &- &- &- &$1.065\pm0.005$ &561.1/368 &601.1\\
    27 (q) &$5.12\pm5.12$ &$0.71_{-0.18}^{+0.11}$ &$30.65_{-2.79}^{+6.45}$ &- &- &$6.35_{-0.02}^{+0.01}$ &$0.01^\star$ &$(1.27\pm0.07)\times10^{-3}$ &- &- &- &$1.067\pm0.004$ &363.9/378 &403.9\\
    28 &$3.58_{-0.74}^{+0.73}$ &- &- &- &- &$6.47_{-0.04}^{+0.04}$ &$0.32_{-0.05}^{+0.05}$ &$(4.2\pm0.5)\times10^{-4}$ &- &- &- &$1.00_{-0.01}^{+0.01}$ &242.7/252 &268.7\\
    29 &$0.52_{-0.39}^{+0.38}$ &$0.23_{-0.03}^{+0.04}$ &$266.05_{-63.26}^{+59.25}$ &- &- &$6.36_{-0.06}^{+0.06}$ &$0.23_{-0.09}^{+0.10}$ &$(6.2\pm2)\times10^{-4}$ &- &- &- &$1.015_{-0.003}^{+0.003}$ &420.4/340 &460.4\\
    30 &$0.89_{-0.37}^{+0.37}$ &$0.12_{-0.03}^{+0.03}$ &$189.37_{-64.02}^{+152.24}$ &- &- &$6.30_{-0.05}^{+0.05}$ &$0.01^\star$ &$(2.3\pm0.5)\times10^{-4}$ &- &- &- &$1.014_{-0.004}^{+0.004}$ &394.1/333 &430.1\\
    " &$1.98_{-0.88}^{+0.43}$ &- &- &- &- &$6.31_{-0.04}^{+0.04}$ &$0.01^\star$ &$(2.1\pm0.4)\times10^{-4}$ &$9.73_{-0.31}^{+0.28}$ &$2.14_{-0.45}^{+0.51}$ &$0.33_{-0.14}^{+0.22}$ &$1.014_{-0.004}^{+0.004}$ &416.4/331 &454.4\\
    31 &$5.30_{-2.40}^{+2.21}$ &- &- &$0.59_{-0.05}^{+0.11}$ &$(1.02\pm0.32)\times10^{-3}$ &$6.28_{-0.06}^{+0.07}$ &$0.01^\star$ &$(1.03\pm0.5)\times10^{-4}$ &- &- &- &$1.024_{-0.004}^{+0.004}$ &399.1/330 &433.1\\
    " &$0.04_{-0.02}^{+0.03}$ &$0.23_{-0.11}^{+0.06}$ &$158.95_{-72.73}^{+40.23}$ &- &- &$6.27_{-0.07}^{+0.07}$ &$0.01^\star$ &$(9.37\pm3.7)\times10^{-5}$ &$11.52_{-0.31}^{+0.35}$ &$0.59_{-0.32}^{+0.43}$ &$0.04_{-0.02}^{+0.03}$ &$1.024\pm0.004$ &434.5/328 &478.5\\
    32 &$3.35_{-3.35}^{+3.12}$ &- &- &$0.70_{-0.13}^{+0.40}$ &$5.1_{-2.6}^{+6.4}\times10^{-4}$ &$6.35_{-0.07}^{+0.06}$ &$0.01^\star$ &$(9.6\pm2.7)\times10^{-5}$ &- &- &- &$1.04_{-0.01}^{+0.01}$ &370.4/325 &404.4\\
    
    \hline
    \end{tabular}
    }
    \begin{tablenotes}
        \item $^\star$ Frozen parameter.
    \end{tablenotes}
    \caption{The best fitting parameters of the absorption (\texttt{tbabs, tbpcf}), blackbody (\texttt{bbody}), iron fluorescence lines (\texttt{gaussian}) and \textit{TKF} (\texttt{gabs}) models, along with the fit statistic and \textit{AIC} score used for model selection. Details of Obs. Sn.1 to 32 are given in this table.}
    \label{tab:tkf1} 
\end{table*}

\begin{table*}
    \centering
    \tiny
    \rotatebox{90}{
    \begin{tabular}{clc|ccccccccccccccccc}
        \hline
        \hline
         Source &\multicolumn{2}{c}{Observation details} &\multicolumn{6}{c}{Continuum parameters} &\multicolumn{3}{c}{Cyclotron Resonance Scattering Feature}\\
         &Obs. Sn. &Energy range & &$\Gamma_1$ &norm.\textsubscript{1} &norm.\textsubscript{2}$^\dagger$ / T$_0$ (keV) &E\textsubscript{cut} (keV) / kT$_e$ (keV) &E\textsubscript{fold} (keV) / $\tau$ &E\textsubscript{cyc} (keV) &$\sigma$\textsubscript{cyc} (keV) &Strength\\
         \hline
         XTE J1858+034 &33 &5--55 &NPEX &$0.39_{-0.05}^{+0.04}$ &$0.036_{-0.004}^{+0.005}$ &$(1.2\pm0.2)\times10^{-4}$ &$5.48_{-0.15}^{+0.27}$ &- &$50.52_{-2.20}^{+3.88}$ &$9.74_{-1.45}^{+2.41}$ &$27.80_{-7.78}^{+18.35}$\\
         
         4U 1700--37 &34 (ta) &3$-$79 &NPEX &$0.82_{-0.04}^{+0.06}$ &$0.10_{-0.01}^{+0.01}$ &$(2.01\pm1.34)\times10^{-6}$ &$12.20_{-0.96}^{+1.34}$ &- &$16.13_{-0.89}^{+1.10}$ &$4.80_{-1.58}^{+1.31}$ &$0.97_{-0.46}^{+0.58}$\\
         &34 (f) &3$-$79 &NPEX &$0.75_{-0.04}^{+0.06}$ &$0.13_{-0.01}^{+0.01}$ &$(3.79\pm2.05)\times10^{-6}$ &$11.68_{-0.76}^{+1.22}$ &- &$16.12_{-1.00}^{+1.11}$ &$4.8^\star$ &$0.91_{-0.17}^{+0.16}$\\
         &34 (q) &3$-$79 &NPEX &$1.23_{-0.04}^{+0.08}$ &$0.07_{-0.01}^{+0.01}$ &$6.93_{-3.64}^{+4.97}\times10^{-7}$ &$11.48_{-1.16}^{+1.86}$ &- &$16.69_{-0.91}^{+1.07}$ &$3.08_{-1.80}^{+2.00}$ &$0.81_{-0.51}^{+1.16}$\\
         
        LMC X--4 &35 (ta) &3$-$79 &NPEX &$0.60_{-0.08}^{+0.08}$ &$0.05_{-0.01}^{+0.01}$ &$(3.7\pm0.6)\times10^{-5}$ &$6.29_{-0.14}^{+0.16}$ &- &- &- &-\\
        &35 (f) &3$-$65 &NPEX &$0.78_{-0.13}^{+0.14}$ &$0.18_{-0.05}^{+0.08}$ &$(1.30\pm0.21)\times10^{-4}$ &$5.12_{-0.12}^{+0.13}$ &- &- &- &-\\
        &35 (q) &3$-$79 &NPEX &$0.54_{-0.11}^{+0.10}$ &$0.03_{-0.01}^{+0.01}$ &$(3.6\pm0.6)\times10^{-5}$ &$6.33_{-0.12}^{+0.17}$ &- &- &- &-\\
        &" & 3$-$79 &FDC &$0.89_{-0.06}^{+0.04}$ &$0.030_{-0.003}^{+0.004}$ &- &$17.70_{-4.57}^{+3.38}$ &$10.10_{-0.43}^{+0.42}$ &- &- &-\\
        &36 &3$-$60 &FDC &$0.72_{-0.08}^{+0.09}$ &$0.020_{-0.003}^{+0.004}$ &- &$20.18_{-3.16}^{+2.08}$ &$9.96_{-0.31}^{+0.33}$ &- &- &-\\
        &" &3$-$60 &FDC &$0.62_{-0.05}^{+0.04}$ &$0.021_{-0.002}^{+0.002}$ &- &$15.10_{-3.73}^{+2.93}$ &$10.21_{-0.32}^{+0.30}$ &- &- &-\\
        &37 &3$-$68 &FDC &$0.81_{-0.18}^{+0.14}$ &$0.007_{-0.003}^{+0.003}$ &- &$23.02_{-2.66}^{+2.24}$ &$8.31_{-0.38}^{+0.40}$ &- &- &-\\
        &" &3$-$68 &CompTT &- &$0.004_{-0.004}^{+0.006}$ &$0.37_{-0.37}^{+0.86}$ &$5.77_{-0.09}^{+0.09}$ &$24.78_{-2.43}^{+3.58}$ &- &- &-\\
        &38 (ta) &3$-$65 &NPEX &$0.77_{-0.09}^{+0.08}$ &$0.05_{-0.01}^{+0.01}$ &$(2.8\pm0.3)\times10^{-5}$ &$6.47_{-0.12}^{+0.13}$ &- &- &- &-\\
        &38 (f) &3$-$65 &NPEX &$0.68_{-0.07}^{+0.07}$ &$0.18_{-0.02}^{+0.02}$ &$1.35_{-0.33}^{+0.37}\times10^{-4}$ &$4.92_{-0.17}^{+0.20}$ &- &- &- &-\\
        &" &3$-$65 &NPEX &$1.34_{-0.18}^{+0.20}$ &$0.18_{-0.04}^{+0.08}$ &- &$17.01_{-3.58}^{+3.45}$ &$7.76_{-0.52}^{+0.48}$ &- &- &-\\
        &38 (q) &3$-$65 &NPEX &$0.51_{-0.10}^{+0.10}$ &$0.017_{-0.003}^{+0.003}$ &$(2.7\pm0.4)\times10^{-5}$ &$6.52_{-0.13}^{+0.15}$ &- &- &- &-\\
        
        IGR J17329-2731 &39 &3$-$55 &NPEX &$0.48_{-0.13}^{+0.13}$ &$0.011_{-0.003}^{+0.003}$ &$3.02_{-3.00}^{+8.44}$ &$9.49_{-1.17}^{+1.29}$ &- &$22.43_{-0.58}^{+0.64}$ &$4.35_{-0.53}^{+0.63}$ &$3.49_{-0.69}^{+0.95}$\\
        
        SMC X--1 &40 &3$-$79 &HEC &$1.32_{-0.03}^{+0.04}$ &$0.08_{-0.01}^{+0.01}$ &- &$19.50_{-0.60}^{+0.70}$ &$11.16_{-0.28}^{+0.29}$ &- &- &-\\
        &41 &3$-$79 &FDC &$1.02_{-0.08}^{+0.08}$ &$0.054_{-0.005}^{+0.004}$ &- &$16.18_{-2.50}^{+2.54}$ &$8.52_{-0.26}^{+0.25}$ &- &- &-\\
        &42 &3$-$50 &FDC &$1.22_{-0.27}^{+0.23}$ &$0.01_{-0.01}^{+0.01}$ &- &$19.79_{-2.22}^{+1.98}$ &$5.80_{-0.40}^{+0.42}$ &- &- &-\\
        &43 &3$-$65 &NPEX &$0.35_{-0.01}^{+0.01}$ &$0.064_{-0.002}^{+0.001}$ &$(6.7\pm0.7)\times10^{-5}$ &$5.70_{-0.08}^{+0.09}$ &- &- &- &-\\
        &" &3$-$65 &FDC &$1.01_{-0.07}^{+0.07}$ &$0.09_{-0.01}^{+0.01}$ &- &$13.88_{-2.53}^{+2.39}$ &$8.90_{-0.21}^{+0.21}$ &- &- &-\\
        
        GRO J1008--57 &44 &3$-$79 &NPEX &$0.48_{-0.02}^{+0.02}$ &$0.46_{-0.01}^{+0.03}$ &$(1.78\pm0.12)\times10^{-4}$ &$8.23_{-0.09}^{+0.11}$ &- &$78^\star$ &$9.10_{-1.46}^{+1.68}$ &$12.61_{-2.78}^{+3.02}$\\
        
        GX 304--1 &45 &3$-$25 &PL &$1.75_{-0.23}^{+0.20}$ &$(9.04\pm0.63)\times10^{-4}$ &- &- &- &- &- &-\\
        
        1A 0535+26 &46 &3$-$79 &FDC &$1.05_{-0.01}^{+0.13}$ &$0.137_{-0.004}^{+0.016}$ &- &$16.23_{-4.55}^{+9.60}$ &$15.73_{-0.91}^{+0.85}$ &$44.88_{-0.75}^{+0.82}$ &$7.64_{-0.67}^{+0.92}$ &$10.68_{-1.65}^{+2.59}$\\
        &47 &3$-$50 &PL &$1.39_{-0.04}^{+0.03}$ &$0.05_{-0.01}^{+0.01}$ &- &- &- &- &- &-\\
        &48 &3$-$50 &compTT+compTT &- &$(3.4\pm0.3)\times10^{-4}$ &$0.64_{-0.56}^{+0.06}$ &$2.00_{-2.00}^{+0.09}$ &$17.44^{+0.65}_{-1.20}$ &$43.88_{-3.49}^{+2.74}$ &$7.67^\star$ &$27.95_{-5.35}^{+5.83}$\\
        &&&&&$(2.0\pm0.5)\times10^{-4}$ &" &$11.12_{-0.42}^{+0.20}$ &"\\
        
        GRO J2058+42 &49 &3$-$79 &CompTT &- &$0.041_{-0.004}^{+0.132}$ &$0.40_{-0.40}^{+0.12}$ &$7.99_{-0.07}^{+0.08}$ &$12.83_{-0.35}^{+0.39}$ &- &- &-\\
        &50 &3$-$79 &CompTT &- &$0.043_{-0.003}^{+0.014}$ &$0.41_{-0.12}^{+0.07}$ &$8.11_{-0.05}^{+0.05}$ &$13.26_{-0.33}^{+0.41}$ &- &- &-\\
        &51 &3$-$79 &NHC &$1.39_{-0.07}^{+0.06}$ &$0.015_{-0.002}^{+0.002}$ &- &$14.41_{-1.31}^{+1.28}$ &$27.54_{-2.42}^{+2.85}$ &- &- &-\\
        1E 1145.1--6141 &52 &3$-$60 &CPL &$0.88_{-0.07}^{+0.06}$ &$0.018\pm0.002$ &- &- &$17.03_{-0.96}^{+1.10}$ &- &- &- &-\\
        
        OAO 1657--415 &53 &5--70 &CPL &$0.56_{-0.05}^{+0.05}$ &$0.017_{-0.002}^{+0.002}$ &- &$16.27_{-0.48}^{+0.50}$ &- &- &-\\
        
        EXO 2030+375 &54 &3$-$55 &CPL &$1.20_{-0.02}^{+0.02}$ &$1.11_{-0.04}^{+0.04}$ &- &$18.93_{-0.24}^{+0.25}$ &- &- &- &-\\
        &55 &3$-$60 &NPEX &$1.02_{-0.04}^{+0.04}$ &$0.018_{-0.002}^{+0.002}$ &$(1.5\pm0.3)\times10^{-6}$ &$7.88_{-0.34}^{+0.40}$ &- &- &-\\
        &56 &3$-$79 &NPEX &$0.79_{-0.03}^{+0.02}$ &$0.20_{-0.01}^{+0.01}$ &$(7.5\pm2.1)\times10^{-6}$ &$10.80_{-0.43}^{+0.53}$ &- &- &- &-\\
        
        IGR J19294+1816 &57 &3$-$25 &NPEX &$1.35_{-0.34}^{+0.41}$ &$0.001_{-0.001}^{+0.002}$ &- &$5.57_{-0.85}^{+1.93}$ &- &- &- &-\\
        &58 &3$-$50 &FDC &$1.14_{-0.03}^{+0.03}$ &$(8.2\pm0.6)\times10^{-3}$ &- &$28.23_{-0.49}^{+0.46}$ &$4.80_{-0.23}^{+0.23}$ &- &- &-\\
        \hline
        \hline
    \end{tabular}
    }
    \begin{tablenotes}
            \item $^\dagger$ Normalization of \texttt{cutoffpl} model with power-law index -2 in \texttt{NPEX} continuum model.
            \item $^\ddagger$ Parameters for the \texttt{CompTT} continuum model.
            \item $^\S$ Optical depth related to strength (line depth) by $\tau_{\rm{cyc}}=\frac{\rm{Strength}_{\rm{cyc}}}{\sqrt{2\pi}\sigma_{\rm{cyc}}}$.
            \item $^\star$ Frozen parameter.
    \end{tablenotes}
    \caption{Table~\ref{tab:cyc1} contd. The best fitting continuum model and Cyclotron resonance scattering feature model parameters from spectral fitting for Obs. Sn.33 to 58.}
    \label{tab:cyc2}
\end{table*}

\begin{table*}
    \centering
    \tiny
    \rotatebox{90}{
    \begin{tabular}{l l||cccccccccccccccc}
       \hline
	 \hline
  Obs. Sn. &tbabs &\multicolumn{2}{c}{tbpcf} &\multicolumn{2}{c}{Blackbody} &\multicolumn{3}{c}{Atomic line} &\multicolumn{3}{c}{\textit{10 keV feature}} &C\textsubscript{FPMB} &Fit &AIC Score\\
	 &nH$_1$ &$f$ &nH$_2$ &kT\textsubscript{BB} &norm.\textsubscript{BB} &E (Fe K$\alpha$) &$\sigma$ (Fe K$\alpha$) &norm. &E\textsubscript{gabs} &$\sigma$\textsubscript{gabs} &strength\textsubscript{gabs} & &$\chi^2$/dof &$\chi^2+2n$\textsubscript{par}\\
    \hline
    33 &$8.45_{-2.46}^{+2.43}$ &- &- &- &- &$6.48_{-0.03}^{+0.03}$ &$0.25_{-0.06}^{+0.05}$ &$(5.0\pm0.1)\times10^{-3}$ &- &- &- &$1.017_{-0.003}^{+0.003}$ &412.4/337 &440.4\\
    34 (ta) &$0.5^\star$ &$0.77_{-0.01}^{+0.01}$ &$30.22_{-1.55}^{+1.53}$ &- &- &$6.36_{-0.01}^{+0.01}$ &$0.01^\star$ &$(9.2\pm0.6)\times10^{-4}$ &$10.61_{-0.42}^{+0.50}$ &$1.66_{-0.63}^{+0.77}$ &$0.10_{-0.06}^{+0.18}$ &$1.013_{-0.003}^{+0.003}$ &481.1/467 &521.1\\
    34 (f) &$0.5^\star$ &$0.76_{-0.01}^{+0.01}$ &$30.02_{-1.65}^{+1.64}$ &- &- &$6.36_{-0.01}^{+0.01}$ &$0.01^\star$ &$(1.40\pm0.08)\times10^{-3}$ &$10.54_{-0.49}^{+0.54}$ &$1.83_{-0.58}^{+0.79}$ &$0.11_{-0.06}^{+0.13}$ &$0.928_{-0.002}^{+0.002}$ &511.6/465 &551.6\\
    34 (q) &$0.5^\star$ &$0.84_{-0.02}^{+0.02}$ &$33.06_{-3.22}^{+3.23}$ &- &- &$6.32_{-0.04}^{+0.04}$ &$0.01^\star$ &$(2.1\pm0.5)\times10^{-4}$ &$11.39_{-0.64}^{+0.80}$ &$1.25_{-0.68}^{+0.67}$ &$0.15_{-0.10}^{+0.15}$ &$0.97_{-0.01}^{+0.01}$ &434.7/388 &474.7\\
    35 (ta) &$0.08^\star$ &$0.30_{-0.08}^{+0.07}$ &$89.37_{-17.20}^{+24.24}$ &- &- &$6.57_{-0.06}^{+0.05}$ &$0.32_{-0.10}^{+0.11}$ &$(5.62\pm1.8)\times10^{-4}$ &$10.70_{-0.31}^{+0.30}$ &$2.21_{-0.49}^{+0.64}$ &$0.56_{-0.20}^{+0.35}$ &$1.021_{-0.004}^{+0.004}$ &473.4/417 &507.4\\
    35 (f) &$0.08^\star$ &$0.31_{-0.12}^{+0.11}$ &$52.02_{-16.48}^{+17.46}$ &- &- &$6.59_{-0.10}^{+0.13}$ &$0.46_{-0.14}^{+0.21}$ &$(1.8\pm0.7)\times10^{-3}$ &$10.57_{-0.33}^{+0.36}$ &$1.56_{-0.47}^{+0.45}$ &$0.29_{-0.15}^{+0.21}$ &$1.07_{-0.01}^{+0.01}$ &452.0/340 &486.0\\
    35 (q) &$0.08^\star$ &$0.34_{-0.12}^{+0.10}$ &$89.46_{-24.08}^{+32.62}$ &- &- &$6.53_{-0.07}^{+0.06}$ &$0.25_{-0.16}^{+0.12}$ &$(3.3\pm1.5)\times10^{-4}$ &$11.14_{-0.58}^{+0.53}$ &$2.03_{-0.69}^{+0.87}$ &$0.34_{-0.18}^{+0.34}$ &$1.00_{-0.01}^{+0.01}$ &460.9/403 &494.9\\
    " &$0.08^\star$ &$0.30_{-0.06}^{+0.06}$ &$414.45_{-72.49}^{+66.53}$ &- &- &$6.48_{-0.05}^{+0.05}$ &$0.41_{-0.06}^{+0.07}$ &$(8.6\pm1.0)\times10^{-4}$ &- &- &- &$1.00_{-0.01}^{+0.01}$ &479.0/406 &503.0\\
    36 &$0.08^\star$ &$0.22_{-0.06}^{+0.08}$ &$123.07_{-27.86}^{+123.56}$ &- &- &$6.48_{-0.07}^{+0.07}$ &$0.25_{-0.14}^{+0.13}$ &$(2.7\pm0.2)\times10^{-4}$ &$10.88_{-0.43}^{+0.49}$ &$1.71_{-0.57}^{+0.46}$ &$0.22_{-0.14}^{+0.10}$ &$1.03_{-0.01}^{+0.01}$ &404.9/362 &434.9\\
    " &$0.08^\star$ &$0.27_{-0.04}^{+0.05}$ &$334.87_{-54.41}^{+51.17}$ &- &- &$6.51_{-0.06}^{+0.06}$ &$0.34_{-0.07}^{+0.09}$ &$(5.05\pm0.82)\times10^{-4}$ &- &- &- &$1.03_{-0.01}^{+0.01}$ &415.5/365 &439.5\\
    37 &$0.08^\star$ &$0.39_{-0.16}^{+0.10}$ &$89.01_{-19.13}^{+42.02}$ &- &- &$6.34_{-0.07}^{+0.07}$ &$0.16_{-0.16}^{+0.14}$ &$(1.3\pm0.7)\times10^{-4}$ &$10.42_{-0.70}^{+0.59}$ &$1.89_{-0.51}^{+0.73}$ &$0.37_{-0.16}^{+0.27}$ &$1.02_{-0.01}^{+0.01}$ &442.3/354 &472.3\\
    " &$0.08^\star$ &- &- &$1.34_{-0.15}^{+0.10}$ &$(1.8\pm0.8)\times10^{-4}$
    &$6.35_{-0.06}^{+0.06}$ &$0.45_{-0.09}^{+0.10}$ &$(2.9\pm0.6)\times10^{-4}$ &- &- &- &$1.02_{-0.01}^{+0.01}$ &446.8/357 &472.8\\
    38 (ta) &$0.08^\star$ &$0.31_{-0.11}^{+0.08}$ &$87.59_{-16.32}^{+24.22}$ &- &- &$6.50_{-0.10}^{+0.09}$ &$0.21_{-0.21}^{+0.19}$ &$(2.2\pm1.6)\times10^{-4}$ &$10.45_{-0.26}^{+0.24}$ &$1.60_{-0.25}^{+0.28}$ &$0.40_{-0.10}^{+0.12}$ &$1.02_{-0.01}^{+0.01}$ &394.2/368 &428.2\\
    38 (f) &$0.08^\star$ &- &- &- &- &$6.40_{-0.13}^{+0.14}$ &$0.71_{-0.18}^{+0.17}$ &$0.004_{-0.001}^{+0.002}$ &$10.28_{-0.33}^{+0.36}$ &$1.35_{-0.52}^{+0.62}$ &$0.29_{-0.17}^{+0.25}$ &$1.02_{-0.01}^{+0.01}$ &321.9/322 &349.9\\
    " &$0.08^\star$ &- &- &$1.19_{-0.21}^{+0.13}$ &$0.003\pm0.002$ &$6.42_{-0.22}^{+0.12}$ &$0.89_{-0.17}^{+0.26}$ &$0.007_{-0.002}^{+0.004}$ &- &- &- &$1.02_{-0.01}^{+0.01}$ &336.2/323 &358.2\\
    38 (q) &$0.08^\star$ &$0.35_{-0.08}^{+0.07}$ &$124.03_{-34.64}^{+48.95}$ &- &- &$6.33_{-0.05}^{+0.06}$ &$0.01^\star$ &$(1.37\pm0.45)\times10^{-4}$ &$10.96_{-0.52}^{+0.47}$ &$1.12_{-0.34}^{+0.55}$ &$0.14_{-0.05}^{+0.08}$ &$1.03_{-0.01}^{+0.01}$ &410.0/363 &444.0\\
    39 &$0.3^\star$ &$0.93_{-0.01}^{+0.01}$ &$56.43_{-4.07}^{+4.06}$ &- &- &$6.32_{-0.02}^{+0.03}$ &$0.01^\star$ &$(3.1\pm0.4)\times10^{-4}$ &- &- &- &$1.01_{-0.01}^{+0.01}$ &372.2/318 &406.2\\
    40 &$0.5^\star$ &- &- &$2.10_{-0.07}^{+0.08}$ &$(4.2\pm0.2)\times10^{-3}$ &$6.39_{-0.08}^{+0.07}$ &$0.42_{-0.11}^{+0.14}$ &$7.65_{-0.17}^{+0.23}\times10^{-4}$ &- &- &- &$1.021\pm0.004$ &486.2/416 &514.2\\
    41 &$0.5^\star$ &- &- &$1.87_{-0.21}^{+0.27}$ &$0.001_{-0.002}^{+0.003}$ &$6.30_{-0.07}^{+0.07}$ &$0.45_{-0.10}^{+0.12}$ &$(6.97\pm0.19)\times10^{-4}$ &- &- &- &$1.038_{-0.004}^{+0.004}$ &479.2/408 &501.2\\
    42 &$0.5^\star$ &$0.61_{-0.16}^{+0.10}$ &$66.80_{-10.34}^{+8.98}$ &- &- &$6.34_{-0.05}^{+0.04}$ &$0.20_{-0.10}^{+0.10}$ &$(2.3\pm0.6)\times10^{-4}$ &- &- &- &$1.02_{-0.01}^{+0.01}$ &334.1/260 &358.1\\
    43 &$0.5^\star$ &- &- &- &- &$6.39_{-0.08}^{+0.07}$ &$0.51_{-0.15}^{+0.14}$ &$(9.4\pm0.2)\times10^{-4}$ &$10.87_{-0.46}^{+0.61}$ &$0.92_{-0.35}^{+0.87}$ &$0.04_{-0.03}^{+0.06}$ &$1.027_{-0.003}^{+0.004}$ &387.6/378 &415.6\\
    " &$0.5^\star$ &- &- &$1.83_{-0.12}^{+0.16}$ &$(2.23\pm0.37)\times10^{-3}$ &$6.38_{-0.10}^{+0.09}$ &$0.57_{-0.18}^{+0.19}$ &$(1.03\pm0.42)\times10^{-3}$ &- &- &- &$1.027\pm0.004$ &396.7/379 &418.7\\
    44 &$0.10_{-0.10}^{+0.21}$ &$0.24_{-0.03}^{+0.03}$ &$101.65_{-10.70}^{+16.04}$ &- &- &$6.48_{-0.02}^{+0.02}$ &$0.14_{-0.04}^{+0.03}$ &$(3.7\pm0.5)\times10^{-3}$ &- &- &- &$1.024\pm0.001$ &579.4/510 &619.4\\
    &&&&&&$6.48_{-0.02}^{+0.02}$ &$1.95_{-0.09}^{+0.10}$ &$0.04\pm0.01$\\
    45 &$0.10_{-0.10}^{+0.26}$ &- &- &$1.18_{-0.02}^{+0.02}$ &$(9.23\pm1.01)\times10^{-5}$ &- &- &- &- &- &- &$0.99_{-0.02}^{+0.02}$ &1057.3/1019 &1069.3\\
    46 &$0.10_{-0.10}^{+0.71}$ &- &- &$1.55_{-0.11}^{+0.31}$ &$(1.2\pm0.5)\times10^{-3}$ &$6.37_{-0.05}^{+0.05}$ &$0.14_{-0.08}^{+0.07}$ &$(3.8\pm0.8)\times10^{-4}$ &- &- &- &$0.971_{-0.003}^{+0.003}$ &506.2/468 &534.2\\
    47 &$0.81_{-0.49}^{+0.50}$ &- &- &$1.72_{-0.04}^{+0.04}$ &$(1.1\pm0.1)\times10^{-3}$ &$6.30_{-0.15}^{+0.18}$ &$0.17_{-0.17}^{+0.23}$ &$(8.9\pm6.2)\times10^{-5}$ &- &- &- &$0.985\pm0.004$ &345.5/331 &369.5\\
    48 &$0.4^\star$ &- &- &- &- &- &- &- &- &- &- &$0.96\pm0.01$ &313.0/278 & NA\\
    49 &$0.60^\star$ &$0.25_{-0.07}^{+0.05}$ &$86.70_{-28.47}^{+32.55}$ &$2.40_{-0.10}^{+0.16}$ &$0.005\pm0.001$ &$6.39_{-0.05}^{+0.06}$ &$0.11_{-0.11}^{+0.12}$ &$(5.2\pm2.2)\times10^{-4}$ &$10.87_{-0.24}^{+0.23}$ &$1.46_{-0.30}^{+0.37}$ &$0.20_{-0.06}^{+0.10}$ &$0.986\pm0.003$ &557.8/468 &595.8\\
    50 &$0.60^\star$ &$0.18_{-0.08}^{+0.05}$ &$73.59_{-25.37}^{+24.44}$ &$2.56_{-0.10}^{+0.15}$ &$0.006\pm0.001$ &$6.39_{-0.04}^{+0.04}$ &$0.24_{-0.09}^{+0.12}$ &$(6.7\pm0.4)\times10^{-4}$ &$10.90_{-0.22}^{+0.21}$ &$1.71_{-0.28}^{+0.32}$ &$0.22_{-0.07}^{+0.08}$ &$0.997\pm0.002$ &592.3/500 &630.2\\
    51 &$0.60^\star$ &- &- &$1.88_{-0.13}^{+0.15}$ &$(5.1\pm0.5)\times10^{-4}$ &$6.09_{-0.31}^{+0.32}$ &$0.49_{-0.44}^{+0.34}$ &$7.3_{-4.9}^{+7.2}\times10^{-5}$ &- &- &- &$0.97_{-0.01}^{+0.01}$ &417.8/411 &441.8\\
    52 &$4.51_{-0.64}^{+0.63}$ &- &- &$2.44_{-0.15}^{+0.17}$ &$(6.1\pm0.7)\times10^{-4}$ &$6.34_{-0.04}^{+0.04}$ &$0.26_{-0.06}^{+0.06}$ &$(2.6\pm0.4)\times10^{-4}$ &- &- &- &$0.993_{-0.004}^{+0.004}$ &369.3/372 &389.3\\
    53 &$1.81^\star$ &$0.79_{-0.03}^{+0.03}$ &$60.99_{-5.41}^{+5.20}$ &- &- &$6.33_{-0.01}^{+0.01}$ &$0.01^\star$ &$(7.13\pm0.58)\times10^{-4}$ &- &- &- &$0.977\pm0.003$ &559.8/424 &587.8\\
    54 &$0.8^\star$ &$0.14_{-0.02}^{+0.02}$ &$85.04_{-5.02}^{+6.29}$ &- &- &$6.51_{-0.01}^{+0.01}$ &$0.24_{-0.02}^{+0.02}$ &$0.058_{-0.003}^{+0.003}$ &$10.86_{-0.12}^{+0.12}$ &$2.06_{-0.15}^{+0.16}$ &$0.25_{-0.03}^{+0.03}$ &$0.984_{-0.001}^{+0.001}$ &623.1/421 &651.1\\
    55 &$0.8^\star$ &$0.53_{-0.10}^{+0.53}$ &$11.32_{-7.02}^{+7.13}$ &- &- &$6.35_{-0.13}^{+0.13}$ &$0.01^\star$ &$(1.8\pm1.2)\times10^{-5}$ &- &- &- &$1.01_{-0.01}^{+0.01}$ &358.5/344 &386.5\\
    56 &$0.8^\star$ &$0.19_{-0.03}^{+0.03}$ &$59.13_{-13.50}^{+19.99}$ &$2.01_{-0.08}^{+0.14}$ &$0.004\pm0.001$ &$6.43_{-0.03}^{+0.03}$ &$0.20_{-0.05}^{+0.05}$ &$(1.2\pm0.2)\times10^{-3}$ &$10.57_{-0.28}^{+0.28}$ &$0.99_{-0.24}^{+0.38}$ &$0.06_{-0.02}^{+0.03}$ &$1.031\pm0.002$ &535.9/483 &573.9\\
    57 &$12.08_{-5.61}^{+6.11}$ &- &- &- &- &- &- &- &- &- &- &$1.10_{-0.06}^{+0.06}$ &107.8/128 &123.8\\
    58 &$6.09_{-0.77}^{+0.77}$ &- &- &- &- &$6.45_{-0.07}^{+0.07}$ &$0.18_{-0.18}^{+0.13}$ &$(7.5\pm2.5)\times10^{-5}$ &- &- &- &$1.02_{-0.01}^{+0.01}$ &303.4/296 &321.4\\

    \hline
    \end{tabular}
    }
    \begin{tablenotes}
        \item $^\star$ Frozen parameter.
    \end{tablenotes}
    \caption{Table~\ref{tab:cyc1} contd. The best fitting parameters of the absorption (\texttt{tbabs, tbpcf}), blackbody (\texttt{bbody}), iron fluorescence lines (\texttt{gaussian}) and \textit{TKF} (\texttt{gabs}) models, along with the fit statistic and \textit{AIC} score used for model selection. Details of Obs. Sn.33 to 58 are given in this table.}
    \label{tab:tkf2} 
\end{table*}

\bsp	
\label{lastpage}
\end{document}